\documentclass{jfm}

\usepackage{natbib}
\usepackage{amssymb,latexsym}
\usepackage{graphicx,color}
\usepackage{psfrag}
\def\spacce#1{\hskip #1pt}
\def\drawline#1#2{\raise 2.5pt\vbox{\hrule width #1pt height #2pt}}
\def\solid{\drawline{24}{.5}\nobreak}

\def\bdashli{\hbox{\drawline{5}{.5}\spacce{2}}}
\def\bdash{\hbox{\drawline{5.8}{.5}\spacce{2}}}

\def\dashed{\bdash\bdash\bdash\nobreak}

\def\bdot{\hbox{\drawline{1}{.5}\spacce{2}}}

\def\dotted{\hbox{\leaders\bdot\hskip 24pt}\nobreak}

\def\chndot{\hbox%
{\drawline{4.6}{.5}\spacce{2}\drawline{1}{.5}\spacce{2}\drawline{4.6}{.5}\spacce{2}\drawline{1}{.5}\spacce{2}\drawline{4.6}{.5}}\nobreak }
\def\circle{$\circ$\nobreak }
\def\linecir{\hbox%
{\drawline{8}{.5}\spacce{2}\circle\spacce{2}\drawline{8}{.5}}\nobreak}
\def\trian{\raise 1.25pt\hbox{$\scriptstyle\triangle$}\nobreak}

\def\dtrian{\raise 1.25pt\hbox%
{$\scriptscriptstyle\bigtriangledown$}\nobreak}

\def\squar{\raise 1.25pt\hbox{$\scriptstyle\Box$}\nobreak}

\def\diamon{\raise 1.25pt\hbox{$\scriptstyle\diamond$}\nobreak}

\newcommand{\soliddtrian}{$\blacktriangledown$\nobreak}

\def\dashcir{\bdashli\bdashli\circle\spacce{2}\bdashli\bdashli}
\def\dashtrian{\bdashli\bdashli\trian\spacce{2}\bdashli\bdashli}

\def\chndottrian{\hbox%
{\drawline{4.6}{.5}\spacce{2}\drawline{1}{.5}\spacce{1}\trian\spacce{1}\drawline{4.6}{.5}\spacce{2}\drawline{1}{.5}\spacce{2}\drawline{4.6}{.5}}\nobreak }

\def\linedtri1{\hbox{\bdash\hspace{-1.6mm}$\bigtriangleup$\hspace{-0.8mm}\bdash}\nobreak}
\def\soliddtrian1{$\blacktriangledown$\nobreak}
\def\solidrtrian2{$\blacktriangleright$\nobreak}
\def\solidltrian3{$\blacktriangleleft$\nobreak}
\def\dd{{\, \rm{d}}}
\def\dr{{\rm{d}}}
\def\Dr{{\rm{D}}}
\def\bra{\langle}
\def\ket{\rangle}
\def\p{\partial}
\def\ra{\rightarrow}
\def\beq{\begin{equation}}
\def\eeq{\end{equation}}
\def\la{\label}
\def\ii{{\rm i}}

\def\degree{$^{\rm o}$}

\def\hu{\widehat{u}}

\def\hchi{\widehat{\chi}}

\def\tu{\widetilde{u}}
\def\tx{\tilde{x}}
\def\tomega{\widetilde{\omega}}

\def\kvec{\mbox{\boldmath $k$}}
\def\xvec{\mbox{\boldmath $x$}}
\def\yvec{\mbox{\boldmath $y$}}
\def\phivec{\mbox{\boldmath $\phi$}}
\def\dthvec{\mbox{\boldmath $\delta\theta$}}
\def\thetavec{\mbox{\boldmath $\theta$}}

\def\uvec{\mbox{\boldmath $u$}}
\def\phimat{\mbox{\boldmath $\mathsfbi \Phi$}}
\def\Umat{\mbox{\boldmath $\mathsfbi U$}}
\def\Amat{\mbox{\boldmath $\mathsfbi A$}}
\def\Rmat{\mbox{\boldmath $\mathsfbi R$}}
\def\Imat{\mbox{\boldmath $\mathsfbi I$}}
\def\Mmat{\mbox{\boldmath $\mathsfbi M$}}
\def\Mmath{\mbox{\boldmath $\overline{\mathsfbi M}$}}
\def\phimath{\mbox{\boldmath $\bar{\mathsfbi \Phi}$}}
\def\Umath{\mbox{\boldmath $\overline{\mathsfbi U}$}}
\def\Rmath{\mbox{\boldmath $\overline{\mathsfbi R}$}}
\def\utau{u_\tau}
\def\retau{h^+}
\def\dis{\varepsilon}
\def\aaa{{\it a}}
\def\bbb{{\it b}}
\def\ccc{{\it c}}
\def\ddd{{\it d}}
\def\eee{{\it e}}
\def\fff{{\it f}}
\def\ggg{{\it g}}
\def\hhh{{\it h}}
\def\iii{{\it i}}
\def\arpath{./}
\def\r#1{(\ref{#1})}

\begin{document}
\title[Structures in wall-bounded turbulence]{Coherent structures in wall-bounded turbulence}
\author[J. Jim\'enez]{Javier Jim\'enez$^{1,2}$}
\affiliation{$^1$ School of Aeronautics, Universidad Polit\'ecnica de Madrid, 28040 Madrid, Spain\\
$^2$ Kavli Inst. Theoretical Physics, U. California Santa Barbara, Santa Barbara, CA 93106, USA 
}
\date{\today}	
\maketitle											 

\begin{abstract}
This article discusses the description of wall-bounded turbulence as a deterministic
high-dimensional dynamical system of interacting coherent structures, defined as eddies with
enough internal dynamics to behave relatively autonomously from any remaining incoherent
part of the flow. The guiding principle is that randomness is not a property, but a
methodological choice of what to ignore in the flow, and that a complete understanding of
turbulence, including the possibility of control, requires that it be kept to a minimum.
After briefly reviewing the underlying low-order statistics of flows at moderate Reynolds
numbers, the article examines what two-point statistics imply for the decomposition of the
flow into individual eddies. Intense eddies are examined next, including their temporal
evolution, and shown to satisfy many of the properties required for coherence. In
particular, it is shown that coherent structures larger than the Corrsin scale are a natural
consequence of the shear. In wall-bounded turbulence, they can be classified into coherent
dispersive waves and transient bursts. The former are found in the viscous layer near the
wall, and as very-large structures spanning the entire boundary layer. Although they are
shear-driven, these waves have enough internal structure to maintain a uniform advection
velocity. Conversely, bursts exist at all scales, are characteristic of the logarithmic layer, and
interact almost linearly with the shear. While the waves require a wall to determine their
length scale, the bursts are essentially independent from it. The article concludes with a
brief review of our present theoretical understanding of turbulent structures, and with a
list of open problems and future perspectives.
%
%
\end{abstract}
\hspace*{\fill}{\it Chance is the name we give to what we choose to ignore (Voltaire)}

\section{Introduction}\la{sec:intro}

Turbulence is often treated as a random process in which important questions are posed in
terms of statistics. In addition, whenever the equations of motion are explicitly invoked,
they are often seen as `filters' modifying the effect of random noise \citep{landaufm}. 
This
article takes the alternative view that randomness is an admission of ignorance that should
be avoided whenever possible \citep{voltaire}, and that turbulence is a dynamical system
which satisfies the Navier--Stokes equations and can be treated deterministically over
time intervals and spatial domains that, even if limited, are of theoretical and practical relevance.
Specifically, we will be interested in whether the description of the flow can be simplified
by decomposing it into `coherent' structures that can be extracted by observation and
predicted from theoretical considerations. In this sense, we continue a tradition of `eddy
chasing' that, as we shall see, has been pursued in the past few decades with as much vigour
as the purely statistical approach.
  
However, any attempt to simplify complexity has to be treated with caution, because it
usually implies neglecting something that may be important. For example, the motion of the
molecules in a gas cannot be simplified without cost. Thermodynamics follows
simple rules, but only at the expense of hiding the instantaneous motion of individual
molecules, preventing us from building `Maxwell demons'. This does not mean that
simplification should not be pursued. It may be the only way of making the system
tractable, but it should be undertaken with proper care to distinguish between what is
important for the system and what is convenient for us.

The structural view of turbulence is based on the hope that at least part of its dynamics
can be described in terms of a relatively small number of more elementary processes than the
full Navier--Stokes equations.

There are several ways of approaching this goal. Reduced-order models seek to project the
equations of motion onto a smaller set of variables that approximate the solution in some
global sense, typically a linear subspace or a few Fourier modes \citep[see, for
example,][]{row:arfm17}. The key word in this sentence is `global', and becomes less
justified as the system becomes more extended. Consider a turbulent boundary layer over a
wing, where the wall-parallel dimensions can be several hundred times the flow thickness. We
can expect to find several thousands of `largest' eddies in such systems, many of which will
be so far apart from each other as to be essentially independent. Global definitions fail in
those cases, essentially because they treat together unrelated quantities.

Our approach will rather be to acknowledge that the evolution of the flow is largely local,
and to look for solutions that are intense enough to evolve on their own, relatively
independently from other solutions far away. We will refer to these putative solutions,
different from global modes, as `coherent structures'. The two outlooks are in some ways
similar to the wave and particle representations in quantum mechanics.

It would be useful for our purpose that structures, if they exist, depend predominantly on 
other structures and on themselves, at least for some time, with relatively small contributions from
the `unstructured' background (see figure \ref{fig:concept}\aaa). Such `self' dependence
suggests several properties that structures should possess. In the first place, they should
be strong with respect to their surroundings and have some dynamics of their own,
e.g., a vortex would do, but a blob of red dye may not. They should also be relevant enough to the
flow not to be considered trivial, even if it should be noted that relevance is a subjective
property more related to the observer than to the flow, and that it depends on the
application. For example, a strong Kolmogorov-scale vortex is not very relevant to the
overall energy balance of the flow, but it might be important for the behaviour of a
premixed flame. Finally, to be useful, structures should be observable, or at least computable.

These requirements generally imply that coherent structures should either be `engines' that
extract energy from some relevant forcing, `sinks' that dissipate it, or `repositories' that
hold energy long enough to be important for the general energy budget of the flow.

It follows from the above arguments that the first questions to be addressed should be
whether structures satisfying these requirements are possible, which simplifications they
imply, and whether the parts that can be described structurally are relevant for the flow as a
whole.

\begin{figure}
\vspace*{3mm}%
\centerline{%
\includegraphics[width=.90\textwidth,clip]{\arpath 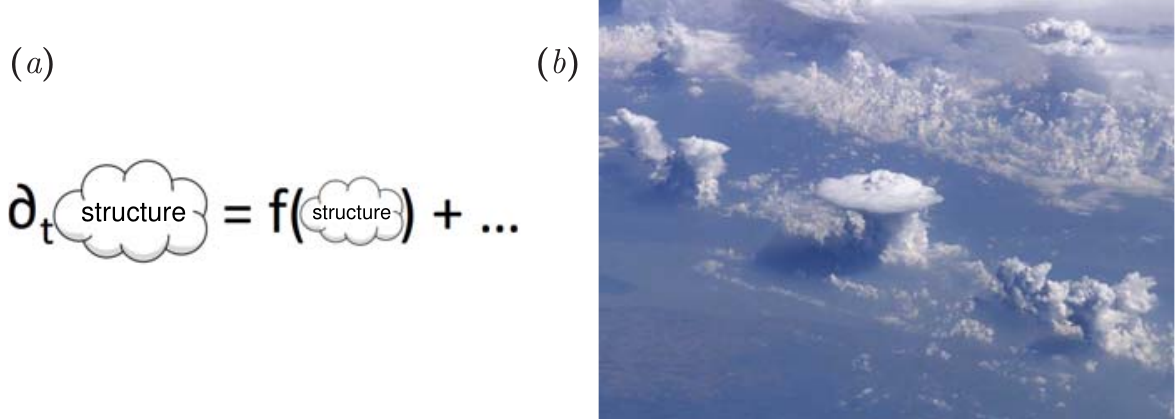}%
}%
\caption{(\aaa) Conceptual sketch of a possible definition of coherent structure. The remainder 
implied by the dots at the end of the equation is assumed to be `small'.
(\bbb)  Structures need not describe the full flow to be useful (image courtesy of NASA).
}
\la{fig:concept}
\end{figure}

In common with other models, structural descriptions should not conflict with known
evidence, but they should not necessarily be required to explain {\em all} the evidence.
They may be useful even if part of the flow, or even most of it, is structureless. For
example, the tall cumulus cloud at the centre of figure \ref{fig:concept}(\bbb) will probably
result in a strong shower. This is a useful local prediction, even if most of the rest of
the sky is clear.
    
Turbulence, and in particular wall-bounded turbulence, is a complex chaotic system with many
degrees of freedom, with some superficial similarity to the molecular description of a gas.
In fact, entropy considerations have been used to motivate the direction of the turbulence
cascade \citep{onsag49,krai71}. The evolution of the Fourier coefficients of inviscid
turbulence with a bounded set of wavenumbers, $k$, can be written as a Liouville system that
conserves volume in phase space, and which can therefore be expected to evolve towards a
maximum-entropy state in which energy is equipartitioned over the wavenumbers. In
three-dimensional flows, this would result in an energy spectrum, $E(k)\sim k^2$, in which most
of the energy is in the highest wavenumbers, simply because the surface of a sphere
increases with its radius, and Fourier shells with larger wavenumbers contain more Fourier
modes. In this view, the three-dimensional energy cascade of the viscous Navier--Stokes
equations is an attempt by the flow to fill the more numerous wavenumbers at small scales,
frustrated by the vigorous viscous dissipation at those scales.

This would appear to argue against the possibility of representing turbulence in terms of
structures that are `coherent' enough to be identified as `objects', because any amount of
organisation reduces entropy, but there are several reasons why this is not true. The first
one is that turbulence is very far from equilibrium. Even assuming statistical stationarity,
and disregarding viscosity in the inertial range, energy flows on average from its injection
at large scales to dissipation at small ones. The implied model is not so much an
equilibrium gas, but one in which heat flows from a hot wall to a cold one. A succinct
discussion of the relation between nonequilibrium systems and macroscopic structures is
\cite{prigogine:78}.

Another reason is that even equilibrium thermodynamic systems fluctuate, although the
relative magnitude of the fluctuations decreases exponentially with the number of degrees of
freedom. This number is typically large in turbulence, $N_{dof}\sim Re^{9/4}$ for a Reynolds
number $Re$ of the energy-containing eddies, but much smaller than in thermodynamics. Even
for a `high' $Re=10^5$, $N_{dof}\approx 10^{11}$ is ten or twelve orders of magnitude less
than the typical number of molecules in a gas.

But the most compelling reason is that, in the far-from-equilibrium state of \cite{kol41}
turbulence, all degrees of freedom are not equivalent and that, when speaking about structures,
we are typically only interested in a small fraction of modes. The probability of random
fluctuations is controlled by their effect on the entropy \citep{landaus}, which depends on
the number of degrees of freedom involved, but their practical significance
is linked to the energy or to the Reynolds stresses that they contain. In \cite{kol41}
turbulence, energy is associated with a relatively small number of large-scale degrees of
freedom, which are therefore relatively free to fluctuate strongly. These large-scale
fluctuations also tend to maintain coherence over long times which are at least
of the order of their eddy turnover.

We will centre on identifying and characterising such coherent structures. There are many
reasons why we may want to do so, although some of them are probably more relevant to the
kind of understanding that appeals to the human mind than to the flow dynamics. To retain
some connection with applications, and although the issue of control will not be addressed
explicitly in this paper, the question of building a `Maxwell demon' to manipulate the
energy flux in turbulence using a structural representation will always be in the background
of our discussion. The most familiar consequence of thermodynamics is that we cannot extract
work from the thermal motion of the molecules in an equilibrium system \citep{Ear:Nor:98},
but we have argued that equilibrium thermodynamics is not applicable here, and anybody who
has flown a glider in thermals, or manoeuvred a sail boat, knows that it is possible to
extract work from turbulence by taking advantage of its structures. The literature on
turbulence control is far too extensive to be reviewed here, but some idea of the flavour of
the discussions on the relevance of structures for control can be gained from
\cite{LumBro:ARFM98}, for wall-bounded flows, or \cite{ChoiEtal:08}, for free-shear ones.

Coherent structures are often found in transitional flows, where they are typically
described as arising from linear modal instabilities of the base laminar flow. This modal
origin gives rise to well-ordered patterns and wavetrains for which there is a rich and
well-developed theory. A classical description of this line of work is \cite{dra:rei:81}.
However, we are interested here in the asymptotic state of turbulence at high Reynolds
numbers, far from transition, and the ordered patterns of linear instability are soon lost
to nonlinearity and to the chaotic interaction of the large number of degrees of
freedom. Part of the goal of this paper is to inquire whether transitional structures play
any role in fully developed turbulence, in the hope that linearly unstable growth may
provide a framework on which to `hang' nonlinearity, even at the cost of considering the
flow in some smoothed or averaged sense.

Several relatively new developments help us in this task. The first one is the realisation that modal
growth is not the only possible way in which perturbations can grow linearly. When the
evolution operator is non-normal, i.e. when it cannot be expanded in a set of mutually
orthogonal eigenfunctions, even completely stable perturbations can grow substantially
before they eventually decay. The linearised Navier--Stokes equations are highly non-normal
for several reasons, and the modal instabilities of individual eigenfunctions give 
a very partial view of their behaviour. A modern account of these techniques is
\cite{schmid01}, and we will discuss this approach in more detail in \S\ref{sec:theory}.

Another modern development is the computation of fully nonlinear invariant solutions of the
Navier-Stokes equations, either permanent waves or relative periodic orbits, starting with
\cite{Nagata90}. These solutions have often been described as ``exact coherent structures''
\citep{Waleffe2001}, which they strongly resemble \citep{jim:kaw:sim:nag:shi:05}, but the
similarity is only partial. In the first place, the known solutions are typically restricted
to a single structure in a `minimal' flow unit and, although it has been shown that the
temporally chaotic flow in such minimal units shares many properties with turbulence in
extended systems \citep{jim:moi:91}, it fails in important aspects. Most crucially,
minimal units are essentially single-scale systems, which cannot reproduce the multiscale
interactions of high-Reynolds number flows. Most of the above discussion on entropic
behaviour does not carry over to minimal units and, although there is a general feeling that it
should be possible to `synthesise' large-scale turbulence from an ensemble of minimal units
of different size, the details remain unclear. Invariant solutions are typically unstable,
and are not expected to be found as such in real flows. This is probably helpful in
connection with their role as building blocks for multiscale solutions, because it prevents
the system from getting `stuck' in an attractor that is too simple to be considered
turbulent. It has often been noted that an invariant solution is a fixed point (or a
permanent orbit) in the space of possible flow configurations, and that the system could
spend a relatively large fraction of the time in its neighbourhood as it `pinballs' among
different solutions. The properties of invariant solutions could therefore be important for
the overall flow statistics \citep{Ruelle78,jimenez87,cvit:88}, even in the context of fully
developed turbulence. A modern review of this point of view is \cite{KawEtal12}.

The third modern development that will help us in our goal is the direct numerical
simulation (DNS) of turbulence, which, although initially restricted to low Reynolds numbers
\citep{Rogallo81,KMM87}, was later extended to increasingly higher values. Simulations today
span a range of Reynolds numbers that overlaps the range of most experiments, and which
often exceeds that of the experiments for which reasonably complete measurements are
possible. The main advantage of simulations is that they are `observationally perfect'. This
is, in fact, the reason for their high cost. The equations have to be simulated in full to
properly represent the flow and, although some information can be discarded during
postprocessing, it is impossible to restrict direct simulations to partial solutions. As a
consequence, simulations potentially provide the answer to `any' question, and allow us to
see the Navier--Stokes equations as a dynamical system. It is routinely possible to compute
and, up to a point, to store, temporally- and spatially-resolved sequences of
three-dimensional fields of any required variable. These sequences reside in computer
storage, and can be interrogated forward and backwards in time with any desired technique,
and as often as required. The pacing item is not any more how to obtain answers, but how to
pose questions. Turbulence research, in common with other sciences at some point in their
development, has changed from a subject driven by the need for good data, to one driven by
the need for new ideas.

Before continuing, it is useful to make explicit the distinction between eddies and structures, which
are often treated as equivalent but are conceptually very different
\citep{adr:moi:88}. In the sense used in this paper, eddies are statistical representations
of the most probable state of the flow, while structures need dynamics. Going back
to the example in figure \ref{fig:concept}(\bbb), the statistically most probable cloud in
most weather patterns is unlikely to be an active storm cumulus, but cumuli are locally very
significant.

This article is organised in three broad sections. Section \ref{sec:classical} addresses the
classical view of wall-bounded turbulence, considered independently of whether coherent
structures exist or not. Sections \ref{sec:eddies} and \ref{sec:struct} review what is meant
by structures, how they are detected, and what is the experimental evidence for their
existence, and \S\ref{sec:theory} summarises some of the models that have been developed to
explain them. Two appendices collect details of the methods of analysis, and a short initial
\S\ref{sec:lorenz} explores the interplay between coherence and chaos. Roughly
speaking, the three parts of the paper deal with what happens, how it happens, and why it
happens. None of them should be considered a full review, and readers are encouraged to
consult the original references provided. Most people will feel that important references
are missing. This is unfortunately unavoidable if an article as broad as this one is to stay
within the size limitations. I have tried to incorporate most points of view, albeit
sometimes briefly, but I am obviously biased towards the particular one that I have tried to
make explicit in this introduction. Mostly, I have been interested in inquiring how the
different strands of research are related to each other, and what facts and observations
should be taken into account by any future explanation. I also apologise for using mostly
data from our group. I have them more readily available than those of others, and the
original publications make the necessary comparisons.

\section{An example of coherence in a chaotic system}\la{sec:lorenz}

Before moving to the study of coherent structures in turbulence, it might be
useful to clarify the idea of coherence in a simpler system. The equations
\begin{eqnarray}
\dr \theta_1 /\dr t &=& {\sigma} (\theta_2-\theta_1), \nonumber\\
\dr \theta_2 /\dr t &=& ({\rho}-\theta_3) \theta_1-\theta_2,
\la{eq:lorenz}\\
\dr \theta_3 /\dr t &=& \theta_1\theta_2-{Q} \theta_3, \nonumber
\end{eqnarray}
were introduced by \cite{lorenz:63} as a model for thermal convection in a two-dimensional
box heated from below. The parameter ${\sigma}$ is the Prandtl number, ${\rho}$ is
proportional to the Rayleigh number, and ${Q}$ is related to the box aspect ratio. The
components of the state vector $\thetavec=[\theta_j],\,j=1\ldots 3$ represent, respectively,
the fluid velocity and the horizontal and vertical temperature gradients. The
solutions to \r{eq:lorenz} have been studied extensively, often for the parameters
${\sigma}=10$, ${\rho}=28$ and ${Q}=8/3$ used in this section \citep[e.g., see][\S VI, for
many of the results cited below]{BPV84}. They are chaotic, and trajectories collapse to a
quasi-two-dimensional fractal attractor of dimension approximately 2.06. A sample trajectory is
shown in figure \ref{fig:lorenz}(\aaa), and the corresponding evolution of the horizontal
temperature gradient is in figure \ref{fig:lorenz}(\bbb).

\begin{figure}
\centerline{%
\includegraphics[width=.95\textwidth,clip]{\arpath 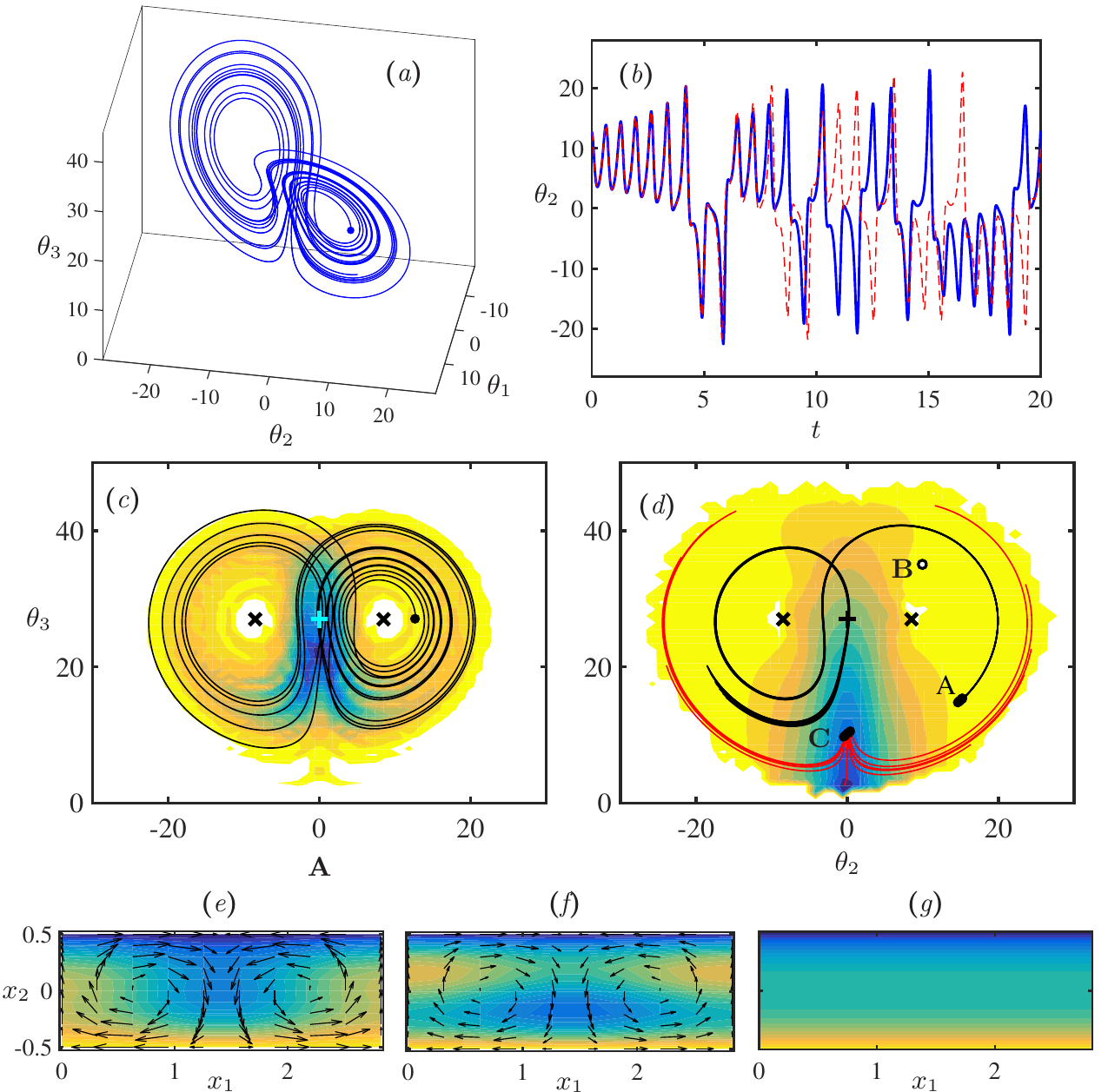}%
}%
\caption{%
Properties of the solutions of the Lorenz system \r{eq:lorenz} for ${\sigma}=10,\, {\rho}=28,\,
{Q}=8/3$.
(\aaa) Sample trajectory in the three-dimensional state space, starting from the initial
condition marked as a solid circle.
(\bbb) Temporal evolution of the horizontal velocity gradient, $\theta_2$. 
\solid, as in (\aaa); \dashed, with initial conditions displaced by $10^{-3}$ in all three variables.
(\ccc) The shaded contours are the joint p.d.f. of the two temperature gradients, ${\rm p.d.f.}(\theta_2,\theta_3)/{\rm p.d.f.}_{max}=0.033(0.066)1$, increasing from light to dark. The solid line is the initial part of the trajectory in (\aaa).
(\ddd) The shaded contours are the leading short-time Lyapunov exponent,
$L_{\tau}=0.1(1)11.1$, increasing from light to dark, with $\tau = 0.3$. The solid lines
are sets of ten trajectories each, randomly initialised within two neighbourhoods
$(|\dthvec|\le 0.5)$ centred on the attractor. The trajectories starting near point `A' barely diverge
over the time displayed $(t=1.5)$. Those starting near `C' bifurcate randomly towards
one of the foci over much shorter times  $(t=0.5)$.
(\eee-\ggg) Flow configuration for the three states marked by letters in (\ddd). The shaded background is the temperature, increasing from dark to light.  Arrows are the velocity field.  Units are arbitrary, but the scale is 
preserved among the three panels. (\eee)  Point `A', $\thetavec= [15, 15, 15]$.
(\fff) `B', $[10, 10, 35]$. (\ggg) `C', $[0, 0, 15]$.
}
\la{fig:lorenz}
\end{figure}

Figure \ref{fig:lorenz}(\bbb) includes a second simulation in which the initial conditions
have been slightly displaced with respect to the first one. Both solutions eventually drift
apart, and this characteristic sensitivity of chaotic systems to small perturbations has
often been used to argue that individual trajectories are not relevant to problems such as
turbulence, which should primarily be studied statistically \citep[see][\S 3]{pope:00}. The
objects of interest in those cases would not be trajectories such as those in figures
\ref{fig:lorenz}(\aaa,\bbb), but long-term probability density functions (p.d.f.) such as
the one in figure \ref{fig:lorenz}(\ccc).

However, inspection of figures \ref{fig:lorenz}(\aaa--\ccc) suggests a more nuanced
interpretation. There are three unstable equilibrium points: two foci at $\theta_2\approx
\pm 15$, where the p.d.f. is minimum, and a saddle at $\theta_2=0$ where it is maximum. The
trajectories evolve on two time scales: a short one, $T \approx 0.9$, over which the state
vector circles fairly regularly one of the foci, and a longer irregular one over which it
switches from one focus to the other. Only the latter behaviour is chaotic. The short-time
evolution is essentially deterministic, and can be considered coherent in the
sense discussed in the introduction, although embedded in a chaotic system which has to
be treated statistically over longer times.

This idea of coherence can be quantified. The sensitivity of the system to a linearised
infinitesimal perturbation of the initial conditions, $\dthvec(0) =[\delta \theta_j(0)]$,
can be measured over a time interval $\tau$ by the leading short-time Lyapunov exponent,
\beq
L_{\tau} = \tau^{-1} \max_{\delta\theta(0)} 
     \log\left(\|\dthvec(\tau)\|/\|\dthvec(0)\| \right), 
\la{eq:lyap1}
\eeq
where the maximum is taken over all possible orientations of $\dthvec(0)$, and which
reflects the exponential growth of the norm of the perturbation. This short-time exponent
depends on the initial conditions, and any given state can be considered predictable for
times such that $\tau L_{\tau}\lesssim 1$. In addition, an overall measure of the
predictability of the system is the global Lyapunov exponent,
\beq
L_\infty = \lim_{\tau\to \infty} L_\tau,
\la{eq:lyap2}
\eeq
which is typically independent of the initial conditions. In this global sense, any system
with a positive Lyapunov exponent should be treated statistically, at least for times longer,
on average, than $\tau\gtrsim 1/L_\infty$. It can be shown that $L_\infty\approx 0.9$
for the case in figure \ref{fig:lorenz}(\aaa), suggesting that no useful predictions can be
done for times longer than $\tau \approx 1$.

However, the global Lyapunov exponent hides a wide variation in the predictability of the
individual states of the system. This is seen in figure \ref{fig:lorenz}(\ddd), which maps
the short-time exponent, for $\tau=0.3$, as a function of the initial conditions on the
attractor. It ranges from $L_\tau\approx 0.1$ in the neighbourhood of the foci, to
$L_\tau\approx 10$ near the central saddle, implying a range of predictability times from
$\tau=10$ to $\tau=0.1$. This is confirmed by the two sets of trajectories included in
figure \ref{fig:lorenz}(\ddd). Both sets are initialised with the same initial scatter, but
the trajectories originating near the low-$L_\tau$ point `A' complete several orbits
around the foci with very little scatter, while those initialised around point `C', near
the saddle, exit randomly towards one or the other wing of the attractor.

The flow structures associated with different points of state space are displayed in figures
\ref{fig:lorenz}(\eee--\fff). The flow at the point marked as `A' in figure
\ref{fig:lorenz}(\ddd), which was shown above to be predictable, is displayed in figure
\ref{fig:lorenz}(\eee). The uniformity of the warm layer near the bottom wall has been
broken, and part of the warmer fluid is being carried upwards by the convection vortices.
The evolution of the flow is predictable because buoyancy and advection reinforce each
other. Eventually, as in figure \ref{fig:lorenz}(\fff) for point `B', advection overshoots and
carries too much warm fluid towards the upper cold plate. The convective vortices weaken and
eventually disappear near point `C' (figure \ref{fig:lorenz}\ggg), which is close to the
unstable conduction equilibrium. The indeterminacy in the location of any subsequent
instability of `C' is substituted in this simplified model by the ambiguity in the direction
of rotation of the convective vortices, which is the property that distinguishes the two
foci.

It is interesting to remark, in view of our discussion in the introduction, that the
predictable (coherent) structures at points `A' and `B' are both far from equilibrium. In
addition, given our underlying interest in flow control, it may be useful to note that,
if control were to be applied to \r{eq:lorenz}, the optimum moment would be near the
unpredictable point `C', rather than when the flow is already committed to circle one of the
foci.

\section{The mean-field theory of wall-bounded turbulence}\la{sec:classical}

We may now abandon general considerations and centre on the problem of wall-bounded
turbulence. Although this article is mainly concerned with the search for structures, we
first consider the geometric and scaling aspects of the flow, in what is usually described
as the `mean field' approximation. This is the classical view of turbulence, and defines the
framework within which structures may or may not exist. Textbook accounts are \cite{tenn},
\cite{tow:76} or \cite{pope:00}, and modern reviews can be found in \cite{SmitMcKMar11} and
\cite{jim12_arfm,jim:13b}.

We mostly use supporting data from equilibrium wall-bounded turbulent flows driven by mild
pressure gradients, such as channels and circular pipes, or from undriven
zero-pressure-gradient boundary layers, which evolve slowly downstream. The channel
half-height, the pipe radius, or the boundary layer thickness, are denoted by $h$. The
streamwise, wall-normal and spanwise coordinates and velocity components are $x_i$ and
$\tu_i$, respectively, with $i=1\ldots 3$ and $x_2=0$ at one wall. Vorticities are
$\tomega_i$, and repeated indices generally imply summation. Ensemble averages are
$\bra\cdot\ket$, usually implemented as averages over all homogeneous directions and time.
More restricted averages are distinguished by subindices. For example, the average along
$x_1$ is $\bra\cdot\ket_1$, and is a function of $x_2$, $x_3$ and time. Mean values are
denoted by capitals, $U=\bra \tu\ket$, and fluctuations with respect to these averages by
lower-case letters, as in $\tu =U+u$. Primes are reserved for root-mean square values,
$u'^2=\bra u^2\ket$. The fluid density is assumed to be constant and equal to unity, and is
dropped from all equations.

Whenever Fourier or other expansions are used, the expansion coefficients are denoted by
carats, as in $u(\xvec) = \sum_k \hu(\kvec) \exp(\ii k_j x_j)$. Wavelengths are defined from
wavenumbers, $\lambda_j=2\pi/k_j$, and spectra are often presented as spectral densities, as
in $\phi_{aa}(k_1) =k_1E_{aa}(k_1) \sim k_1 \bra|\widehat{a}|^2\ket$, or their
two-dimensional counterparts, $\phi_{aa}(k_1,k_3) =k_1 k_3E_{aa}(k_1,k_3)$. These are
proportional to the energy per unit logarithmic band of wavenumbers (or wavelengths), and
therefore give an intuitive graphical representation of the predominant wavelength of a
given quantity when displayed in a semilogarithmic plot. They are normalised so that $\bra
a^2\ket=\int E_{aa} \dd k =\int \phi_{aa} \dd (\log k)$.

We occasionally make reference to statistically stationary uniform shear turbulence
\citep{pumir:96,sek:don:jim:15}, which shares with the wall-bounded case the role of
shear as the ultimate source of energy, but without the walls. It thus allows us to distinguish
between the effects of the wall and those of the shear. Other wall-less shear flows, such as
free-shear layers or jets, are less relevant to our discussion because they extract their
energy from a Kelvin--Helmholtz modal instability of the mean velocity profile
\citep{brownr,gkw85}, which is not present in wall-bounded turbulence \citep{reytied67} or
in the uniform-shear case. 

\begin{table}
\centering
\begin{tabular}{llcccl}
Abbreviation & Flow type & $\retau$  & $L_1/h$ & $L_3/h$ & Reference \\[.5ex]
CH950 & Plane Poiseuille & 935  & $8\pi$  &  $3\pi$ & \Citet{ala:jim:zan:mos:04}\\
CH2000 & Plane Poiseuille & 2000  & $8\pi$  &  $3\pi$ & \cite{hoy:jim:06} \\
CH5200 & Plane Poiseuille & 5200  & $8\pi$  &  $3\pi$ & \cite{lee:moser:15} \\
BL6600 & Boundary layer  & 1500--2000  & $21\pi$ & $3.2\pi$ & \cite{sillero13}\\
HSF100 & Homogeneous shear & $Re_\lambda=105$  & 3 & 1 & \cite{sek:don:jim:15}\\
HSF250 & Homogeneous shear & $Re_\lambda=245$  & 3 & 1 & \cite{sek:don:jim:15}\\
\end{tabular}
\caption{Summary of the cases most often used in the article as sources of data. $L_1$ and $L_3$ are 
numerical box sizes in the streamwise and spanwise direction, respectively. In all cases, $L_2=2h$, except in
the zero-pressure-gradient boundary layer BL6600, where $h$ is the boundary layer thickness
at the middle of the box.  
}
\la{tab:cases}
\end{table}

To save repetition in figure captions, Table \ref{tab:cases} summarizes the data sets
most commonly used in the paper. For their details, the reader is directed to the original
references.

When used without subindices, $S\equiv \p_2 U_1$ is the shear of the mean velocity profile.
Although shear-driven flows are generally not isotropic, we will define an `isotropic'
velocity fluctuation intensity, $q^2=u'_i u'_i$, and an enstrophy $\omega'^2=\omega'_i
\omega'_i$. The `isotropic' dissipation rate for the fluctuating kinetic energy is
$\dis=2\nu s_{ij} s_{ij}$, where $\nu$ is the kinematic viscosity, $s_{ij}=(\p_i u_j + \p_j
u_i)/2$ is the fluctuating rate-of-strain tensor, and $\p_j=\p/\p x_j$. Spectra of the energy or
vorticity norm are defined as the sum of the spectra of the three respective components, as
in $\phi_{qq}=\sum_i \phi_{u_i u_i}\equiv \phi_{ii}$. Based on these quantities, we
define the integral length scale $L_\dis =q^3/\bra\dis\ket$, the Kolmogorov viscous length
$\eta=(\nu^3/\bra\dis\ket)^{1/4}$, and the `integral' Reynolds number, $Re_L=qL_\dis/\nu$. The
Taylor-microscale Reynolds number is $Re_\lambda= (5 Re_L/3)^{1/2}$.

Wall units are denoted by a `+' superscript. They are defined in terms of $\nu$ and of the
friction velocity $\utau$, which measures the total momentum transfer in the cross-shear
direction and can be expressed in terms of the shear at the wall, $S_w$, as $\utau^2=\nu
S_w$. Lengths expressed in these units, such as $x^+=x\utau/\nu$, are Reynolds numbers, and
the flow thickness $h^+$ is used as the characteristic Reynolds number of wall-bounded
flows. In this normalisation, $\nu^+=1$ and can be left out of the equations. Because we
will see that $q\sim \utau$, and that the largest energy-containing eddies have sizes
$O(h)$, $\retau$ is roughly equivalent to the integral Reynolds number $Re_L$. Wall units,
which depend on viscosity, are essentially the same as the Kolmogorov length and velocity
scales, which are also based on viscosity. Although the exact correspondence depends on the
flow, $\eta^+\approx 2$ at the wall, and $\eta^+\approx 0.8 (x_2^+)^{1/4}$ in the
logarithmic range of wall distances defined in the next section \citep{jim:13b}. It is
useful fact that, although not strictly equivalent, $\bra\dis\ket\approx \nu\omega'^2$
within 2\% at all wall distances, so that dissipation and enstrophy can be used
interchangeably for most purposes.

\subsection{Length scales and the classification into layers}\la{sec:scales}

\begin{figure}
\vspace*{3mm}%
\centerline{%
\includegraphics[width=.95\textwidth,clip]{\arpath 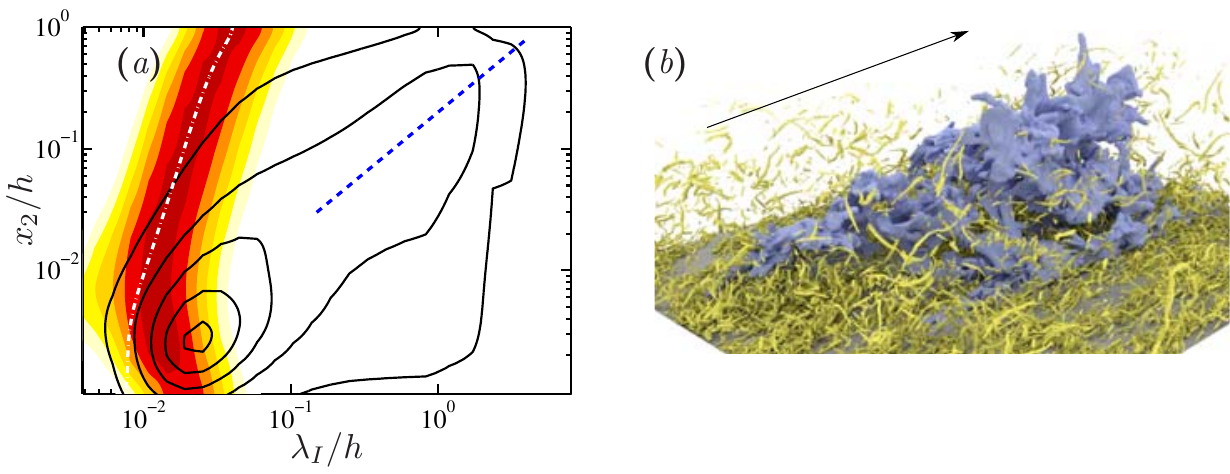}%
}%
\caption{%
(\aaa) The line contours are the premultiplied spectrum of the kinetic energy
$\phi_{qq}/\utau^2 = 0.5(1)4.5$, as a function of the wall-parallel wavelength, $\lambda_I$,
and of the distance from the wall. The shaded contours are the normalised premultiplied
spectrum of the vorticity magnitude $\nu \phi_{\omega\omega}/\dis = 0.5(1)4.5$. Channel
CH5200. \dashed, $\lambda_I=5x_2$; \chndot, $\lambda_I=25\eta$.
(\bbb) Turbulent boundary layer BL6600 at $\retau\approx 1800$. The large
central object is an isosurface of the streamwise-velocity $(u_1^+=2)$. It is approximately
$2.5h$ long, and spans most of the thickness of the layer. The smaller objects are vortices
visualised as isosurfaces of the discriminant of the velocity gradient. Flow is from left to
right (picture credits, J.A. Sillero).
}
\la{fig:specky}
\end{figure}

The length scales of turbulence range from a small limit of the order of a few viscous
Kolmogorov lengths, $\eta$, to a large one of the order of the integral length, $L_\dis$,
and it follows from the definitions in the previous section that $L_\dis/\eta = Re_L^{3/4}$.
This ratio is usually large, and there is an intermediate `inertial' range in which none of
the two scales is important, and where eddies can only be self-similar.

The best-known self-similar range is the \cite{kol41} inertial energy cascade, but more
relevant to our discussion is the logarithmic layer of wall-bounded flows. For an
equilibrium shear flow that is statistically homogeneous in the streamwise and spanwise
directions, the conservation of streamwise momentum can be written as \citep{tenn},
\beq
0=\p_t U_1 = -\p_2\bra u_1 u_2\ket -\p_1P +\nu \p_{22} U_1 = 
-\p_2\bra u_1 u_2\ket + \utau^2/h +\nu \p_{22} U_1 ,
\la{eq:mom1}
\eeq
where $P=\bra \widetilde{p}\ket$ is the ensemble-averaged kinematic pressure. 
Far enough from the wall for viscosity to be unimportant, $x^+_2\gg 1$, but close enough for
pressure gradients and other streamwise nonuniformities to be negligible, $x_2\ll h$,
\r{eq:mom1} requires that the tangential Reynolds stress satisfies $-\bra u_1 u_2\ket
\approx \utau^2 (1-x_2/h)\approx \utau^2$. This implies that the `correlated' parts of $u_1$
and $u_2$ scale with $\utau$, and suggests that the same should be true for the full
intensities, $u'_j \sim q\sim \utau$. In addition, neither the flow thickness nor viscosity
can be relevant in this range of wall distances, and there is no length scale. The result is
that structures in this intermediate region can only have a characteristic aspect ratio, but
not a characteristic size, and that the size of the largest momentum- and energy-carrying
eddies has to grow linearly with $x_2$.

It follows from these considerations that the possible structures of wall-bounded flows are
stratified in scale space by their size, and in position by their distance
from the wall, and that the flow can be approximately classified into three layers: a viscous or
buffer layer, where all eddy sizes scale in wall units; an outer layer, where the length
scale of the energy and of the energy production is the flow thickness $h$; and a scale-less
intermediate layer in which the length scale of the energy production is
proportional to $x_2$. Everywhere, the velocity scale is $\utau$, and the
dissipation length is $\eta$.

Spectra of the energy and enstrophy are presented in figure \ref{fig:specky}(\aaa) in terms
of an `isotropic' wall-parallel wavelength
\beq
\lambda_I=2\pi/k_I,\, \mbox{where}\quad k_I^2=k_1^2+k_3^2,
\la{eq:lambdai}
\eeq
which represents the size of the eddy, but neglects for the moment the possible anisotropies
of the flow. Each horizontal section of this figure is a spectral density at a given
distance from the wall and, as expected, the peak of the enstrophy spectrum is everywhere
near some small multiple of the Kolmogorov scale, $\lambda_I^+\approx 25\eta^+ \approx 20
(x_2^+)^{1/4}$. For these small scales, most of the enstrophy is concentrated near $k_1\approx
k_3$, so that $\lambda_I \approx \lambda_3/\sqrt{2}$ and the above relation is equivalent to
$\lambda_1\approx \lambda_3\approx 35\eta$ \citep{jim:13b}. On the other hand, the scale of
the energy-containing eddies grows linearly away from the wall. We will see later that
$k_1\ll k_3$ for these eddies, so that $\lambda_I\approx \lambda_3$.

The boundary between the intermediate and outer layers, conventionally taken to be
$x_2/h\approx 0.2$, is defined by the end of the linear growth of the size of the
energy-containing eddies. The transition between the intermediate and buffer layers is
defined by the level, $x_2^+\approx 80$ ($x_2\approx 0.015 h$ in figure
\ref{fig:specky}\aaa), at which the length scales of the energy and of the enstrophy become
comparable. Below this point, viscosity is important for all eddies, including the
energy-containing ones. Above it, the vortices contain very little energy, and we will see
below that they do not participate in the energy production.

A flow snapshot displaying the separation between the energy and dissipation scales is
figure \ref{fig:specky}(\bbb), where a perturbation velocity isosurface is shown together
with the much smaller vortices. Even at this moderate Reynolds number $(\retau=1800,\,
Re_\lambda \approx 110)$, the range of lengths is $L_\dis/\eta \approx 150$, and it is hard
to avoid the conclusion that the two types of eddies can only interact indirectly across an
intermediate range in which neither the small nor the large length scale are relevant. A
more extreme example is the atmospheric surface layer, where $\eta \approx 30\,\mu$m, and
$h=O(100\,$m). The two scales are then separated by a factor of $10^6$.

It is shown in appendix \ref{sec:appA} that the functional relation between two variables
can often be derived from their scaling properties. For example, the power law in the \cite{kol41}
inertial energy spectrum is a consequence of the lack of both a velocity and a length scale.
The intermediate layer in wall-bounded turbulence lacks a length scale but not a velocity
scale, and its mean velocity profile is bound to be logarithmic (see appendix \ref{sec:appA}),
\beq
U_1^+ = \kappa^{-1} \log x_2^+ + A,
\la{eq:loglaw}
\eeq
from where the layer takes its common name. The constants $A$ and $\kappa$ have to be determined
experimentally. The K\'arm\'an constant, $\kappa\approx 0.4$, reflects the dynamics of
turbulence in the logarithmic layer and is approximately universal, but $A$ is not.

The reason for the latter is that equation \r{eq:loglaw} is only a particular self-similar
solution of the equations of motion, to which other solutions tend in the range of wall
distances where boundary conditions can be approximately neglected. Typically, this
self-similar range only exists in some limiting case ($\retau\gg 1$ in wall-bounded
turbulence), outside which \r{eq:loglaw} is only an approximation that requires additional
adjustable parameters. For example, the assumptions leading to \r{eq:loglaw} do not hold
near the wall, and $A$ substitutes for the missing no-slip boundary condition. It depends on
the details of the wall and of the viscous layer, and is $A\approx 5$ for smooth walls.
Additional boundary-condition surrogates have been proposed, such as a virtual origin for
$x_2$ \citep{oberlack.01}. They improve the agreement in experiments at moderate Reynolds
numbers \citep{miz:jim:11}, but can be neglected as the Reynolds number increases. In the
case of a truly asymptotic logarithmic layer, even $A$ becomes negligible compared to $\log
x_2^+\gg 1$.

\subsection{The energy balance}\la{sec:balances}

\begin{figure}
\vspace*{3mm}%
\centerline{%
\includegraphics[width=.95\textwidth,clip]{\arpath 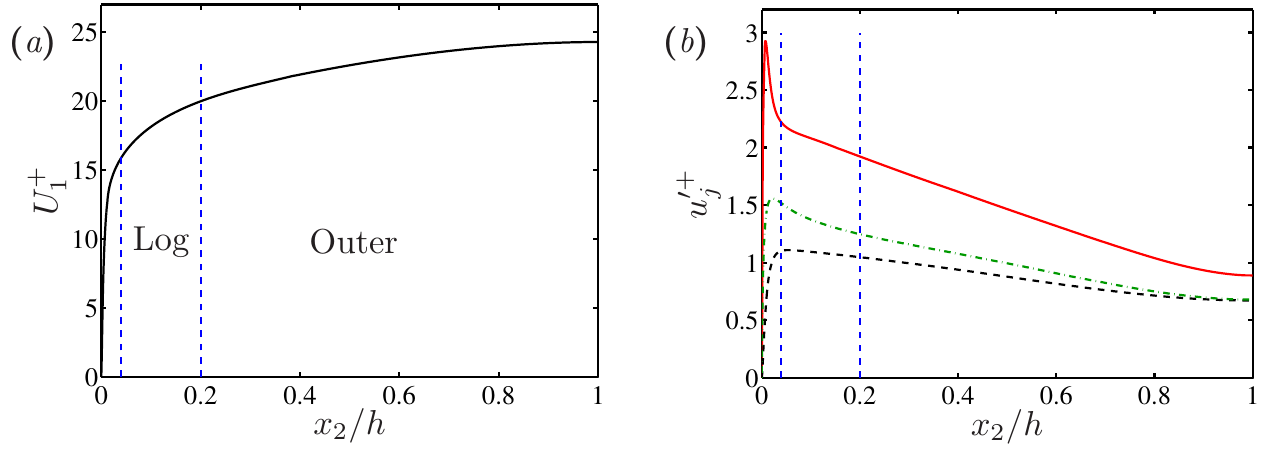}%
}%
\caption{%
(\aaa) Mean velocity profile for the channel CH2000. 
(\bbb) Velocity fluctuation intensities 
\solid, $u'_1$; \dashed, $u'_2$; \chndot, $u'_3$. The dashed vertical lines are conventional upper limits for the buffer $(x_2^+=80)$, and logarithmic  $(x_2/h=0.2)$ layers.
}
\la{fig:uprof}
\end{figure}

Even if they are relatively thin, the viscous and logarithmic layers are important for the flow as a
whole. The thickness of the buffer layer with respect to the total thickness is $80/\retau$,
which ranges from 40\% for barely turbulent flows $(\retau\approx 200)$ to $10^{-4}$ for
large water mains $(\retau\approx 5\times 10^5)$, but it follows from \r{eq:loglaw} that, even in
the latter case, 40\% of the velocity drop takes place below $x_2^+=80$ (see figure
\ref{fig:uprof}\aaa). The maximum turbulence intensity is also found in the buffer layer,
and the fluctuations decay away from the wall (see figure \ref{fig:uprof}\bbb). The mean
shear derived from \r{eq:loglaw}, $S =\utau/\kappa x_2$, which is the energy source for the
turbulence fluctuations, is also maximum near the wall.

\begin{figure}
\vspace*{3mm}%
\centerline{%
\includegraphics[width=.95\textwidth,clip]{\arpath 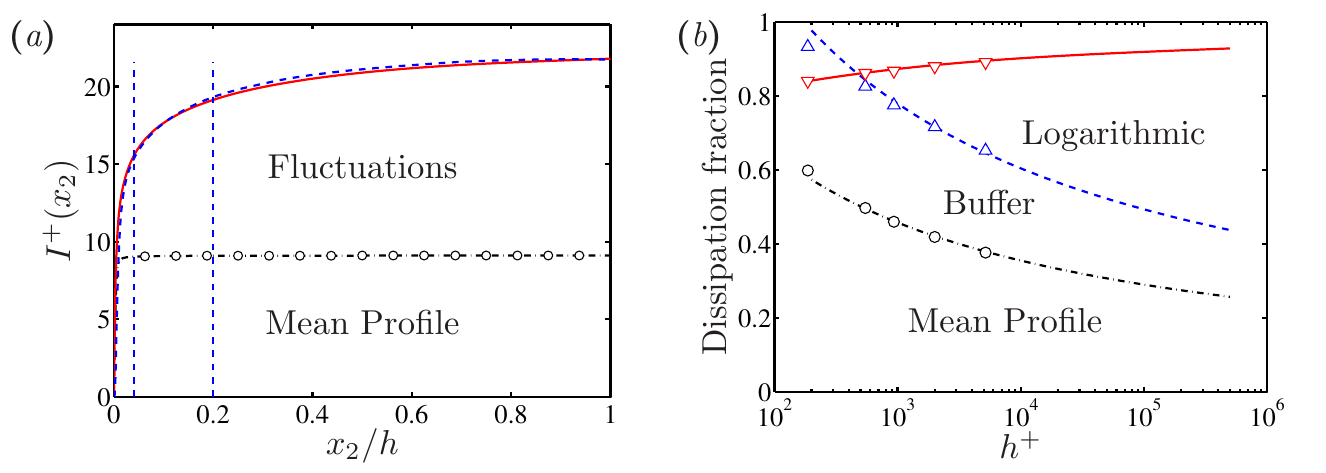}%
}%
\caption{%
(\aaa) Contribution to the energy dissipation below a given wall distance from the different
terms in the right-hand side of \r{eq:chan_enerpar}. \linecir, Dissipation due to the mean
shear; \solid, total dissipation; \dashed, total energy input on the left-hand side of
\r{eq:chan_enerpar}. The dashed vertical lines are conventional upper limits for the buffer
$(x_2^+=80)$, and logarithmic $(x_2/h=0.2)$ layers. Channel CH2000.
(\bbb) Fraction of the dissipation due to different layers in channels, versus the Reynolds
number. \circle, from the mean shear; \trian, below $x_2^+=80$; \dtrian, below $x_2/h=0.2$.
The trend lines are the semiempirical fits discussed in the text: \chndot, $9.1/U_b^+$; \dashed,
$15.5/U_b^+$; \solid, $1-2.5/U_b^+$.
}
\la{fig:chan_disip}
\end{figure}

Consider the energy balance in a turbulent channel \citep{tenn}. Energy enters
the system through the work of the pressure gradient $\p_1 P=-\utau^2/h$ on the
volumetric flux $2h U_b$, where $U_b=h^{-1}\int_0^h U_1 \dd x_2$ is the bulk velocity. This
energy input has to balance the total dissipation if the flow is to be statistically
stationary. In wall units, this is expressed as
\beq
h^+U_b^+= \int_0^{h^+} U_1^+ \dd x^+_2 = 
              \int_0^{h^+} (\dis^+ +{S^+}^2 ) \dd x_2^+,
\la{eq:chan_enerplus}
\eeq
where $\dis$ in the last integral is the `turbulent' dissipation due to the gradients of the
velocity fluctuations, and ${S^+}^2$ is the dissipation due to the effect of the viscosity
on the mean velocity profile. When the balance leading to \r{eq:chan_enerplus} is applied to
a layer stretching from the wall to $x_2$ it gives an idea of the contributions to the
dissipation from the different parts of the flow. The equation takes the form,
\beq
I^+(x_2)=-\bra u_1 u_2\ket^+ U_1^+  + \int_0^{x_2^+} U_1^+ \dd \xi = 
              \int_0^{x_2^+} (\dis^+ +{S^+}^2 ) \dd \xi +\ldots
\la{eq:chan_enerpar}
\eeq
where the extra term in the energy input in the left-hand side is the work of the tangential
Reynolds stress against the mean profile. The two terms in the right-hand side of
\r{eq:chan_enerpar} are plotted in figure \ref{fig:chan_disip}(\aaa). The small difference
between the energy input (dashed) and the dissipation (solid line) in this figure is the effect
of the small internal energy fluxes represented by the trailing dots in \r{eq:chan_enerpar}.
They are negligible at high Reynolds numbers.

In shear flows without walls, the dissipation due to the mean velocity profile is $O(\dis/Re)$, 
and can be neglected. In the wall-bounded case, figure \ref{fig:chan_disip}(\aaa)
shows that both contributions are of the same order, although the dissipation due to the
mean profile resides almost exclusively below $x_2^+=20$. Because the shear in that region
scales well in wall units, this part of the dissipation is very nearly independent of the
Reynolds number, $\int_0^h {S^+}^2 \dd x_2 \approx 9.1$. Figure \ref{fig:chan_disip}(\aaa) also
shows that a relatively large fraction of the turbulent dissipation, denoted as $\dis_{80}$,
takes place below $x_2^+=80$. The velocity gradients in this part of the flow are also
approximately independent of the Reynolds number, and $\dis_{80}^+ \approx 6.4$.

Most of the remaining dissipation takes place within the logarithmic layer, $80\nu/\utau <
x_2<0.2 h$, and can be estimated from \r{eq:loglaw}. The energy balance of the fluctuations,
averaged over wall-parallel planes, takes the form \citep{tenn}
\beq
0=\Dr_t \bra q^2/2\ket = -S\bra u_1 u_2\ket -\bra\dis\ket
 + \ldots ,
\la{eq:enertot}
\eeq
where $\Dr_t=\p_t +U_1 \p_1$ is the mean advective derivative. The trailing dots stand for
transfer terms that are not important in the logarithmic layer, where the dissipation is
almost exclusively due to $\dis$, and has to be approximately
balanced by the local energy production, $-S\bra u_1 u_2\ket$. This can be written as
\beq
\dis^+ \approx  - \bra  u_1 u_2\ket^+ S^+   \approx  (1-x_2/h) /\kappa x^+_2,
\la{eq:chan_prodplus}
\eeq
which can be integrated to 
\beq
\dis_{log}^+\approx \int_{80}^{0.2\retau} \dis^+ \dd x_2^+ \approx
  \kappa^{-1} \log \retau -15.5.
\la{eq:chan_AA}
\eeq
A similar estimate for the total energy input shows that
\beq
U_b^+ \approx  \kappa^{-1} \log \retau + 2.5
\la{eq:chan_Ubplus}
\eeq
also grows logarithmically with $\retau$, so that the remaining dissipation above the
logarithmic layer is approximately independent of the Reynolds number, $\dis_{out}^+\approx
2.5$. As a consequence, the relative contributions of the buffer and outer layers to the
dissipation decrease logarithmically as the Reynolds number increases. This is shown in
figure \ref{fig:chan_disip}(\bbb), which includes simulation results from channels at
several Reynolds numbers, and logarithmic fits based on the arguments above. In the
asymptotic limit of an `infinite' Reynolds number, most of the dissipation resides in the
logarithmic layer, but figure \ref{fig:chan_disip}(\bbb) shows that the fraction of the
dissipation due to the mean and fluctuating velocities in the buffer layer is still of the
order of 50\% of the total at the largest `realistic' Reynolds numbers, $\retau=O(10^6)$.

Because of this `singular' nature, the near-wall layer is not only important for the rest of
the flow, but it is also essentially independent from it. This was shown by
\cite{jim:pin:99} using `autonomous' simulations in which the outer flow was artificially
removed above a certain distance, $\delta$, from the wall. The dynamics of the buffer layer
was unaffected as long as $\delta^+\gtrsim 60$. The same conclusion can be drawn from the
minimal-flow experiments by \cite{jim:moi:91}, who simulated channels in numerical boxes
small enough for no large flow scales to be possible above $x_2^+ \approx 100$. Again,
the buffer layer remained essentially unaffected. Minimal flows were extended to the
logarithmic layer by \cite{flo:jim:10} with similar results: the dynamics of the
higher-shear region near the wall is essentially independent from outside influences. These
observations should not be interpreted to mean that there are no interactions between the
inner and outer layers. These interactions will be documented below, but they can mostly be
expressed as modulations or superpositions, which are not required for the maintenance of the
flow.

Understanding the structure of this near-wall region has practical implications. Energy
dissipation by turbulence is the root cause of hydrodynamic friction drag, which is
estimated to be responsible for 5\% of the total energy expenditure of humankind
\citep{jim:13b}. The energy input, $U_b^+= (2/c_f)^{1/2}$, determines the friction
coefficient $c_f$, and it follows from figure \ref{fig:chan_disip} that any attempt to
understand or to control wall friction has to take into account the buffer and logarithmic
layers.

\subsection{The scales of the energy production}\la{sec:production}

The above discussion says little about how turbulence extracts energy from the mean flow. We
saw in \r{eq:enertot} that the average energy production in a parallel shear flow is $-S
\bra u_1 u_2\ket$, which depends on the coupling of the Reynolds-stress with the mean shear.

\begin{figure}
\vspace*{3mm}%
\centerline{%
\includegraphics[width=.95\textwidth,clip]{\arpath 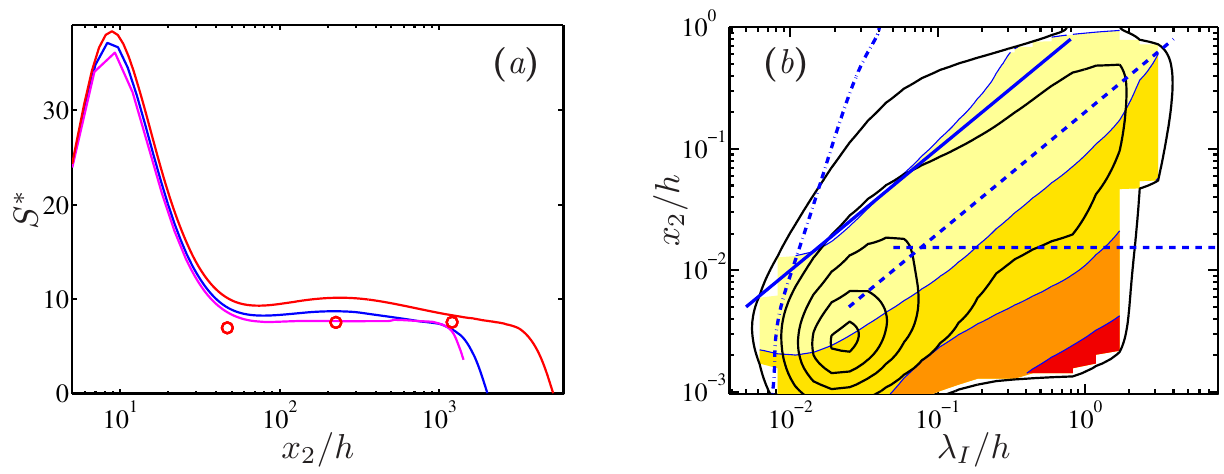}%
}%
\caption{%
(\aaa) \cite{corrsin58} integral shear parameter $S^*$. \solid, Channels and boundary layers
$\retau=2000-5200$ (various sources); \circle, statistically stationary homogeneous shear
turbulence, plotted at the equivalent $Re_\lambda$ \citep{sek:don:jim:15}.
(\bbb) The line contours are the spectral energy density of the kinetic energy, as in figure
\ref{fig:specky}(\aaa). The shaded contours are the spectral Corrsin shear
parameter $s^*(\lambda_I) = 2(\times 10) 2000$, from light to dark. Channel CH5200.
\solid, $\lambda_I=x_2$; \dashed, $\lambda_I=5 x_2$; \chndot,
$\lambda_I=25\eta$. The horizontal dashed line is $x_2^+=80$
}
\la{fig:specstar}
\end{figure}

The coupling criterion was established by \cite{corrsin58}. The turnover time over which the
nonlinear self-interaction of an eddy of size $\ell$ and characteristic velocity $u_\ell$
changes its energy is of order $\tau_{to} = \ell/u_\ell$, while its deformation by the shear
takes place in $\tau_s = S^{-1}$. The Corrsin parameter is the ratio of these times,
$s^*(\ell)=\tau_{to}/\tau_s=S\ell/u_\ell$. If $s^*(\ell)\ll 1$, the eddy evolves
independently of the shear and there is little or no energy production. This is the regime
of the inertial cascade. If $s^*(\ell)\gg 1$, the eddy is controlled by the shear, and can
extract (or lose) energy from (or to) it. This is the range of the energy production. Note
that this implies that the energy-producing eddies of a shear flow are quasi-linear, in the
sense that they are controlled by the interaction of the fluctuations with the mean flow,
with only slower nonlinear effects. The inertial cascade is fully nonlinear.

When this criterion is applied to the integral scales, where $\ell=L_\dis$ and
$u_\ell=q$, we obtain an integral Corrsin parameter, $S^*=Sq^2/\dis$, which determines
whether some part of the flow is involved in the production of turbulent energy (if
$S^*\gg 1$), or just transfers or dissipates it (if $S^*\ll 1$).

Figure \ref{fig:specstar}(\aaa) shows that equilibrium shear flows tend to have $S^*\approx
10$, at least above the buffer layer in the wall-bounded case. This is a moderately large
number that implies a quasilinear interaction of the energy-containing eddies with the mean
flow throughout the whole logarithmic layer. Linear processes do not have an intrinsic
amplitude, and the fact that $S^*$ is similar for fairly different flows suggests that the
`linear' energy production eventually saturates by means of a relatively universal mechanism that
drains its energy to the cascade, roughly equivalent to an eddy viscosity
\citep{ala:jim:06}. This recalls the engineering rule of thumb that the Reynolds
number based on the eddy viscosity of turbulent flows is always of order 10--30
\citep{tenn}. The very high value of $S^*$ in the buffer layer is not an indicator of
extremely sheared flow in the sense just discussed. The shear in this region is the highest
in the flow, but the argument of \cite{corrsin58} assumes that the balance is between
nonlinearity and shear, while the fastest evolution time near the wall is viscous. For
eddies of size $\ell$ the viscous decay time is $T_\nu=\ell^2/\nu$, and the relevant
shear parameter is $S_\nu^*=S\ell^2/\nu$. In fact, the transition of $S^*$ from its
near-wall peak to the outer plateau can be used as a convenient definition of the upper
edge of the buffer layer, and is the origin of the value used in this article, $x_2^+\approx 80$.

The spectral shear parameter for individual wavelengths, $s^*(\lambda_I,x_2)$, can be
estimated by identifying the eddy size $\ell$ with the wavelength $\lambda_I$, and the eddy
velocity scale with the spectral energy density $u_\ell = \phi_{qq}^{1/2}$. The resulting
$s^*=S\lambda_I/\phi_{qq}^{1/2}$ is overlaid in figure \ref{fig:specstar}(\bbb) to the
energy spectrum from figure \ref{fig:specky}(\aaa). It increases sharply towards the longer
wavelengths near the wall, and falls below $s^*=2$ to the left of $\lambda_I \approx L_c =
x_2$. The Corrsin length $L_c$ defined in this way represents the scale of the smallest
eddies that interact directly with the shear, and is typically a fixed small fraction of the
integral length, $L_c/L_\dis \approx (S^*)^{-3/2}$. As in figure \ref{fig:specstar}(\aaa),
the shear-dominated region below $x_2^+\approx 80$ should be interpreted as viscous,
including the highest $s^*$ in the very long near-wall region in the lower right-hand corner
of figure \ref{fig:specstar}(\bbb).

Note that, as expected, the length scale of the vorticity, $\lambda_I\approx25\eta$, is
below the Corrsin scale for $x_2^+\gtrsim 50$, equivalent to $x_2/h \approx 10^{-2}$ in the
case of figure \ref{fig:specstar}(\bbb). Except in the buffer layer, these viscous vortices
do not interact with the shear, and do not participate in the turbulence-production process.

\subsection{Anisotropy}\la{sec:anisotropy}

Although figure \ref{fig:specky}(\aaa) is drawn in terms of an isotropic wavelength and of
the kinetic energy, shear flows are not isotropic. Figure \ref{fig:uprof}(\aaa) shows that
the intensities of the three velocity components are different. The largest share of the
kinetic energy is contained in $u'^2_1$. This is especially true in the buffer layer, but
the contribution of this component is at least half of the total at all wall distances. The
other two velocity components split the rest of the energy approximately evenly, at least
far from the wall. This is most easily understood in terms of the energy equation for
individual components \citep{tenn}, although we will see later how this difference is
implemented in detail by the energy-production mechanisms. In a channel,
\beq
0=\Dr_t \bra u_1^2/2\ket = -S\bra u_1 u_2\ket +\bra p\, \p_1 u_1\ket -\nu \bra|\nabla u_1|^2\ket
 + \ldots ,
\la{eq:eneru1}
\eeq
for the streamwise component, and
\beq
0=\Dr_t \bra u_m^2/2\ket = \bra p\, \p_m u_m\ket -\nu \bra |\nabla u_m|^2\ket + \ldots,
\quad m=2,3,
\la{eq:eneru2}
\eeq
for each of the two transverse ones, where the repeated indices in the pressure term do not
imply summation. The streamwise component is special because, on average, it is the only one
that receives energy directly from the shear through the production term, $-S\bra u_1
u_2\ket$. Approximately half of this energy is dissipated to viscosity, and the rest is
transferred to the other two velocity components by the pressure-strain term, $\bra p\, \p_1
u_1\ket$. This is a redistribution term, because it follows from continuity that $\bra p\,
\p_j u_j\ket =0$, so that the net effect of the pressure on the kinetic energy vanishes. Its
effect on the velocities is approximately isotropic, and each of the two transverse
components receives roughly equal amounts of energy \citep{hoyas08}. Roughly speaking, the
kinetic energy of the streamwise velocity is twice that of the other two components because
it receives twice as much energy as they do.

The details of the distribution of the kinetic energy among the three velocity components
depend on the flow. The buffer layer is approximately universal, but the outer layers are
not. The contribution of the transverse velocities to the kinetic energy is somewhat larger
in boundary layers than in channels, most probably due to the intermittency at the
turbulent-nonturbulent interface \citep{sillero13}. In contrast, the streamwise
component is substantially stronger in Couette flow than in either boundary layers or
channels, reflecting the presence of strong persistent streamwise `rollers' which are not
found in other flows \citep{PirBerOrl14}.

\begin{figure}
\vspace*{3mm}%
\centerline{%
\includegraphics[width=.95\textwidth,clip]{\arpath 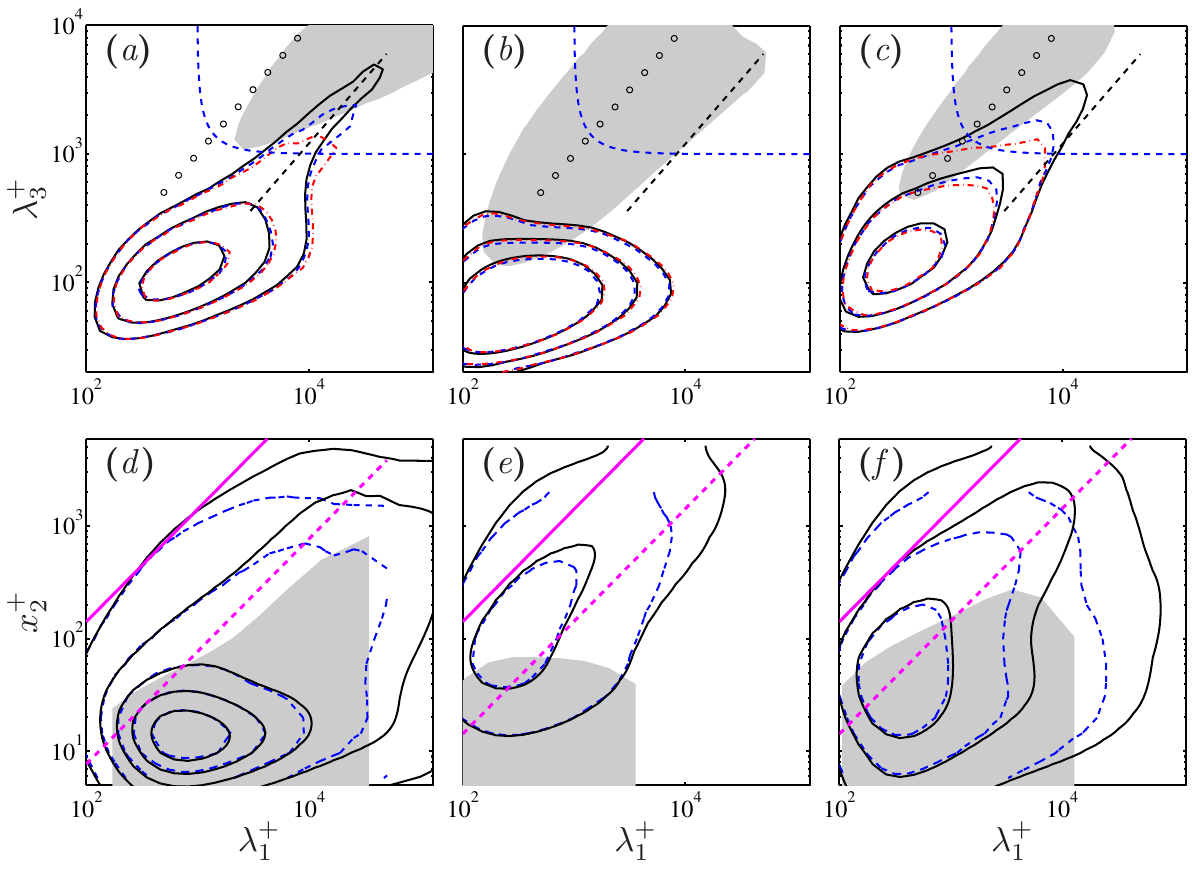}%
}%
\caption{%
(\aaa--\ccc) Two-dimensional spectral densities of the velocity components at $x_2^+=15$.
The dashed diagonal is $\lambda_1= 8 \lambda_3$, the hyperbolic curve is $\lambda_I^+=1000$,
and the circles are $\lambda_1=\lambda_3$. The grey patch is the integrated energy density,
$\Phi =(1/h)\int_0^h \phi \dd x_2$ for $\retau=5200$.
(\aaa) $\phi_{11}^+=(0.25, 0.5, 1.0)$.  $\Phi_{11}^+=0.1$
(\bbb) $\phi_{22}^+=(0.006, 0.012, 0.025)$. $\Phi_{22}^+=0.02$
(\ccc) $\phi_{33}^+=(0.05, 0.1, 0.2)$. $\Phi_{33}^+=0.05$
(\ddd--\fff\/) One-dimensional streamwise spectral densities as functions of wall distance.
The two diagonals are approximately: \solid, $\lambda_I=x_2$; \dashed, $\lambda_I=5x_2$,
converted to $\lambda_1$ by taking into account the aspect ratio of the outer-layer
spectra. 
(\ddd\/) $\phi_{11}^+=0.2 (0.4) 1.8$. The grey patch is the vertically correlated region
near the wall for $\retau=2003$, from figure \ref{fig:corrheight}(\eee).
(\eee)  As in (\ddd\/), for $\phi_{22}^+=0.1 (0.2) 0.9$.
(\fff\/)  $\phi_{33}^+=0.1 (0.2) 0.9$.
Channels: \chndot, CH950; \dashed, CH2000; \solid, CH5200.
}
\la{fig:sp2}
\end{figure}

Figures \ref{fig:sp2}(\aaa--\ccc) display wall-parallel two-dimensional spectral densities
for the three velocity components at the location of the buffer-layer peak of $u'_1$. The
streamwise-velocity spectrum in figure \ref{fig:sp2}(\aaa) has two well differentiated parts
\citep{hoy:jim:06}: a near-wall one for $\lambda^+_I\lesssim 10^3$, which scales in wall
units; and a ridge along $\lambda_1\approx 8\lambda_3$, which gets longer as $\retau$
increases, and is therefore associated with the outer flow. Figures \ref{fig:sp2}(\aaa--\ccc)
also contain an isocontour of the vertically integrated energy densities, $\Phi
=(1/h)\int_0^h \phi \dd x_2$, showing that the large-scale energy in the buffer region is
generally only associated with the long-wavelength edge of the integrated spectrum. Most of
the kinetic energy, which resides in the outer part of the flow, does not reach near the
wall.

The wall-normal structure of these spectra is displayed in figures
\ref{fig:sp2}(\ddd--\fff\/) as spanwise-integrated one-dimensional spectral densities,
plotted as functions of the distance from the wall and of the streamwise wavelength. The
first conclusion from figure \ref{fig:sp2} is that the wall-normal velocity is damped in the
neighbourhood of the wall, and that the damping is strongest for the largest eddies. There
is no large-scale $\phi_{22}$ in figure \ref{fig:sp2}(\bbb), even if the contours in this
figure are forty times weaker than those for $\phi_{11}$ in figure \ref{fig:sp2}(\aaa).
Farther from the wall, the transverse velocities in figures \ref{fig:sp2}(\eee) and
\ref{fig:sp2}(\fff\/) are weaker than $u_1$, but only by a factor of two, which is also the
ratio of their overall intensities. Moreover, the comparison of the different spectra, using
as reference the various trend lines in the figures, shows that $u_2$ and $u_3$ tend to
occur at similar scales far from the wall, suggesting that they may be part of a common structure
in that region, while $u_1$ and $u_3$ are similarly paired near it. This is also suggested
by the shaded grey patches in figures \ref{fig:sp2}(\ddd--\fff\/), which mark the depth of
the near-wall vertically coherent layer for the three velocities. This information cannot be
obtained from the spectra, and will be discussed in detail in \S\ref{sec:correl}, but the
results have been added to figures \ref{fig:sp2}(\ddd--\fff\/) to aid in their
interpretation. There is a near-wall layer in which all the velocity components are
vertically correlated, but the coherent layer of $u_2$ does not extend beyond
$\lambda_1^+\approx 5000$ and $x_2^+\approx 80$, scaling in wall units. The coherent layer
of $u_1$ and $u_3$ extends up to $\lambda_1\approx 7h$ and $x_2/h\approx 0.1 \mbox{--} 0.2$,
scaling in outer units. Beyond $\lambda_1\approx 7h$, the near-wall coherent layer of $u_3$
disappears, but $u_1$ continues to get taller until it fills most of the channel at very
long wavelengths.

The grey patches in figures \ref{fig:sp2}(\aaa--\ccc) show that the spectra of the three
velocities are very different above the buffer layer. The two transverse components are
approximately equilateral, $\lambda_1\approx\lambda_3$, but the spectrum of the streamwise
velocity is longer. Its short-wavelength edge, $\lambda_1\approx2 \lambda_3$, is only
slightly more elongated than for the transverse velocities, but its longest wavelengths are at
least ten times longer than for the two transverse velocities, and extend to the longest
dimension of the computational box. This long-wavelength behaviour will be discussed in the
next section, and suggests that, if there are coherent structures in the flow, there are at
least two kinds: very elongated `streaks' of $u_1$, and shorter structures of $(u_2, u_3)$.

\subsection{Correlations}\la{sec:correl}

\begin{figure}

\centerline{%
\includegraphics[width=.85\textwidth,clip]{\arpath 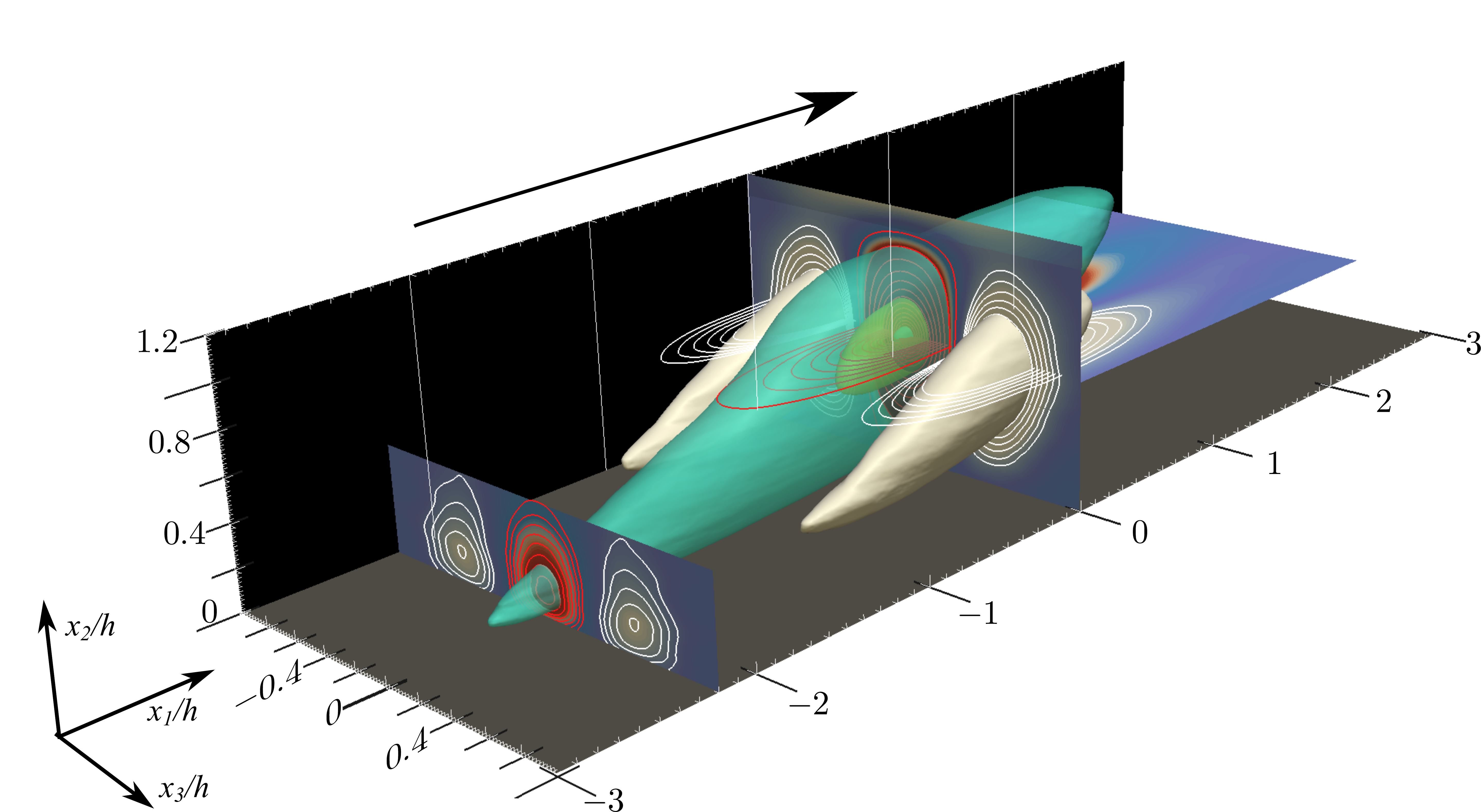}%
}%
%
\caption{%
Three-dimensional representation of the correlation function of the streamwise velocity
fluctuations, $C_{11}$, with respect to the reference point $\tx_2/h\!=\!0.6$.  Boundary layer
BL6600 at $\retau\!=\!1530$. The flow goes from left to right. Several isosurfaces are shown, with
$C_{11}\!=\!-0.09$ (white), +0.09 (turquoise), +0.4 (yellow) and +0.8 (blue). In the
different planar cross sections, the contour lines of positive and negative values are
coloured red and white, respectively. Reproduced with permission from \cite{sil:jim:mos:14}.
}
\la{fig:corr3d}
\end{figure}
 
While spectra describe the size of eddies along homogeneous directions, non-homogeneous
directions have to be analysed using two-point covariances, defined for $u_i$ as
\beq
R_{ii} (\xvec, \tilde{\xvec}) = \bra u_i(\xvec) u_i(\tilde{\xvec})\ket,
\la{eq:corr1}
\eeq
where the repeated index does not imply summation. The covariance is
symmetric in its two arguments, but we will define $\tilde{\xvec}$ as the reference point,
and $\xvec$ as the variable argument. Along homogeneous directions, $R_{ii}$ depends only on
the coordinate increment, $\Delta x_j = x_j- \tx_j$, and forms a Fourier-transform pair with the
power spectrum (see appendix \ref{sec:MCE}). For example, in channels, $R_{ii}=
R_{ii}(\Delta x_1, x_2, \tx_2, \Delta x_3)$. At the reference point, the covariance reduces
to the variance $R_{ii}(\tilde{\xvec},\tilde{\xvec})= u'^2_i(\tilde{\xvec})$, and the
dimensionless version of the covariance is the two-point correlation,
\beq
C_{ii} (\xvec, \tilde{\xvec}) = \frac{\bra u_i(\xvec) u_i(\tilde{\xvec})\ket}%
           {u'_i(\xvec)u'_i(\tilde{\xvec})},
\la{eq:corr2}
\eeq
which is unity at $\xvec=\tilde{\xvec}$. Correlations and covariances are high-dimensional
quantities. In channels, the correlation is four-dimensional. In boundary layers, where the
only homogeneous direction is the span, it is five-dimensional. This means that, except for
relatively low Reynolds numbers, it is difficult to compute and store correlations or
covariances for more than a few reference points. For a channel simulation using a moderately
sized grid with $m=1000^3=10^9$ points, the correlation is an $(m\times m)$ matrix, which
can be reduced, using homogeneity, to a block-diagonal form of $m^{2/3}=10^6$ sub-matrices
of size $(m^{1/3}\times m^{1/3})$. The total number of non-zero elements is
$m^{4/3}=10^{12}$, although the rank is only $m$. The problem of using empirical correlations is more
fundamental than a practical question of computer storage. It is shown in appendix \ref{sec:appB} that
the covariance can be represented as a matrix $\Rmat=\Umat \Umat^*$, where $\Umat$ is an
$(m\times n)$ matrix whose columns are the $n$ observations (`snapshots'), and the asterisk
denotes Hermitian transpose. It is clear from this definition that the rank of $\Rmat$
constructed in this way is at most $n$. The consequence is that the computation of the
covariance requires as many independent snapshots as degrees of freedom, which increases
with the cube of the grid diameter. Any smaller number of independent samples only provides
an approximation to the covariance. In practice, this means that covariances are typically
only used in the form of incomplete approximations of deficient rank, usually through some variant of
the method of snapshots \citep[][see appendix \ref{sec:POD}]{sirov87}.

In spite of these limitations, correlations give useful information about the
three-dimensional structure of the flow variables. Figure \ref{fig:corr3d} is an example of
$C_{11}$ for a boundary layer. It shows an inclined central positive lobe surrounded by
smaller negative ones. The cross-flow sections are approximately circular, while the
streamwise ones are elongated. Most of the data in this subsection are drawn from
\cite{sil:jim:mos:14}, which should be consulted for further details. As in spectra, it is
striking that the correlations of the different velocity components have very different
geometries. While the inclination angle of the streamwise velocity correlation in figure
\ref{fig:corr3d} is approximately 10\degree\ with respect to the wall, the correlation of
the wall-normal velocity is vertical, and that of the spanwise velocity is tilted about
30\degree. These values remain approximately constant away from the buffer and outer layers,
apply to boundary layers and channels, and are independent of the Reynolds number within the
range in which they have been studied. The correlation of the pressure fluctuations is also
vertical, broadly similar to that of $u_2$.

\begin{figure}
\centerline{%
\includegraphics[width=.98\textwidth,clip]{\arpath 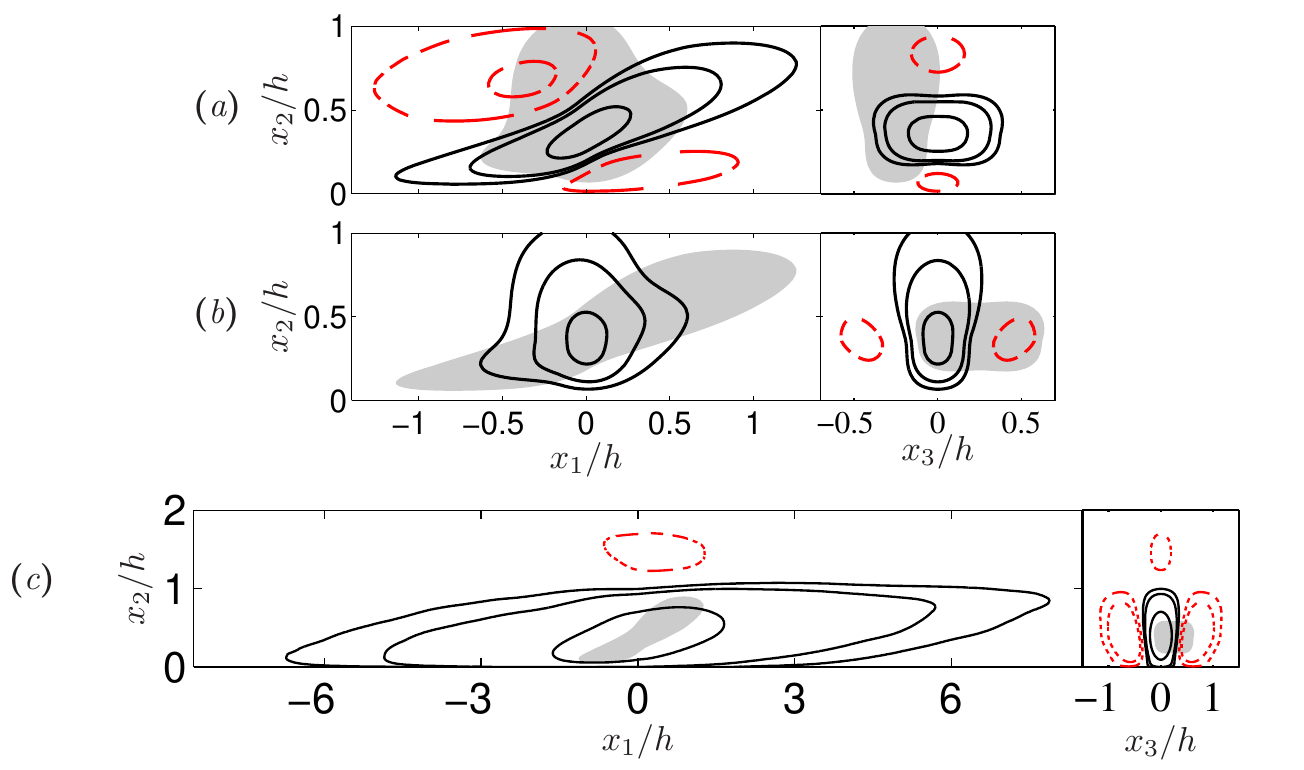}%
}%
%
\caption{%
Two-dimensional sections through the reference point $(\tx_2/h\!=\!0.35)$ of the two-point
autocorrelation function of the three velocity components. Channel CH2000.
(\aaa) Spanwise velocity. (\bbb) Wall-normal velocity. (\ccc) Streamwise velocity. Contours
are $C=[-0.1, -0.05, 0.05, 0.1, 0.3]$. Negative contours are dashed. The grey patches in
(\aaa--\bbb) are the $C=0.05$ contour of the other transverse velocity component. It has
been shifted horizontally by $0.25h$ in the cross-flow sections to suggest a possible
completion of the streamwise roller. The grey patch in (\ccc) is $C_{33}$.
}
\la{fig:corrcuts}
\end{figure}

Streamwise and cross-flow sections of the velocity autocorrelation function of the three
velocity components are given in figure \ref{fig:corrcuts} for a reference point in the
outer edge of the logarithmic region, $\tx_2/h=0.35$. This is approximately where the
correlations of $u_1$ are longest in channels. The correlations of the two transverse
velocities in figures \ref{fig:corrcuts}(\aaa,\bbb) have comparable dimensions. As mentioned
above, the correlation of the spanwise velocity in figure \ref{fig:corrcuts}(\aaa) is tilted
at approximately 30\degree\ to the wall. It has relatively strong negative lobes which are
tilted at approximately the same inclination, located above and below the main positive one.
The cross-flow section in the right-hand part of figure \ref{fig:corrcuts}(\aaa) shows that
the positive correlation lobe is flat and wide with respect to its height. Conversely, the
correlation of the wall-normal velocity in figure \ref{fig:corrcuts}(\bbb) is relatively
isotropic in the streamwise section, slightly tilted backwards, and tall and narrow in the
cross-flow plane.

It should be noted that the centres of the two correlations do not necessarily correspond to
the same location in the homogeneous wall-parallel directions. Correlations describe the
relation between the same velocity component at two points. For example, a positive $C_{22}$
means that $u_2 (x_2)$ has, on average, the same sign as $u_2 (\tx_2)$, but their common
sign can be positive or negative. Similarly, $C_{22}$ and $C_{33}$ say little about the
relative position of $u_2$ with respect to $u_3$. However, if figures
\ref{fig:corrcuts}(\aaa,\bbb) are taken together, they are consistent with
a `roller', oriented approximately streamwise and inclined at 30\degree\ to the wall. The
positive lobe of $C_{33}$ would correspond to the top or to the bottom of the roller, where
the transverse velocity is directed spanwise. Its negative lobe corresponds to the other
edge of the roller, where $u_3$ has opposite sign. The positive lobe of $C_{22}$ would
correspond to the two `side walls' closing the circuit. Grey patches of $C_{22}$ have been
added to the $C_{33}$ plot, and vice-versa, to suggest the relation implied by this model
between the two components. The two correlations have been shifted by $0.25h$ in the
spanwise direction, to make $C_{22}$ coincide with a possible lateral edge of the roller.
This offset would imply a roller diameter of approximately $0.5h$, of the same order as the
distance from the wall to the reference point, $\tx_2=0.35h$, and consistent with the
vertical distance between the positive and negative lobes of $C_{33}$. The negative lobes in
the transverse cross-section of $C_{22}$ in figure \ref{fig:corrcuts}(\bbb) are also at the right
distance and position to represent the opposite lateral wall of the roller. Note that
the symmetry of $C_{22}$ with respect to $x_3=0$ is statistical, and does not imply the
symmetry of individual eddies. The presence of both an upper and a lower negative lobe in $C_{33}$ is
also statistical. Some rollers are detected at their upper edge, and others at their lower
one.

Inclined `vortices' have often been mentioned in descriptions of boundary layer eddies
\citep{adr07}, and detected in shear flows by stochastic estimation \citep{adr:moi:88}. We
shall see later that they appear as parts of conditional Reynolds-stress structures
\citep{dong17}, but it should be stressed that the dimensions of the correlations in figures
\ref{fig:corrcuts}(\aaa--\bbb) are much larger than those of individual vortices. The diameter of the
roller implied by them is approximately 1000 wall units, or $300\eta$, and it can be shown that
these dimensions scale with $h$ as the Reynolds number changes. On the other hand, figure
\ref{fig:specky}(\aaa) shows that the size of the vorticity scales in viscous units, and is
always approximately $30\eta$.
  
The correlation of $u_1$ in figure \ref{fig:corrcuts}(\ccc) is much longer streamwise than
those of the transverse velocities, although not much wider in the cross plane. The shadow
of $C_{33}$ included in figure \ref{fig:corrcuts}(\ccc) emphasises the
relation of the respective sizes. It is interesting that, even at this relatively small
distance from the wall, there is a negative lobe of $C_{11}$ near the opposite wall. The
streamwise-velocity eddies are large enough to be `global', involving the whole channel.
Although the energy considerations in \S\ref{sec:balances} imply that the $u_1$ streak and
the cross-flow rollers have to be related, and we shall see later that they appear together
in conditional flow fields, the difference in their size makes it difficult to describe them
as parts of a single eddy. At the very least, each streak must contain several rollers.
It should also be noted at this point that, because the correlations are second-order quantities,
they favour the locations and times where the velocity is strongest, which need not 
occur simultaneously for all the velocity components.

The correlation functions for the cross-flow velocities are relatively independent of the
Reynolds number, and vary little among the flows in which they have been studied
\citep{sil:jim:mos:14}. They are good candidates for `universal' features of shear flows.
Those of $u_1$ are not, and are known with less certainty. The lengths implied by figure
\ref{fig:corrcuts}(\ccc) for channels $(\lambda_1\approx 20h)$ are, unfortunately,
comparable to the length of the computational box $(L_1=25 h)$, and vary by $\pm 20\%$ among
different simulations, including those performed within the same research group using similar codes
\citep{sil:jim:mos:14}. Experiments are also ambiguous because it is difficult to deduce
long spatial dimensions from temporal data, or to collect spatial measurements over very
long distances. Nevertheless, the lengths in figure \ref{fig:corrcuts}(\aaa) are probably
approximately correct. A simulation in a very large box $(L_1=190 h)$ was performed by
\cite{lozano14} at $\retau=550$, and did not result in longer eddies. The
length-to-width ratio, $\lambda_1/\lambda_3\approx 8$, of the spectral ridge in figure
\ref{fig:sp2}(\aaa) is consistent with $\lambda_1=16h$ for streaks of width $\lambda_3=2h$,
which is the maximum value implied by the cross-flow section in figure
\ref{fig:corrcuts}(\ccc). This width is much narrower than the spanwise dimensions of the
computational box, and there is no reason to believe that it is numerically constrained.
The shorter end of the spectral ridge in figure \ref{fig:sp2}(\aaa) is also short with
respect to the computational box, and its aspect ratio should not be numerically
constrained. The fact that it does not change with the Reynolds number, which modifies the
ratio between the length of these streaks and the size of the box, and that the aspect ratio
of the shorter streaks is similar to that of the longer ones, gives some confidence on the
results for the longer eddies.

On the other hand, the correlation length of $u_1$ in boundary layers is shorter than in
channels, $\lambda_1\approx 4h$ \citep{sil:jim:mos:14}, while that in Couette flows appears
to be longer than any of the experiments or simulations that have been performed; it might
be infinite \citep{PirBerOrl14}. The reason for these differences is unclear, although
reasonable models can be advanced in some cases. For example, very long streaks presumably
require very long times to organise (see the discussion of figure \ref{fig:tgrowthis} in
\S\ref{sec:linear}). Visually tracking the large scales in simulations of channels confirms
that they evolve extremely slowly, with evolution times of the order of many turnovers,
$t \utau/h\gg 1$. Assuming that the relevant deformation velocity is $\utau$, the time needed to
organise an eddy of length $\lambda_1$ is $\lambda_1/\utau$, during which time the flow is
advected approximately by $U_b\lambda_1/\utau \approx 30 \lambda_1$. In a uniform channel,
very long advection lengths are available, but, in a boundary layer, a streak of length
$\lambda_1=4 h$ would be advected by approximately $100 h$ during its formation. Over that
distance, the thickness of the boundary layer grows by a factor of two. In essence, streaks
in boundary layers do not have time to become very long before the mean velocity profile
changes enough to require them to adapt to a new size. A similar argument could explain
the scatter among the measured correlation lengths of different channel simulations, because
not all of them are run for the same time. The streaks that we see in channels could still
be slowly growing in some cases. This problem also applies to the development length of
laboratory experiments.

\cite{hutmar07} have proposed that real streamwise-velocity eddies may be very long,
meandering on a shorter scale. They note that the meanders hide the real length of the
streaks in spectra and correlations, and they present visual evidence of lengths of $20h$ in
boundary layers, meandering at the spectrally measured length of $6h$. Unfortunately, their
argument does not explain why channels or Couette flows should meander less than boundary
layers, and thus appear longer in the spectra. Neither do the estimates of the evolution
time in the previous paragraph explain the differences between Couette flow and channels.
The question of the real length of the large velocity streaks in wall-bounded turbulence,
and the reasons for it, remain at the moment unsettled. This question will be revisited when
discussing individual velocity structures in \S\ref{sec:streaks}.

\subsection{Filtered correlations}\la{sec:filtercorr}

\begin{figure}
\vspace*{3mm}%
\centerline{%
\includegraphics[width=.95\textwidth,clip]{\arpath 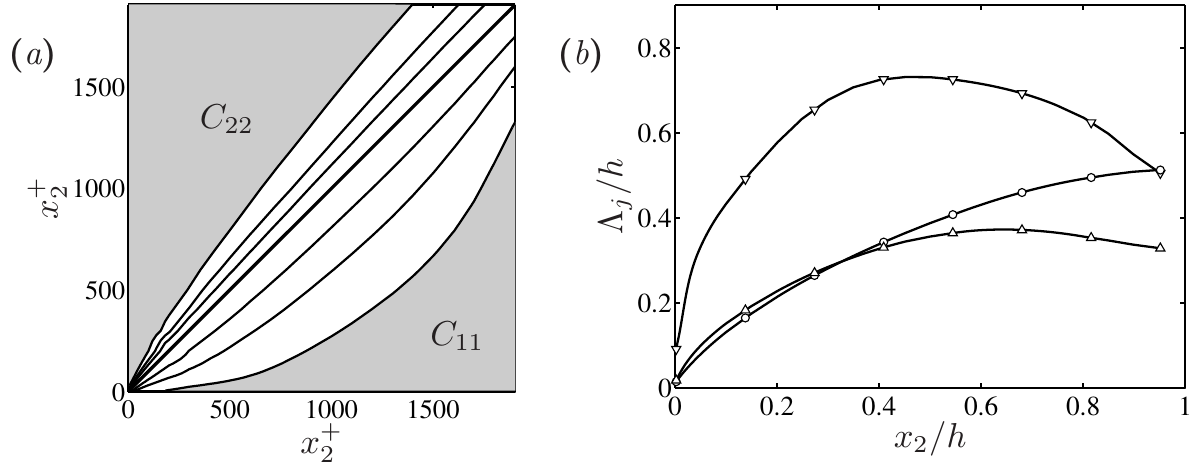}%
}%
%
\caption{%
(\aaa) One-dimensional two-point correlation, $C_{(2)}(x_2,\tx_2)$, as a function of $x_2$.
Above the diagonal, $C_{(2)22}$. Below the diagonal, $C_{(2)11}$.
Contours are $C_{(2)}=[0.3, 0.5, 0.7, 1]$.
(\bbb) Correlation depth at $C_{(2)}=0.3$ of the different velocity components, as function
of the distance from the wall to the reference point. \dtrian, $u_1$; \circle, $u_2$; \trian,
$u_3$.
Channel CH2000. 
}
\la{fig:corrunfilter}
\end{figure}

Consider next the question of defining the correlation depth of the velocities at different
distances from the wall. In channels, we can define a one-dimensional vertical correlation profile,
\beq
C_{(2)ii}(x_2, \tx_2) = \max_{\Delta x_1} C_{ii}(\Delta x_1, x_2, \tx_2),
\la{eq:1dcorr}
\eeq
where the value at $\Delta x_1=0$ is substituted by the maximum over $\Delta x_1$ to take
into account the inclination of the correlations. This quantity is represented in figure
\ref{fig:corrunfilter}(\aaa) for $u_1$ and $u_2$, using the symmetry of $C_{(2)}$ with respect
to its two arguments to include the two velocity components in the same figure.

There are many ways of defining the correlation depth, $\Lambda_j(\tx_2)$, of $u_j$ at a
given distance from the wall, but the simplest one is to measure the width at a given level
of the correlation profile centred at $\tx_2$. Figure \ref{fig:corrunfilter}(\bbb) displays
the correlation depth defined at $C_{(2)ii}=0.3$ for the three velocity components. It is
evident that $u_1$ is deeper than $u_2$ or $u_3$, and that all the velocities are more
deeply correlated away from the wall than near it. But it is difficult to extract other
conclusions from the figure. In particular, since the spectra in figures
\ref{fig:sp2}(\ddd--\fff\/) show that all the velocity components get longer as they move
away from the wall, it is unclear from figure \ref{fig:corrunfilter}(\bbb) whether the
growth in depth is due to the differences in wall distance, or to the longer eddies. An
interesting related question, when the reference point $\tx_2$ is chosen in the buffer
layer, is how far from the wall are the eddies responsible for the spectral ridge in figure
\ref{fig:sp2}(\aaa).

\begin{figure}
\centerline{%
\includegraphics[width=.98\textwidth,clip]{\arpath 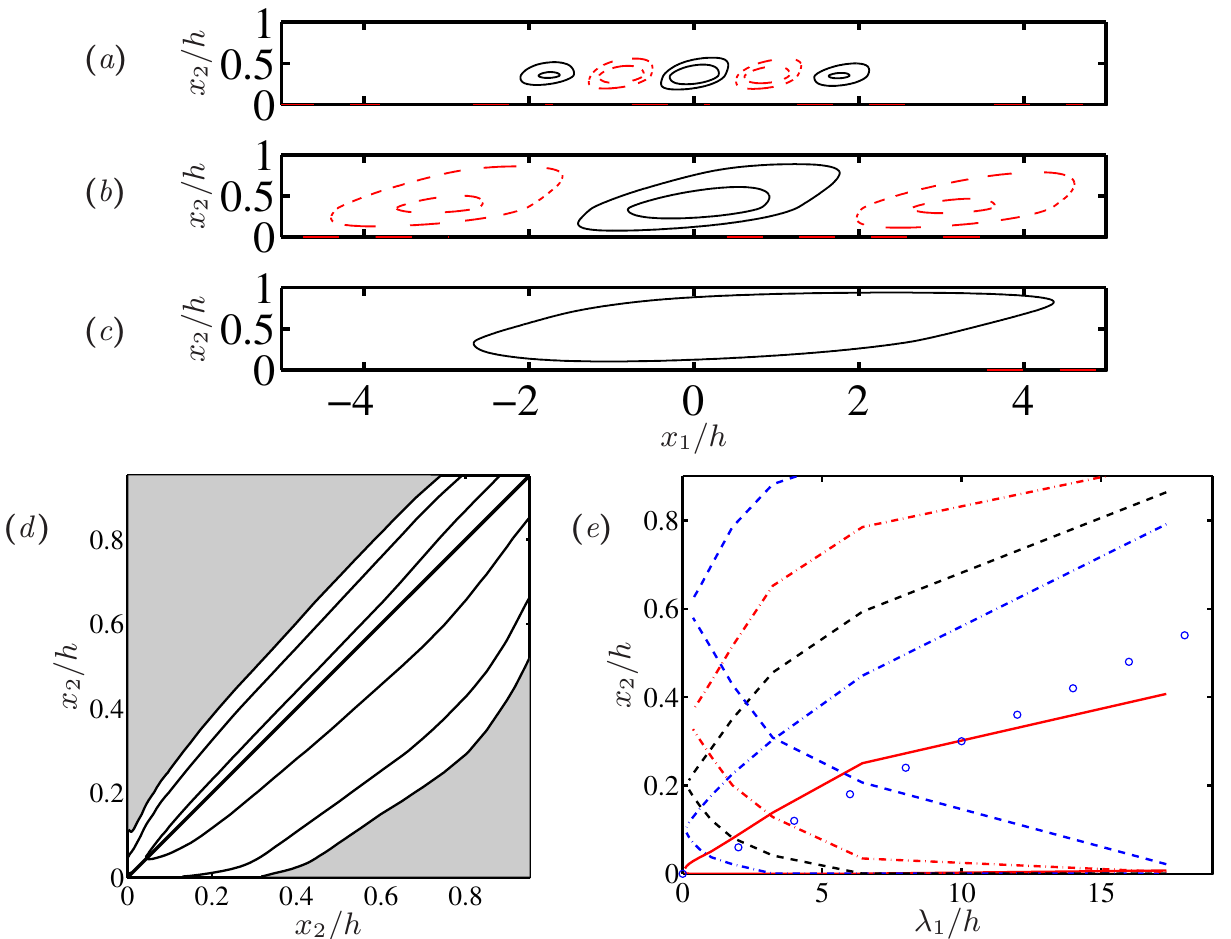}%
}%
%
\caption{%
Two-dimensional sections through the reference point $(\tx_2/h\!=\!0.35)$ of the two-point
filtered autocorrelation function of the streamwise velocity. 
(\aaa) $\lambda_1/h\in (1.4,2.2)$ (\bbb) $(5.0,8.2)$ (\ccc) $(13.6, 22.0)$. Contours
are $C_{11}=\pm [1, 2]$, and negative contours are dashed. 
(\ddd\/) One-dimensional filtered correlation, $C_{(2)11}(x_2,\tx_2;\lambda_1)$, of the
streamwise velocity, as a function of the distance from the wall. Above the diagonal, as in
(\aaa). Below the diagonal, as in (\bbb). Contours are $C_{(2)11}=[0.3, 0.5, 1]$.
(\eee) Correlation depth of $u_1$, as a function of the central filter wavelength. Each pair of lines
represents the upper and lower limits within which $C_{(2)11}>0.4$. 
 $\tx_2=[15/\retau,0.1,0.2,0.35,0.6]$, as marked by their common origin at $\lambda_1=0$.  
The circles are $x_2=0.03 \lambda_1$. Channel CH2000.
}
\la{fig:corrheight}
\end{figure}

Fortunately, the Fourier duality between covariances and spectra along homogeneous
directions allows us to filter the covariance to only include a restricted range of
wall-parallel wavelengths.

Consider a generic variable, $u$, whose covariance is $R_{uu}(\Delta x_1, x_2,
\tx_2)$. Its spectrum, $E_{uu}(k_1, x_2, \tx_2) = \bra \hu(k_1, x_2) \hu^*(k_1, \tx_2)\ket$,
is the Fourier transform of $R_{uu}$ (see appendix \ref{sec:MCE}). This is
true for any spectrum. In particular, if we apply to $\hu$ a spectral filter  $G(k_1)$ that
restricts it to some range of wavenumbers, $\overline{\hu}(k_1) = \hu(k_1) G(k_1)$, its
spectrum is similarly filtered to $E_{\bar u \bar u}(k_1, x_2, \tx_2) = E_{uu}(k_1, x_2,
\tx_2) |G(k_1)|^2$. The covariance of the filtered variable can then be obtained as the inverse
Fourier transform of the filtered spectrum. A similar procedure can be applied to $k_3$ and
to three-dimensional covariances. 

Filtered correlation functions can be defined by normalising the covariance with the
standard deviation of the filtered variables, but it is more useful to normalise them with
the total fluctuation intensities,
\beq
C_{\bar{u}\bar{u}} (\xvec, \tilde{\xvec}) = \frac{\bra \overline{u}(\xvec) \overline{u}(\tilde{\xvec})\ket}%
           {u'(x_2)u'(\tilde{x}_2)}\, (N_{bands}).
\la{eq:corrfil}
\eeq
This has the advantage of keeping some spectral information about the relative intensity of
the filtered variable in the different wavenumber ranges, but at the price of lacking a
well-defined maximum value. The value of $C_{\bar{u}\bar{u}}$ at $\xvec=\tilde{\xvec}$ is
the relative energy of the filtered variable with respect to the total. In \r{eq:corrfil},
and in the figures in the rest of this section, we have assumed that the filter is used to
separate the flow into $N_{bands}$ approximately equal logarithmic bands,
and we have multiplied the filtered correlations by $N_{bands}$ to get maxima of order
unity. Even so, the filtered correlations of the different variables reach different maximum
levels, and the threshold used to determine correlation depths has to be adjusted to
some fraction of the empirical maximum of each case.

Using these definitions, the effect of filtering the $C_{11}$ correlation in figure
\ref{fig:corrcuts}(\ccc) is shown in figure \ref{fig:corrheight}(\aaa--\ccc). The streamwise
coordinate is filtered with a family of self-similar sharp spectral box filters in which the
maximum and minimum wavelengths differ by a factor of 1.6. The central filter wavelength
increases from figure \ref{fig:corrheight}(\aaa) to \ref{fig:corrheight}(\ccc), and the
longest filter spans roughly half of the length of the computational box. It can be shown
that the streamwise average of any correlation which does not include $k_1=0$ has to vanish,
and all the filtered correlations in figure \ref{fig:corrheight} are oscillatory. As the
central wavelength is made shorter, the depth of the correlation decreases, until it
eventually separates from the wall.

This is made clearer by the correlation profile $C_{(2)11}(x_2,\tx_2;\lambda_1)$ in figure
\ref{fig:corrheight}(\ddd\/), built from the filtered correlations in figures
\ref{fig:corrheight}(\aaa,\bbb), which should be compared with the unfiltered figure
\ref{fig:corrunfilter}(\aaa). The main difference between the two figures is that the depth
of the filtered correlations in figure \ref{fig:corrheight}(\ddd\/) is more independent of
$\tx_2$ than it was in figure \ref{fig:corrunfilter}(\aaa). Moreover, different filter
wavelengths produce very different depths, answering the question of whether the correlation
depth is linked to the length of the eddies or to their distance from the wall.

Figure \ref{fig:corrheight}(\eee) displays the minimum and maximum distances from the wall at
which the filtered correlation profile exceed the arbitrary threshold $C_{(2)11}=0.4$. The
depth at short wavelengths is very small, and the maximum and minimum heights coincide. As
the central wavelength of the filter increases, so does the depth, and the lower limit
reaches the wall at some wavelength that depends on the reference height $\tx_2$. Eddies
longer than that limit `attach' to the wall and link directly the inner and outer flow
regions. The solid line starting near the origin in figure \ref{fig:corrheight}(\eee)
corresponds to the upper correlation limit of eddies centred at $\tx_2^+=15$, and marks the
depth of the correlated wall layer as a function of the streamwise wavelength. It grows
approximately linearly as $x_2 \approx \lambda_1/30$, until it saturates to $x_2\approx
0.3h$ for very long wavelengths. The slope of this line depends on the correlation level
used to define the depth, but it is clear from figure \ref{fig:corrheight}(\eee) that other
thresholds behave similarly. For short wavelengths, $\lambda_1^+\lesssim 2000$, the
correlation depth settles to about $x_2^+\approx 30$, and the vertically correlated layer
does not extend above the buffer layer. Similarly defined coherence limits for the three
velocity components have been incorporated as grey patches to the spectra in figures
\ref{fig:sp2}(\ddd--\fff\/).

It is interesting to note that a large coherence depth is a property of long eddies,
independently of whether they are attached to the wall or not. For example, the pair of
lines corresponding to $\tx_2=0.6$ in figure \ref{fig:corrheight}(\eee) define a correlation
depth spanning half of the channel for wavelengths that are much shorter than the one at
which this particular correlation attaches to the wall, $\lambda_1\approx 15h$. The message
of figure \ref{fig:corrheight}(\eee) is that long eddies are also deep. They attach to the
wall when they grow to be too deep to do otherwise, but they do not appear to originate from
the wall.

\subsection{The effect of the Reynolds number}\la{sec:attached}

\begin{figure}
\vspace*{3mm}%
\centerline{%
\includegraphics[width=.95\textwidth,clip]{\arpath 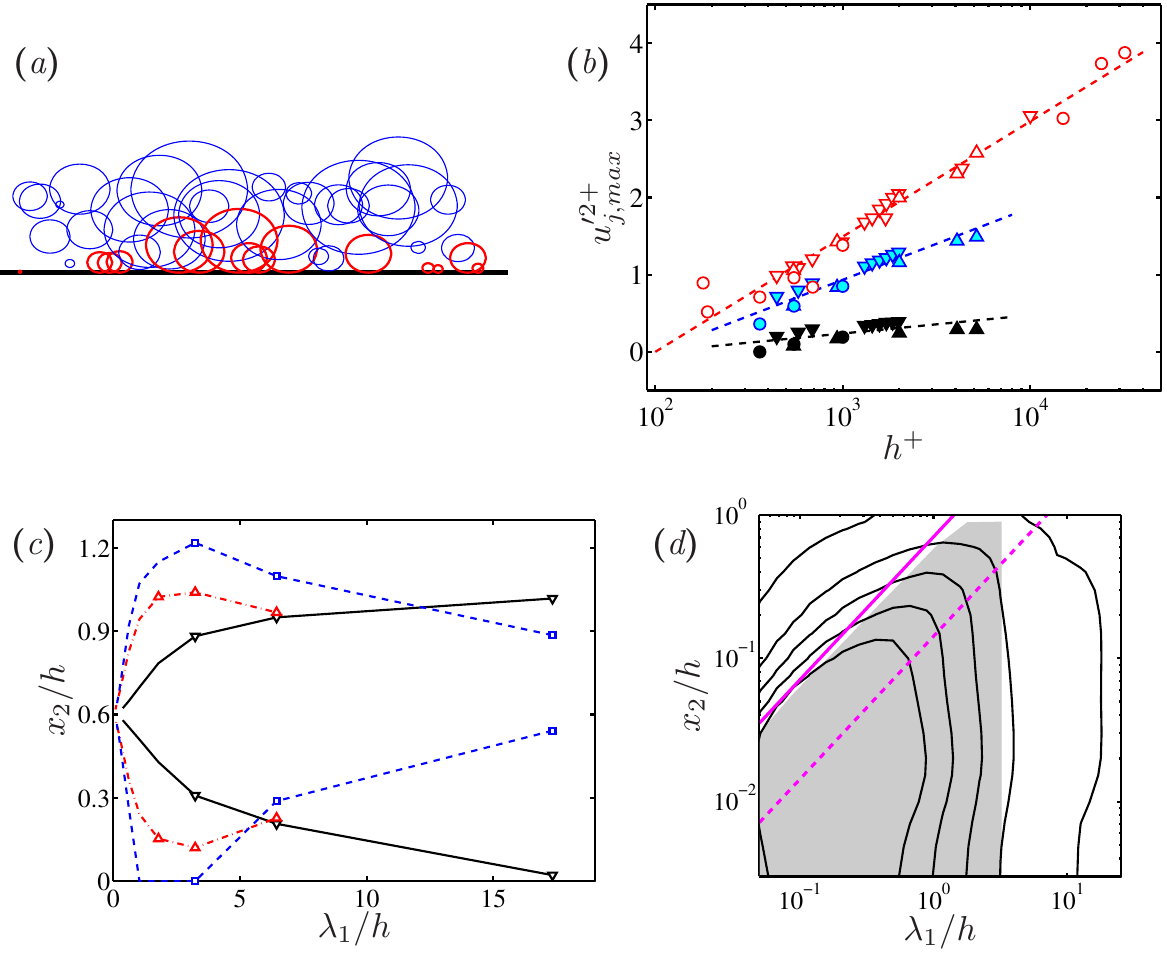}%
}%
\caption{%
(\aaa) Sketch of attached and detached eddies in a channel, created with independent random
sizes and positions. The thicker (red) objects are attached.
(\bbb) Maximum value of the velocity fluctuation intensities. Various experimental and
numerical flows: Open symbols, ${u'^2_1}^+$; closed, ${u'^2_2}^+$; light blue fill, ${u'^2_3}^+$.
\dtrian, boundary layers; \trian, channels; \circle, pipes. Each velocity component has been
offset by a different constant, so that the trend lines cross zero at $\retau=100$. The
trend lines are: 
${u'^2_1}^+=3.58+0.65\log\retau$,  
${u'^2_2}^+=0.51+0.1\log\retau$, and 
${u'^2_3}^+=-0.60+0.4\log\retau$.
(\ccc) Correlation limits for $\tx_2/h=0.6$, as functions of the central filter wavelength. \dtrian,
$C_{(2)11}=0.4$; \trian, $C_{(2)33}=0.25$; \squar, $C_{(2)pp}=0.25$. 
Each pair of lines represents the upper and lower limits within which $C_{(2)}$ exceeds the
specified level.
(\ddd) One-dimensional streamwise spectral density of the pressure, as a function of the wall
distance. The two diagonals are approximately: \solid, $\lambda_I=x_2$; \dashed, $\lambda_I=5x_2$,
reduced to $\lambda_1$ assuming $\lambda_1\approx \lambda_3$. The
grey patch is the region vertically correlated with the wall.
Channel CH2000.
}
\la{fig:uprimemax}
\end{figure}

The resulting organisation of the velocity eddies is sketched in figure
\ref{fig:uprimemax}(\aaa). Eddies of all sizes can be found at all heights, but they cannot
grow larger than their distance from the wall. Statistically, this means that the size of the
largest eddies scales linearly with the wall distance.
 
The relevance of attached eddies was first stressed by \cite{tow:61} and developed by
\cite{per:hen:cho:86}. They noted that, if the intensity of eddies
centred within the logarithmic layer scales with $\utau$, and if those eddies
retain their intensity down to the wall, the fluctuation energy near the wall should
increase with the logarithm of the Reynolds number. The argument hinges on the approximately
uniform long-wavelength spectral `skirt' of $\phi_{11}$ and $\phi_{33}$ in figures
\ref{fig:sp2}(\ddd,\fff\/). It does not apply to $\phi_{22}$, where the damping of $u_2$ by
the impermeability condition ensures that the near-wall coherent layer in figure
\ref{fig:sp2}(\eee) contains no energy at long wavelengths.

It can be shown that the longest wavelengths of the spectra in figure \ref{fig:sp2} scale
in outer units, while their short-wavelength end scales in wall units. The range of
energy-containing eddies near the wall thus scales like $\lambda_{max}/\lambda_{min} \sim
\retau$, and the total fluctuation energy is $\int \phi \dd(\log\lambda_1)\sim
\log(\retau)$. Figure \ref{fig:uprimemax}(\bbb) shows that this is true for the maxima of
$u'^{2+}_1$ and $u'^{2+}_3$, but not (or much more weakly) for $u'^{2+}_2$.

Figure \ref{fig:uprimemax}(\ccc) shows the vertical correlation limits of $u_1$ and $u_3$
for eddies centred at one particular location in the outer layer $(\tx_2=0.6h)$. The main
difference between the two variables in this figure is that, while the depth of $u_1$
increases with the wavelength over the whole range of the figure, that of $u_3$ only grows
up to $\lambda_1\approx 3h$, beyond which the energy of $u_3$ decreases, and the filtered
correlation becomes too weak to show at the selected threshold. As a consequence, the
wall-attached eddies of $u_1$ eventually fill the whole channel in figure
\ref{fig:sp2}(\ddd), but those of $u_3$ never reach above $x_2/h\approx 0.2$ in figure
\ref{fig:sp2}(\fff\/).
 
More surprising is the correlation depth of the pressure, which is included in figure
\ref{fig:uprimemax}(\ccc) and in the pressure spectrum in figure
\ref{fig:uprimemax}(\ddd\/). The pressure is an attached variable, with deep eddies that
span a large fraction of the channel, but only at the comparatively short wavelengths of the
$(u_2, u_3)$ eddies discussed in figure \ref{fig:corrcuts}. Given the irrotational character
of the longest $u_1$ eddies, whose vorticity is very low, one would expect their dynamics to
be mainly controlled by the pressure. Indeed, the pressure correlation and the pressure
spectrum are long, but the correlation in figure \ref{fig:uprimemax}(\ccc) gets thinner for
the longer wavelengths, and detaches from the wall. Figure \ref{fig:uprimemax}(\ddd\/) shows
that only a very weak vertical correlation of the pressure reaches the wall beyond
$\lambda_1\approx 5h$. The implication is that the near-wall long and wide eddies in the
spectral ridge of figure \ref{fig:sp2}(\aaa) are mostly maintained by weak but persistent
Reynolds stresses. The pressure appears to be mostly associated with the rollers discussed
in figure \ref{fig:corrcuts}.
  
Figures \ref{fig:sp2}(\ddd,\fff\/) show that arguments similar to the ones above imply
that the fluctuation profile above the buffer layer should be $u'^{2+}_j\sim \log(h/x_2)$,
for $u_1$ and $u_3$. Both predictions have been confirmed observationally: by numerical
simulations in the case of $u_3$ \citep{jim:hoy:08}, and experimentally in the case of $u_1$
\citep{mar:etal:JFMR13}.

It should be mentioned at this point that there is some controversy about the behaviour of
even basic flow statistics at very high Reynolds numbers $(\retau \gtrsim 10^4)$. This range
is still inaccessible to simulations, and hard to measure experimentally. For example, the
presence of an `outer' kinetic energy peak in the middle of the logarithmic layer has been
variously reported \citep{SmitMcKMar11} as an outer maximum of the total fluctuation
intensity, or as a peak in the spectral energy density. The two claims are different, and
both have proved hard to confirm by high-resolution experiments \citep{Ciclope17}. The
second claim, a peak in the spectral density $\phi_{11}$, could be due either to a peak in
the total energy (the first claim), or to the concentration of the same energy into a
narrower range of wavelengths. The latter would be a natural consequence of the narrowing of
the spectrum as the flow gets closer to a single scale far from the wall (see figure
\ref{fig:sp2}\eee), presumably because, as discussed in \S\ref{sec:scales}, this layer has a
natural unit of length. However, in the absence of unambiguous experimental confirmation, it
is possible that a different kind of eddies than those described here exists at high
Reynolds numbers, and that they are not captured by the simulations used in this article.

Summarising the results up to now, we have shown that the one- and two-point statistics of
wall-bounded turbulent flows suggest the existence of two types of eddies: a self-similar
family of inclined large-scale rollers that mostly involve the transverse velocity
components, and which are restricted to the logarithmic layer; and the much longer streaks of
the streamwise velocity, also self-similar in the spanwise direction, that exists at all
heights from the wall. The largest of these streaks fill most of the width of the flow. For
sizes $\lambda_I\gtrsim x_2$, both types of eddies interact directly with the shear, and are
therefore presumably involved in the energy-generation cycle, but their very different sizes
suggest that they are only indirectly related to one another. Long eddies are deep in the
wall-normal direction, but only attach to the wall if they become so large that they do not
fit into the flow otherwise. Examples of structures of the transverse velocities are given
in figures \ref{fig:condQ} and \ref{fig:condatt} below. Examples of streaks are 
figure \ref{fig:streaksall}.

\section{Eddies and coherence}\la{sec:eddies}

We noted in the introduction that our definitions of eddies and structures are conceptually
different, because the former are statistical constructs while the latter should also
include dynamics. Most of our discussion up to now applies to eddies rather than to
structures, since we have made very few references to temporal evolution. Even so, being
able to describe eddies as particularly probable states of the flow is useful, and many
techniques have been developed for doing it. A short summary of the methods used in this
review, and of their relation to one another, is in appendix \ref{sec:appB}. In this section we
collect some useful results about eddies, and start investigating their temporal evolution.
We will then be in a position to classify the parameter plane of eddy size and wall distance
in terms of where the different dynamical models are most likely to apply, although we delay
to \S\ref{sec:struct} the consideration of coherent structures.

The definition of eddies depends, to some extent, on the application for which they are
intended. As mentioned in the introduction, the question of whether the flow can be
described in terms of eddies recalls the wave-particle duality of quantum mechanics. The
problem is there how to describe `ostensibly' localised objects, such as particles, in terms
of extended fields, while our problem is how to describe an ostensibly field-like flow in
terms of localised structures. Turbulent flows are often expressed as Fourier expansions
because sines and cosines have well-defined wavelengths, and size is a crucial aspect of the
turbulence problem \citep{rich20}. However, the Fourier basis functions are spatially
homogeneous and do not describe location. Conversely, points are
perfectly localised, but have no size. A packet of several Fourier modes can be localised,
but only at the expense of some spread $\Delta k$ in its wavenumber. This is related to
the spread $\Delta x$ of its position by the uncertainty relation \citep{gasquet98}, $\Delta
k \Delta x \gtrsim O(1)$; but this is only a lower bound. Most superpositions of wavetrains
of different wavelengths have no obvious spatial structure (see figure \ref{fig:eddies} in
appendix \ref{sec:MCE}). A definition that has often been proposed for an `eddy' is a
Fourier packet for which the above inequality is satisfied as tightly as possible
\citep{tenn}. Algorithmic definitions have been given, for example, by \cite{Ber:Hol:Lum:93}, who
approximate the covariance as a superposition of randomly distributed compact eddies, or by
\cite{moi:mos:89}, who construct `most-compact' eddies by adjusting the relative phases of
several proper orthogonal (POD) modes. This last method is described in appendix
\ref{sec:MCE}, and has been used to construct the eddies in figures
\ref{fig:pod}(\aaa--\ddd).

\begin{figure}
\vspace*{7mm}%
\centerline{%
\includegraphics[width=.98\textwidth,clip]{\arpath 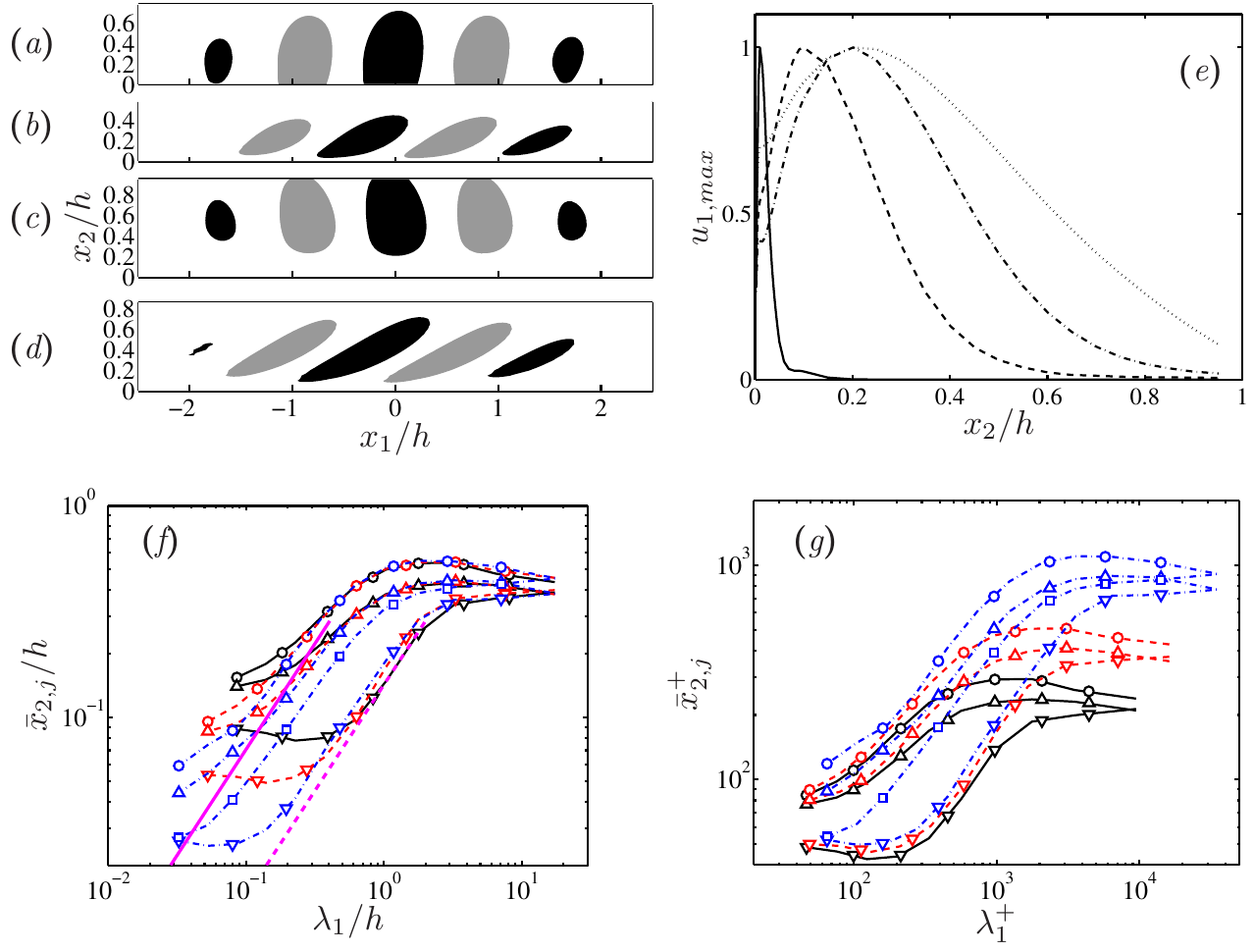}%
}%
\vspace{5mm}
\caption{%
(\aaa--\ddd) Most-compact eddies for pressure and for the individual velocity components, as
discussed in the text, normalised to unit maximum amplitude and computed from correlations
in the half-channel $x_2<h$, filtered in the band $\lambda_1/h=1.4\mbox{--} 2.2$. Black
patches are $u_{j,mc}>0.5$, and grey ones are $u_{j,mc}<-0.5$. Channel CH2000. (\aaa) $p$.
(\bbb) $u_1$. (\ccc) $u_2$. (\ddd) $u_3$.
(\eee) Profiles of the maximum eddy amplitude for $u_1$, as functions of the wall distance.
Filter bands centred at: \solid, $\lambda_1^+=280$; \dashed, $\lambda_1/h=1.0$; \chndot,
$\lambda_1/h=1.8$; \dotted, $\lambda_1/h=6.5$. 
(\fff) Height of the centre of gravity of the compact eddies computed from the leading PODs, in outer
scaling. The two thicker diagonals are approximately: \solid, $\lambda_I=\bar{x}_{2,j}$;
\dashed, $\lambda_I=5\bar{x}_{2,j}$, reduced to $\lambda_1$ assuming $\lambda_1\approx
\lambda_3$. Channels: \solid, $\retau=550$ \citep{ala:jim:zan:mos:04}; \dashed, CH950; \chndot,
CH2000. \squar, $p$; \dtrian, $u_1$; \circle, $u_2$; \trian, $u_3$.
(\ggg) As in (\fff), in wall units. 
}
\la{fig:pod}
\end{figure}

POD modes are optimal basis functions, $\phivec_{(\alpha)}$, which provide an
expansion of the flow that approximates the covariance with as few degrees of freedom as possible,
\beq
\uvec(\xvec) = \sum_\alpha \widehat{\uvec}_{(\alpha)} \phivec_{(\alpha)}(\xvec).  
\la{eq:most0}
\eeq
Unfortunately, it is shown in appendix \ref{sec:POD} that the PODs are Fourier modes along
homogeneous coordinate directions,
\beq
\phivec_{(\alpha,\kvec)}(\xvec)= \widehat{\phivec}_{(\alpha)}(x_2) \exp[\ii(k_1 x_1+k_3 x_3)],  
\la{eq:most01}
\eeq
and are therefore poor representations of coherent structures.

The POD modes are eigenfunctions of the two-point covariance. The eigenvalue of the
eigenfunction \r{eq:most01} measures the average contribution to the variance of the
corresponding term of the expansion \r{eq:most0}, and it is typically true that the first
few eigenvalues account for most of the variance. The eddies in figures
\ref{fig:pod}(\aaa--\ddd) are constructed using the leading (i.e. highest eigenvalue) PODs,
$\phi_{(1)j}$, of the filtered correlations in figure \ref{fig:corrheight}. For this case,
the most-compact eddy of $u_1$ contains about 30\% of the total energy of the filtered
$u_1$. Those of $u_2$ and $u_3$ contain about 45\% in both cases. The corresponding
percentages for the sum of the first five PODs are 80\% for $u_1$, and 90\% for the two
transverse components. These percentages increase for filters with longer wavelengths, and
decrease for shorter ones.

The most-compact eddies do not form an orthogonal basis, as the PODs do, nor are they
optimum in the sense of generating a most efficient expansion, because they are sums of
eigenfunctions that are only optimal for a single wavenumber. Therefore, they are not very
useful for constructing global reduced-order models of the flow. They represent the most
localised eddies whose covariance approximates the two-point covariance of the flow, and
they are useful in constructing the local models discussed in the introduction as
one of the reasons to consider coherent structures. Their streamwise shape depends on the
window used to filter the covariance from which they are obtained. The eddies in figures
\ref{fig:pod}(\aaa--\ddd) could be made to contain fewer side-bands by choosing a wider
filter, but only at the expense of making their vertical structure less representative of
the flow, because they would mix PODs with more diverse vertical structure. Conversely,
they could be made more representative of a single POD by choosing a narrower filter, but
only at the expense of increasing the number of side bands. Filters such as those used in
figure \ref{fig:pod}(\aaa--\ddd), in which the wavenumber interval is approximately half the
central wavenumber are probably a good compromise in most cases.

Figure \ref{fig:pod}(\eee) shows the wall-normal structure of the maximum over $x_1$, 
\beq
u_{1,max}(x_2; \lambda_1) = \max_{x_1} u_{1,mc}(x_1, x_2; \lambda_1),
\la{eq:most1}
\eeq
of the most-compact eddies, $u_{1,mc}$. They are constructed from correlations which have
been filtered over bands with various central wavelengths and uniform relative width in
Fourier space $(\lambda_{1,max}/\lambda_{1,min}\approx 1.6)$. The profile defined
in \r{eq:most1} plays the same role as the one-dimensional correlation profile defined in
\r{eq:1dcorr} and, as in that case, longer wavelengths are vertically deeper. This is
quantified in figures \ref{fig:pod}(\fff,\ggg) by the centre of gravity of the wall-normal
distribution,
\beq
\bar{x}_{2,j} =\frac{\int_0^h x_2 u_{j,max} \dd x_2}{\int_0^h u_{j,max} \dd x_2}.
\la{eq:correh}
\eeq
Figure \ref{fig:pod} shows that eddies separate into a short family (for $u_2$, $u_3$ and
$p$) in which $ \lambda_I\approx\bar{x}_{2,j}$, and a longer one for $u_1$ in which
$\lambda_I\approx 5\bar{x}_{2,1}$. These two trend lines, plotted either in terms of
$\lambda_1$ or of $\lambda_I$, have been incorporated in all the figures containing spectra
in this article, for reference. The shorter one, $\lambda_I\approx \bar{x}_2$, is the
Corrsin length, below which the flow decouples from the energy-containing eddies. The
longer one coincides with the approximate position of the core of the $\phi_{11}$ energy
spectrum. Most of the energy of the two transverse velocity components is contained between
these two lines. Figures \ref{fig:pod}(\fff\/) and \ref{fig:pod}(\ggg) show that short and
shallow eddies scale in wall units, while long and tall ones scale in outer units. Although
not shown in the figure, the eddies of $u_1$ and $p$ remain attached to the wall at all
wavelengths (see figure \ref{fig:pod}\eee), those of $u_2$ have negligible energy near the
wall and are always detached, and those of $u_3$ detach for $\lambda_1/h\gtrsim 2$.

\subsection{Advection velocities}\la{sec:advect}

We have not discussed up to now the temporal evolution of eddies, but whether they can be
considered coherent or not depends on whether they are able to keep their shape for 
dynamically significant times. This depends, among other things, on how their propagation
velocity changes with the wavenumber and with the location along inhomogeneous directions.

Consider a variable $\chi$ that we wish to approximate as a wave with phase velocity
$c_\chi$ along the direction $x_1$. A simple definition of the phase velocity was introduced by
\cite{jcadvel} by minimising $\bra (\p_t \chi+c_\chi\p_1 \chi)^2\ket$. The result is
\beq
c_\chi=-\frac{\bra\p_t \chi\p_1 \chi\ket}{\bra (\p_1 \chi)^2\ket},
\la{eq:advel1}
\eeq
which can be expressed as
\beq
c_\chi(k)=-\frac{\Im \bra k_1 \hchi^* \p_t \hchi\ket}{\bra k_1^2 |\hchi|^2\ket} 
\la{eq:advel2}
\eeq
for individual Fourier modes, where the asterisk stands for complex conjugation, and $\Im$
for the imaginary part.

Denoting by $c_j$ the advection velocity of the $u_j$ velocity component, the spectral phase
velocity of the streamwise velocity component, $c_{1}$, is shown in figure
\ref{fig:advelvsxz} at three heights in the channel, normalised with the mean flow velocity
at each height. \Citet{jcadvel} showed that the phase velocity agrees with the local
velocity for most wavelengths, except for very large scales, where $c_1\approx U_b$
independently of $x_2$, and in the buffer layer, where $c_1^+\approx 11$. Figures
\ref{fig:advelvsxz} shows that this remains approximately true in the present case, at a
higher Reynolds number than in \cite{jcadvel}, and that the scale dependence is reasonably well
described as a function of the wall-parallel isotropic wavelength $\lambda_I$.

Figure \ref{fig:advelvsy}(\aaa) summarises the wall-normal dependence of $c_1$, and shows
that the transition to the large-scale behaviour in which the eddies do not follow the flow
takes places at $\lambda_I\approx 5 x_2$ above $x_2^+\approx 100$. Closer to the wall, the
transition wavelength saturates at $\lambda_I^+\approx 10^3$, which is also the boundary in figure
\ref{fig:sp2}(\aaa) between the near-wall spectral energy component and the self-similar
outer ridge. A similar behaviour is found for the other two velocity components, although
there is very little large-scale energy in the case of $u_2$.

\begin{figure}
\centerline{%
\includegraphics[width=.98\textwidth,clip]{\arpath 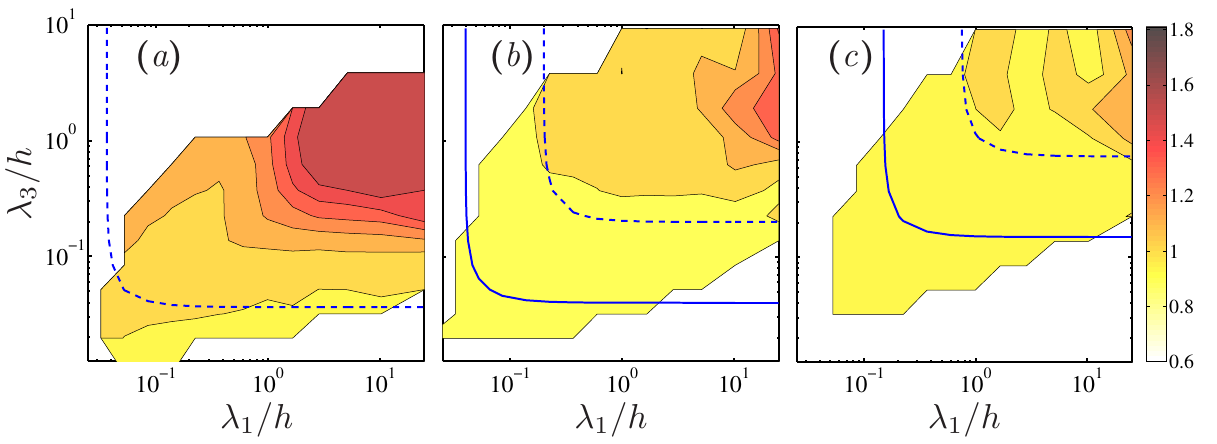}%
}%
\caption{%
Phase velocity for $u_1$ normalised with the mean velocity at each height,
$c_{1}/U_1(x_2)$.
Channel CH2000. Contours are spaced by 0.1, drawn only where the spectra density $\phi_{11}$
is greater than 2\% of its maximum. (\aaa) $x_2^+=15$. (\bbb) $x_2^+=80$; (\ccc)
$x_2/h=0.15\, (x_2^+=300)$. The hyperbola-like thicker lines are: \solid, $\lambda_I=x_2$;
\dashed, $\lambda_I=5x_2$.
}
\la{fig:advelvsxz}
\end{figure}

\begin{figure}
\vspace*{3mm}%
\centerline{%
\includegraphics[width=.98\textwidth,clip]{\arpath 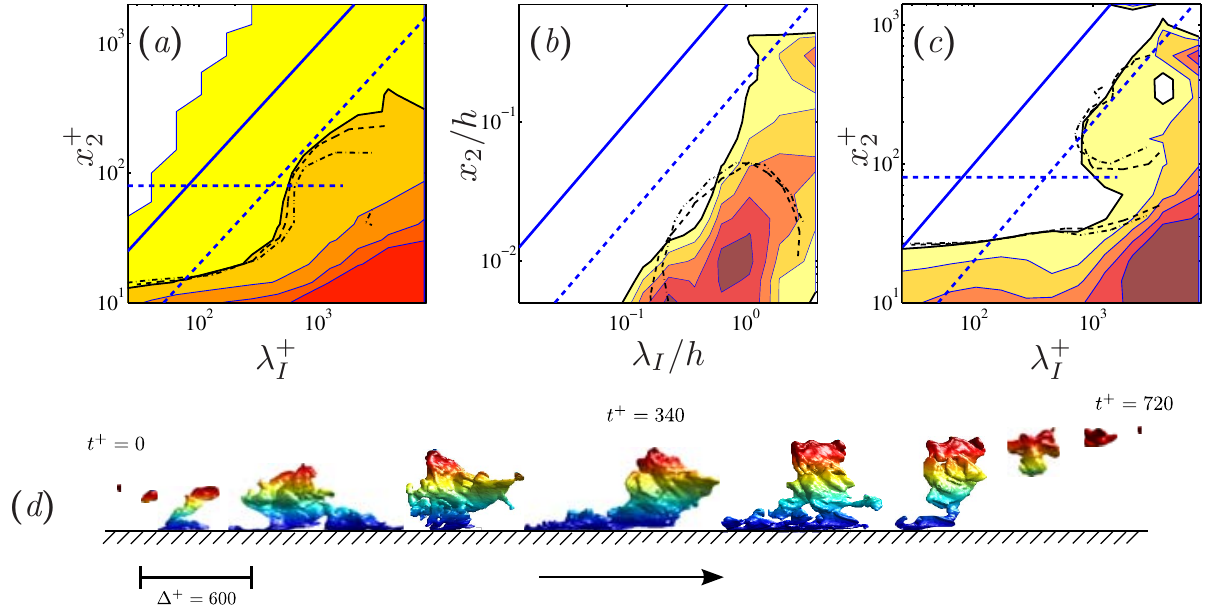}%
}%
\caption{%
Advection velocity parameters for $u_1$ in channels, as a function of $x_2$, averaged over
shells of constant $\lambda_I$. Shaded contours are CH2000. The thicker lines are:
\chndot, $\retau=550$ \citep{ala:jim:zan:mos:04}; \dashed, CH950; \solid, CH2000.
(\aaa) Phase velocity normalised with the mean flow velocity. Shaded contours are 
$c_1/U_1(x_2)=0.9(0.1)1.4$, and thick lines are  $c_1/U_1(x_2)=1$.
(\bbb) Wavenumber dispersion parameter, as defined in \r{eq:cstar}. 
Shaded contours are $c^*=1.0(0.3)2.2$, and thick lines are $c^*_1=1$.
(\ccc) Shearing parameter, as defined in \r{eq:advel4}. Shaded contours are
$\gamma_{s1} = 0.2(0.2)1.0$, and thick lines are $\gamma_{s1} = 0.2$.
The straight lines in (\aaa--\ccc) are: \solid, $\lambda_I=x_2$; \dashed, $\lambda_I=5 x_2$
and $x_2^+=80$.
(\ddd) Instantaneous snapshots of an attached logarithmic-layer ejection in a channel at
$\retau = 4200$. The flow (and time) goes from left to right, and the streamwise
displacement of the structure has been shortened in order to fit its complete lifetime in 
the figure. The structure is coloured with the distance from the wall. Reproduced with
permission from \cite{loz:jim:14}.
}
\la{fig:advelvsy}
\end{figure}

The propagation velocity of wave packets is the group velocity \citep{whitham}, 
\beq
c_{g\chi}(k)=\p_{k_1} (k_1 c_\chi) = c_\chi +k_1 \p_{k_1} c_\chi.
\la{eq:advelg}
\eeq
The difference between the phase and group velocities is generally small, but significant.
For example, it was shown by \cite{jcadvel} that it influences the reduction of experimental
frequency spectra to their wavenumber counterparts. In the present context, the dispersion
lifetime of a wave packet depends on the difference between the two velocities. Assume that
the  wave packet is chosen such that $\Delta k_1\approx k_1$, as discussed in
the previous section. The packet disperses at a rate $\Delta c \approx |\p_{k_1} c| \Delta
k_1 \approx |\p_{k_1} c| k_1 = |c_g-c|$. We can then define an analogue of the Corrsin
parameter discussed in \S\ref{sec:production},
\beq
c^*_j(\lambda_I, x_2) = |c_{gj}-c_j|/\phi_{qq}^{1/2},
\la{eq:cstar}
\eeq
which compares the dispersion and turnover times. If $c_j^*\ll 1$, the lifetime of the
packet is determined by its nonlinear turnover; otherwise, it is limited by dispersion.
Figure \ref{fig:advelvsy}(\bbb) shows that dispersive eddies of $u_1$ are only found at long
wavelengths, $\lambda_I \gtrsim h$, and near the wall. Approximately the same limits
apply to the other two velocity components. Figure \ref{fig:advelvsy}(\bbb) should be used
in conjunction with the Corrsin parameter in figure \ref{fig:specstar}(\bbb), which measures
the importance of the shear with respect to nonlinearity. A comparison of the contour levels
of the two figures shows that deformation by the shear is faster than wavenumber dispersion
almost everywhere. 

This could suggest that all large eddies are sheared by the mean flow, but this turns out not to be
the case. The actual effect of the shear on the deformation of the wave packets can be
quantified by a `shear-deformation parameter' that compares the shear with the wall-normal
variation of the phase velocity,
\beq
\gamma_{sj}=1-\frac{\p_2 c_{j}}{S}.
\la{eq:advel4}
\eeq
When $\gamma_s\approx 1$, the (nonlinear) self-interaction sustains the shape of the
structure against the effect of the mean flow, and $c_j$ is approximately independent of
$x_2$. When $\gamma_s\ll 1$, eddies are advected and deformed by the shear. This does
not imply that the sheared eddies are not coherent, but suggests that their interaction with
the mean flow is essentially linear, and that they are unlikely to survive much longer than
the shearing time, $ST_s = O(1)$.

The Corrsin parameters $s^*$ and $c^*$ quantify processes acting on the eddies,
while $\gamma_s$ measures the result of those processes. The difference between $s^*$ and
$c^*$ on one side and $\gamma_s$ on the other, encapsulates the difference between
statistically defined eddies, and the internal dynamics of the structures. The shear-deformation
parameter for $u_1$ is shown in figure \ref{fig:advelvsy}(\ccc). Only structures in the
buffer layer, $x_2^+\lesssim 50$, and those longer than $\lambda_I/x_2\approx 5$ are
nonlinearly coherent.

The dispersive behaviour implied by figure \ref{fig:advelvsy}(\bbb) is reasonably well
understood in the case of the buffer layer, and illustrates the relation between coherence
and dispersion. \cite{loz:jim:14} measured the propagation velocity of individual structures
defined by intense isosurfaces of the tangential Reynolds stress, $-u_1u_2$ (see
\S\ref{sec:struct}). They measured the phase velocity at each wall distance by tracking the
motion of small-scale features within individual structures, and the group velocity by
tracking the motion of the rectangular box circumscribing each structure. All the structures
in the sample were relatively large, and attached to the wall in the sense of having roots
well within the buffer layer. They found that the two propagation velocities
were similar above $x^+_2 \approx 100$, but different below that level. An example is given
in figure \ref{fig:advelvsy}(\ddd), which shows the evolution of an ejection $(u_1 < 0, u_2
> 0)$ throughout its lifetime. The upper part of the ejection extends into the logarithmic
layer and moves approximately as a unit. It drags underneath a dispersive viscous root that
moves more slowly, and which keeps getting left behind (second to fourth attached frames),
and reforming at its leading edge (frame five).

In essence, dispersion happens because eddies at a given distance from the wall are
superpositions of eddies of different sizes and heights. Larger eddies, with longer
wavelengths, tend to be centred at higher distances from the wall, and move with the faster
velocity corresponding to their taller vertical range. This dependence of the
phase velocity on wavelength defines dispersion.

\begin{figure}
\vspace*{3mm}%
\centerline{%
\includegraphics[width=.70\textwidth,clip]{\arpath 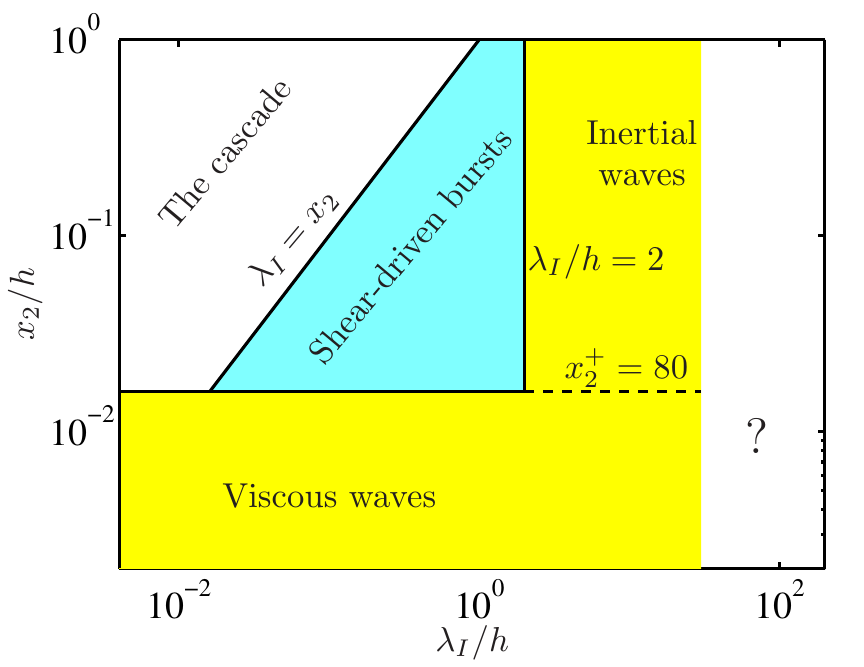}%
}%
\caption{%
Sketch of the expected behaviour of the flow structures in channels, as a function of the
wavelength and of the distance from the wall. All limits should be understood as
approximate.
}
\la{fig:coherent}
\end{figure}

\begin{figure}
%
\centerline{%
\includegraphics[width=.98\textwidth,clip]{\arpath 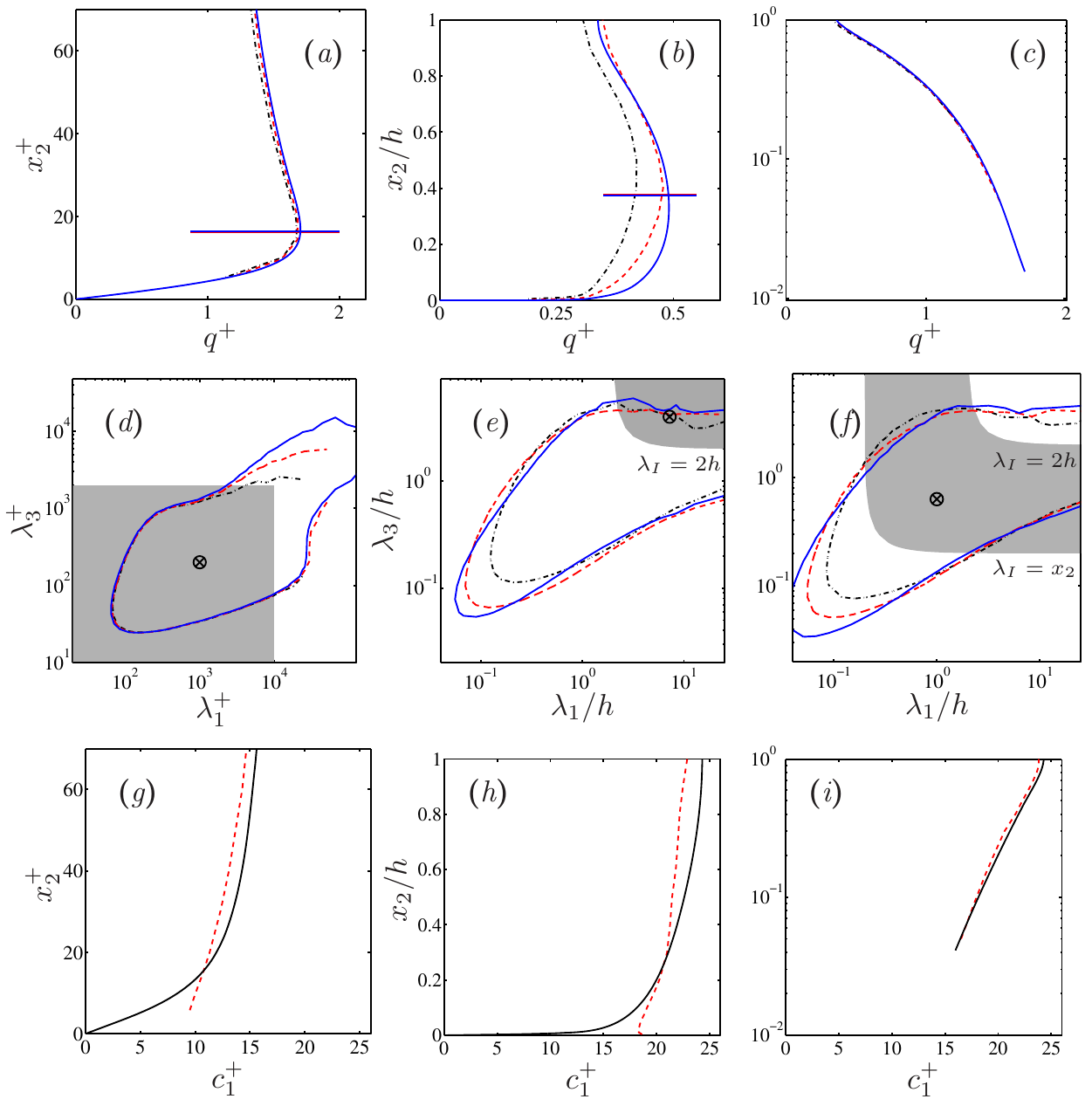}%
}%
\caption{%
Behaviour of the velocity fluctuations integrated over the spectral ranges sketched in figure
\ref{fig:coherent}, as functions of the distance from the wall. \chndot, CH950; \dashed,
CH2000; \solid, CH5200. 
(\aaa,\ddd,\ggg)  Near-wall viscous waves; $\lambda_1^+\le 10^4,\, \lambda_3^+\le 2\times 10^3$.
(\bbb,\eee,\hhh)  Large inertial waves; $\lambda_I\ge 2h$. 
(\ccc,\fff,\iii)  Shear-driven eddies; $x_2\le\lambda_I < 2h$. 
(\aaa--\ccc) Total  integrated fluctuation intensity, $q=\sqrt{u_iu_i/3}$. The horizontal line in (\aaa) is the 
wall distance at which $U_1^+(x_2)=11$; the one in (\bbb) is $U_1(x_2)=U_b$.
(\ddd--\fff) Spectral densities, $\phi^+_{qq}=0.1$. The shaded regions are the   
ranges used in (\aaa--\ccc) to integrate the different intensity profiles. 
(\ddd) Drawn at $x^+_2=15$. (\eee) $x_2/h=0.35$. (\fff) $x_2/h=0.2$.
(\ggg--\iii) \dashed, Profiles of the phase velocity of $u_1$ at the spectral points marked
with a circle in (\ddd--\fff); \solid, mean velocity profile. Channel CH2000.
}
\la{fig:waveform}
\end{figure}

The conclusions from this first part of the paper can be summarised in the sketch in figure
\ref{fig:coherent}. There are two coherent parts of the $(x_2, \lambda_I)$ parameter plane:
the buffer layer, where structures are held together by viscosity, and the very large
structures in the outer region, which presumably survive because they draw their energy from
the high shear near the wall, but are only slowly deformed by the weaker shear in the outer
part of the flow. This agrees with our previous argument that these are the only two parts of the
flow that possess an intrinsic unit of length: the wall unit near the wall, and the flow
thickness for the very large eddies. It also suggests that the structures in these two
ranges could be approximately described as semi-permanent solutions of the Navier--Stokes
equations. They draw their energy from the shear, but possess enough internal dynamics to
maintain a uniform propagation velocity across their wall-normal extent.

The eddies in the trapezoidal region labelled as `shear-driven bursts' in figure \ref{fig:coherent}
possess no such unit of length, and are therefore difficult to describe as equilibrium
solutions. Accordingly, this is also the region in which the propagation velocity is found
to track the mean profile, and where the life of the structures is determined by their
deformation by the shear. Finally, structures to the left of the Corrsin length are too
small to couple with the shear, and constitute the quasi-isotropic Kolmogorov cascade. The
question mark to the right of $\lambda_I/h\approx 30$ reflects the uncertainty in the maximum
length of the large-scale velocity streaks. As discussed in \S\ref{sec:correl}, this limit
probably depends on the flow involved.

The effect of the behaviour of the advection velocity on the kinetic energy profiles is
shown in figure \ref{fig:waveform}. Figures \ref{fig:waveform}(\aaa--\ccc) display the
profiles of the kinetic energy integrated over spectral regions corresponding, respectively,
to the near-wall viscous waves, the large-scale inertial waves, and the logarithmic-layer
sheared structures. The corresponding spectral ranges are plotted in figures
\ref{fig:waveform}(\ddd--\fff\/) at wall-parallel planes characteristic of the corresponding
structural regimes, and the vertical profiles of the phase velocities for the three cases are compared with the mean velocity profile  in
figures \ref{fig:waveform}(\ggg--\iii).
The point where the phase and mean-velocity profiles intersect defines the critical layer at
which a permanent wave resonates with the mean flow and is expected to reach its maximum
amplitude. This resonance is both a linear result \citep{McKSha2010}, and an approximate
nonlinear one \citep{hal:she:10}. It is clearly visible in figures
\ref{fig:waveform}(\aaa,\bbb), where the structures in the viscous and outer layers reach a
maximum close to the level at which their approximately constant phase velocity crosses the
mean velocity profile.

Figures \ref{fig:waveform}(\ccc,\fff,\iii) contain results for the self-similar structures
of the logarithmic layer. For them, figure \ref{fig:waveform}(\iii) shows that the phase
velocity follows very closely the mean profile, and a critical layer cannot be defined. We
have already seen that these are not permanent waves. They have a finite lifetime due to
their deformation by the shear, and display no amplitude maximum in figure
\ref{fig:waveform}(\ccc).

\section{The evidence for structures}\la{sec:struct}

There are several ways to relate coherent structures to statistical eddies, the simplest of
which is to look for objects whose properties suggest that they should evolve more or less
autonomously. This often takes the form of visualisation of some intense property,
whether in the form of interactive graphics or of automatic machine processing. Well-known
examples are the streamwise velocity streaks \citep{kli:rey:sch:run:67}, and the
Reynolds-stress quadrant events \citep{lu:wil:73}, both of which were first observed
experimentally and will be discussed below. Velocity and Reynolds stress are quantities of independent
interest, but the structure of their intense regions had to be discovered by direct
observation.

The second approach is to seek events that are coherent in time, rather than in space, such
as the `bursts' in shear flow. They signal dynamical processes, and can be taken as
indicators of the presence of dynamically relevant structures. The first evidence for bursts
in wall-bounded turbulence was also experimental \citep{kim:kli:rey:71}, but the question of
whether they reflected temporal or spatial intermittency could not be usefully discussed
until temporally resolved complete flow fields began to be available from simulations
\citep{rob:91}. The minimal flow unit in \cite{jim:moi:91} was particularly useful in this
respect because it simplified the temporal tracking of individual structures
\citep{jim:kaw:sim:nag:shi:05}.
 
The third approach is to test the predictions of theoretical models. This is the most
satisfactory of the three, because a good model usually predicts more than what it was
originally developed for. It is also the hardest, specially in the context of chaotic
nonlinear turbulence, but we will see that the situation is not hopeless, and that partial
models are beginning to appear.
 
Even a cursory review of each of these approaches would fill a longer article than  the
present one, and will not be attempted. Instead, we organise our discussion around
individual types of structures and their relations, and point in each case to 
sources where more information can be found.
  
\subsection{Intermittency}\la{sec:interm}

\begin{figure}
\centerline{%
\includegraphics[width=.95\textwidth,clip]{\arpath 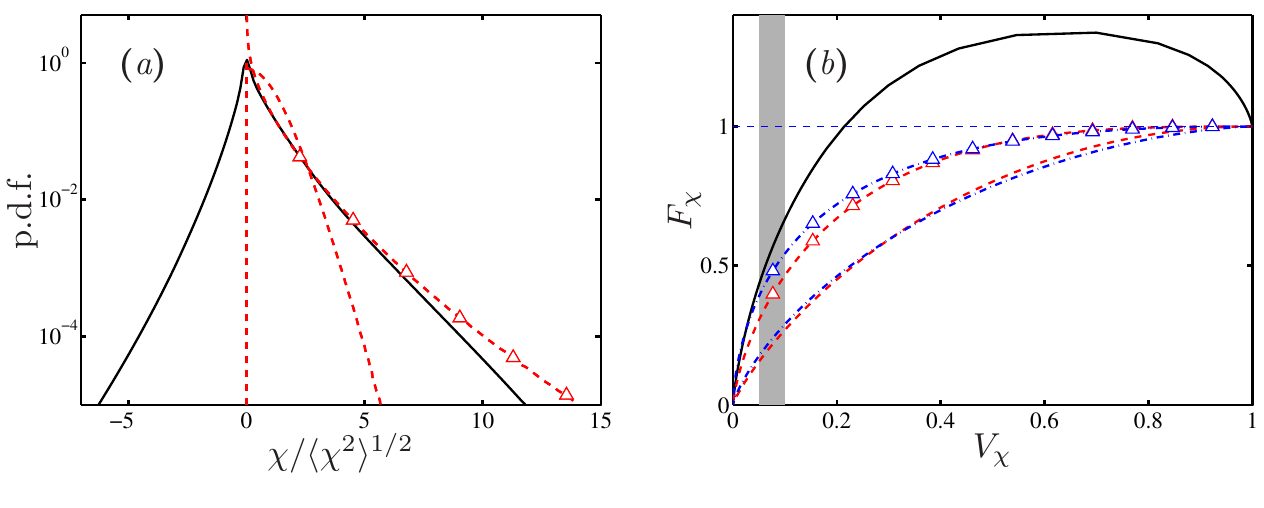}%
}%
\caption{%
(\aaa) Probability density function of several quantities, illustrating the effect of
different representations. (\bbb) Cumulative property fraction \r{eq:enstfrac} against cumulative
probability \r{eq:volfrac}. Channel CH2000.
\solid, $\chi=-u_1u_2$; \dashed, $|u_2|$; \dashtrian, $u_2^2$; \chndot, $|\omega|$;
\chndottrian, $\omega^2$. The shaded patch in (\bbb) is $V_\chi\in(0.05 - 0.1)$.
}
\la{fig:severalpdfs}
\end{figure}

Consider first the intensity as an indicator of coherence. Isolating individually connected
regions of the flow by thresholding their intensity is a classical way of identifying
coherent structures, but it implies the choice of a threshold. This is easiest for
intermittent quantities, for which high intensity is localised. There is a well-developed
theory of the intermittency associated with a singular behaviour in the limit of very high
Reynolds numbers \citep{sreeni91}, but we will use a less rigorous definition. If, for a
quantity $\chi$ with p.d.f. $p(\chi)$, we define the volume fraction of the data above a threshold
$\chi_0$ as
\beq
V_{\chi}(\chi_0)= \int_{\chi_0}^\infty p(\chi)  \dd \chi , 
\la{eq:volfrac}
\eeq
and the fraction of $\chi$ above that threshold as 
\beq
F_{\chi}(\chi_0)=\bra \chi \ket^{-1}\, \int_{\chi_0}^\infty \chi\,  p(\chi)  \dd\chi , 
\la{eq:enstfrac}
\eeq
$\chi$ will be considered to be intermittent if a threshold can be found such that
$F(\chi_0)$ is relatively large for a relative small volume $V(\chi_0)$.

Unfortunately, this definition is physically ambiguous. Consider the problem of isolating
velocity structures in which the absolute value of $u_2$ is above a threshold. The p.d.f. of
the velocity is known to be approximately Gaussian, as shown by the dashed curve in figure
\ref{fig:severalpdfs}(\aaa). Correspondingly, figure \ref{fig:severalpdfs}(\bbb) shows that
the velocity fraction \r{eq:enstfrac} grows relatively slowly with the volume fraction
\r{eq:volfrac}.

However, the distribution of the same quantity becomes more intermittent if we consider
$u_2^2$ (dashed line with symbols in figure \ref{fig:severalpdfs}\aaa). As a consequence,
the fraction of the wall-normal component of the kinetic energy also grows more steeply with
the volume than in the case of $|u_2|$ (figure \ref{fig:severalpdfs}\bbb). The same is true
of the vorticity magnitude, which has been included in figure \ref{fig:severalpdfs}(\bbb).
Even for such a technically intermittent variable, the fraction of $|\omega|$ included in a given
volume fraction of strong vorticity grows approximately as that of $|u_2|$, while that of
the enstrophy, $|\omega|^2$, behaves approximately as $u_2^2$. Any p.d.f. can be made more
intermittent by representing it in terms of a higher power of its variable. Of particular
interest is the product $-u_1u_2$, which we will use in the next section to classify the
flow into regions of active wall-normal transfer of the streamwise momentum. Its p.d.f has been
added to figures \ref{fig:severalpdfs}(\aaa,\bbb) and behaves as a quadratic variable. In
fact, it was shown by \cite{AntAtk73} and \cite{lu:wil:73} that the probability distribution
of $-u_1u_2$ is essentially that of the product of two Gaussian variables. The main reason
why we can partition the flow into discrete intense regions that contain a relatively large
fraction of quantities of interest within a small volume fraction, is that the interesting
quantities in fluid mechanics (energy, enstrophy and Reynolds stresses) are often quadratic.

A systematic way of choosing a threshold to partition the flow into separate intense
regions was introduced by \cite{moi:jim:04}. Consider the enstrophy. A very high
threshold isolates a few very intense vortices which account for a very small fraction of
the total enstrophy. Conversely, a very low threshold includes a larger enstrophy
fraction, but at the price of linking all the vortices into a single large tangle. The
`percolation' transition between the two limits is typically abrupt \citep{stauffer}, and
can be used to define a threshold that includes as much enstrophy as possible while still
segmenting the flow into individual structures. While the percolation threshold depends on
the shape of the individual objects being thresholded, it typically
takes place for volume fractions of the order of 5--10\% in three dimensions.
This range has been added to figure \ref{fig:severalpdfs}(\bbb), and shows that we can
expect to identify intense structures accounting for approximately 40--60\% of the
quadratic quantities of the flow, but for a much smaller fraction (15--25\%) of the linear ones.
Details of the application of this technique to wall-bounded flows can be found in
\cite{del:jim:zan:mos:06} for the vorticity, in \cite{loz:flo:jim:12} and \cite{dong17} for
the Reynolds stresses, and in \cite{juan:phd:14} for individual velocity components. The
resulting structures will be discussed in the rest of this section.

It is clear that the percolation threshold is only one among many possible threshold choices, and
that conclusions derived from the structures thus obtained have to be tested against other
methods of analysis. For example, individual structures should only be treated as
indicative, while the statistical properties of classes of structures are more
meaningful. Also, the percolation analysis described above typically results in a range of
thresholds approximately spanning a decade. Choosing its midpoint as a nominal threshold is
reasonable, but any conclusion should be tested against the results of thresholding above
and below the nominal value. Finally, although it should be taken into account that
different quantities and methods often result in different statistics, any conclusion derived
from thresholded structures should be complemented by the analysis of the statistical eddies
discussed in \S\ref{sec:classical} and \S\ref{sec:eddies}.

\subsection{The tangential Reynolds stress}\la{sec:Qs}

\begin{figure}
\centerline{%
\includegraphics[width=.98\textwidth,clip]{\arpath 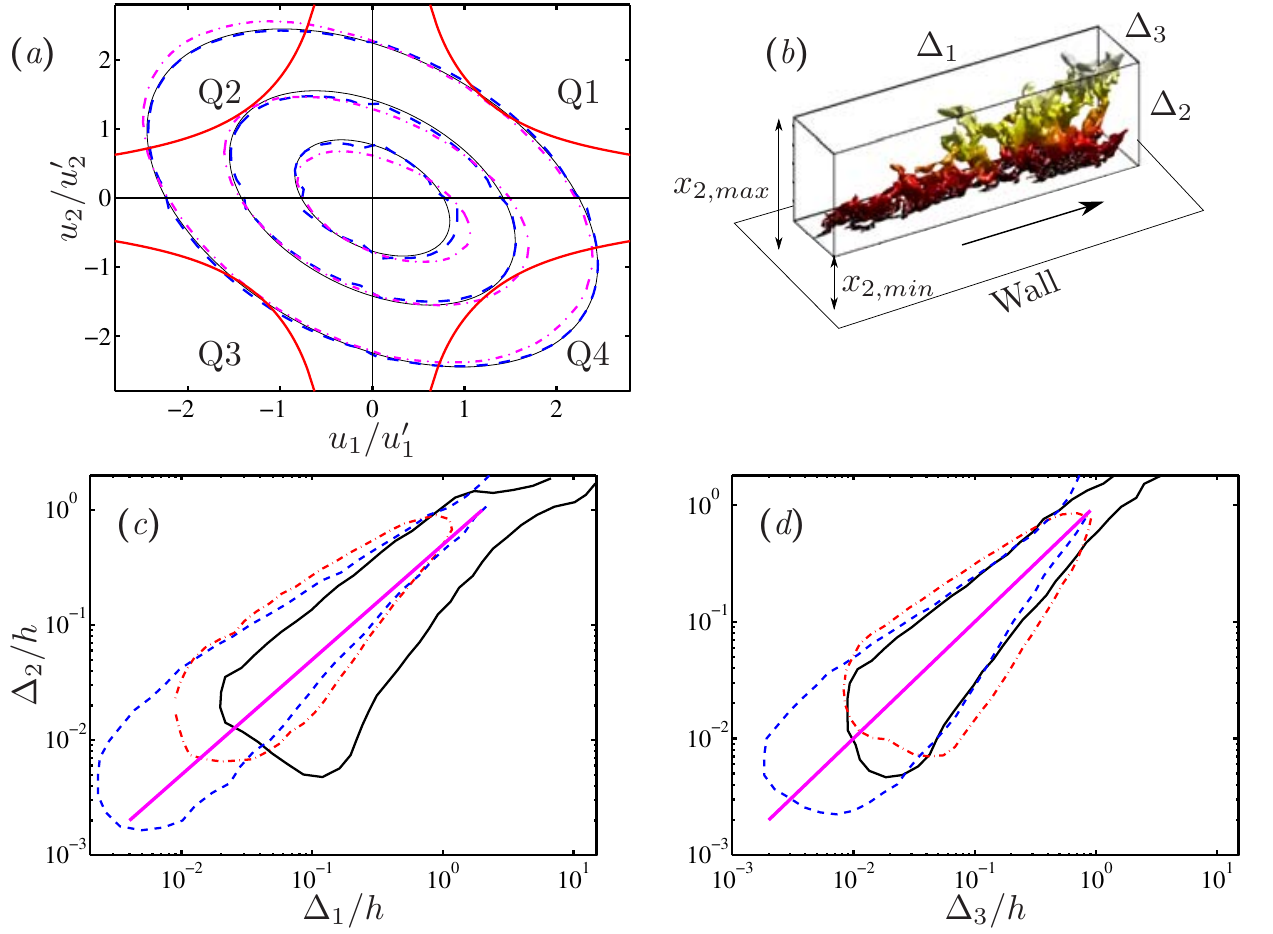}%
}%
\caption{%
(\aaa) Joint p.d.f. of $(u_1, u_2)$. \solid, Joint-Gaussian p.d.f. with correlation
coefficient $\bra u_1 u_2\ket/u_1' u_2'=-0.4$; \dashed, homogeneous shear turbulence, HSF100; \chndot, channel CH950 at
$x_2/h=0.15$. Contours contain 70\%, 30\%, and 5\% of the total mass of the p.d.f. The
hyperbolic lines are $u_1u_2=\pm 1.75 u_1' u_2'$.
(\bbb) Sketch of a Reynolds-stress structure (Q2) within its circumscribed parallelepiped,
coloured by the distance from the wall. The flow is from left to right, and the top of the
structure is at $x_{2,max}=x_{2,min}+\Delta_2$.
(\ccc) Joint p.d.f. of the streamwise and wall-normal dimensions of the Q$^-$ structures.
\solid, Attached Q$^-$s in channel CH2000; \chndot, detached Q$^-$s in CH2000; \dashed,
homogeneous shear turbulence HSF250, with dimensions normalised by the span of the
computational box. The diagonal is $\Delta_1=2\Delta_2$, and contours contain 90\% of the
p.d.f.s.
(\ddd) As in (\ccc), for the spanwise dimension. The diagonal is $\Delta_3=\Delta_2$. 
}
\la{fig:uvpdf}
\end{figure}

The first coherent structures to be treated quantitatively in wall-bounded flows were those
of the tangential Reynolds stress, $-u_1u_2$, which is the quantity associated with mean
momentum transfer in \r{eq:mom1}, and with the mean production of turbulent energy in
\r{eq:enertot}. Consider the joint p.d.f. of the two velocity components in figure
\ref{fig:uvpdf}(\aaa), which can be classified into the four quadrants labelled Q1 to Q4.
Most points for which $|u_1 u_2|$ is large either belong to Q2 (ejections), where positive
$u_2$ carries low streamwise velocity from the wall upwards, or to Q4 (sweeps), where the
opposite is true. \cite{wal:eck:bro:72} and \cite{lu:wil:73} argued that these events are
the dominant contributors to the exchange of streamwise momentum among different layers of
the flow, and ultimately to the generation of turbulent drag. They introduced the analysis
of the flow in terms of strong `quadrant' events, defined by an empirically determined
threshold for $|u_1 u_2|$. These early single-point velocity measurements eventually
evolved into fully three-dimensional structures in direct simulations, denoted here as Qs.
In agreement with our discussion of figure \ref{fig:severalpdfs}(\bbb), the percolation
analysis described in the previous section isolates Qs that fill about 7\% of the volume,
and 60\% of the total momentum transfer. It is encouraging that the threshold defined in
this way agrees approximately with those found by other groups from
single-point temporal signals, using very different methods. A typical value $|u_1
u_2|>1.75\, u_1' u_2'$ has been added to figure \ref{fig:uvpdf}(\aaa) as the four hyperbolic
lines bounding the Qs. 

Qs have been studied for wall turbulence by \cite{loz:flo:jim:12}, and for homogeneous shear
turbulent flows (HSF) by \cite{dong17}. Their temporal evolution, including mergings and splits,
has been documented in \cite{loz:jim:14}. The interested reader should consult these papers
for details. We only summarise here their most salient features.

When structures are circumscribed in a parallelepiped aligned with the coordinate
directions, as in the sketch in figure \ref{fig:uvpdf}(\bbb), they can be classified
according to their position and dimensions. Structures separate into two clearly distinct
classes: those which are attached to the wall, and those which are not. This distinction
appears clearly in the joint p.d.f. of minimum and maximum distances from the wall, and
applies to all the structures that have been studied up to now in wall-bounded turbulence,
not only to the Qs. It is determined by whether the minimum distance from the wall is below
or above $x_2^+\approx 20$ \citep{del:jim:zan:mos:06}. In the example in figure
\ref{fig:advelvsy}(\eee), this corresponds to whether the structures have developed a
viscous dispersive root near the wall. As seen in that figure, a given structure need not
be attached or detached over its full lifetime. Ejections, which tend to move away from the
wall, start their lives as attached, and eventually detach. Sweeps behave the other way
around. Notwithstanding these differences, the properties of Q2s and Q4s are fairly similar,
and we will treat them together from now on as Q$^-$s. Although detached structures are
numerically much more common than attached ones, they are also smaller, and most of the
volume associated with intense Reynolds stress is in attached structures. In the case of Qs
in channels, approximately 60\% of all the intense structures are Q$^-$s, but only 25\% of
them are attached to the wall. In spite of this, attached structures account for
approximately 70\% of the volume of intense structures, and carry approximately 60\% of the
total tangential Reynolds stress \citep{loz:flo:jim:12}.

Figure \ref{fig:uvpdf}(\ccc) shows p.d.f.s of the streamwise and wall-normal dimensions of
Q$^-$s in three classes of structures: attached and detached Q$^-$s in a channel, and Q$^-$s
in HSF, where the absence of walls makes attachment irrelevant. All of them describe
self-similar families, at least above the buffer layer. Detached and HSF Q$^-$s have similar
aspect ratios $(\Delta_1/\Delta_2\approx 2)$, while attached Q$^-$s are slightly more
elongated $(\Delta_1/\Delta_2\approx 3)$. This discrepancy was investigated by
\cite{dong17}, who showed that it is due to `spurious' connections between neighbouring
attached Q$^-$s through their viscous roots. If the points below $x_2^+=100$ are removed
from the identification of connected structures, even objects that would otherwise be
attached have the same aspect ratio as the detached ones. Note that these aspect ratios are
approximately consistent with those in figures \ref{fig:corrcuts}(\aaa,\bbb), where the
correlations of $u_2$ and $u_3$ have heights of order $h$, and streamwise lengths of order
$2h$. They also agree with the `short' POD reconstructions of $u_2$ and $u_3$ in figure
\ref{fig:pod}(\fff\/), but they are much shorter than the `long'  correlations of $u_1$ in
figures \ref{fig:corrcuts}(\ccc) and \ref{fig:pod}(\fff\/). The spanwise aspect ratios in
\ref{fig:uvpdf}(\ddd) are identical for the three families $(\Delta_3/\Delta_2\approx 1)$,
in agreement with the cross-flow correlations in figure \ref{fig:corrcuts}. In all these
cases, there is essentially no difference between the channel and the wall-less HSF,
reinforcing the conclusion in \S\ref{sec:classical} and \S\ref{sec:eddies} that coherent
structures and eddies are a consequence of the shear, not of the wall. Structures attach to
the wall when they become too large to fit in the channel otherwise.

The fact that the attached Q$^-$s in channels are longer than the detached ones, but not
wider, suggests that neighbouring Q$^-$s are arranged streamwise from one another. This was
confirmed by \cite{loz:flo:jim:12}, who showed that attached Q$^-$s of the same kind (e.g.,
a Q2 and its nearest Q2) are preferentially located streamwise from each other, while those
of different kind (i.e., Q2 and Q4) form spanwise pairs. Interestingly, the same is only
true in HSF for Qs whose diagonal dimension is larger than the Corrsin scale. Smaller
structures, whose internal turnover time is too fast to couple with the shear, do not have a
frame of reference from which to determine their orientation, and are statistically isotropic
\citep{dong17}. For these small structures, it is possible to define Q$_{ij}$'s based on the
intensity of the Reynolds stress $-u_i u_j$. Many of their characteristics are the same as
the more classical Qs based on $-u_1 u_2$, but the statistics of their relative positions are
oriented in each case with respect to the velocity components with which they are defined.
In this sense, only larger-than-Corrsin Qs are physically relevant for the energy and
momentum balance of shear turbulence. As discussed in \S\ref{sec:classical}, smaller
structures are part of the isotropic Kolmogorov cascade. Note that, since the Corrsin scale
in channels is $L_c\approx x_2$, all attached structures are larger than $L_c$, and most
structures larger than $L_c$ are attached. 

It is interesting in this respect that there are essentially no large Q1 or Q3 structures in HSF
\citep{dong17}, in the same way that there are very few attached Q1 and Q3 in channels
\citep{loz:flo:jim:12}. Large eddies with $u_1 u_2<0$ draw energy from the shear, while
those with $u_1 u_2>0$ lose it. As a consequence, the former grow while the latter dwindle,
and the only Qs which are approximately evenly distributed among the four quadrants are the
smaller ones uncoupled from the shear. In this sense, attached eddies in wall-bounded
turbulence are equivalent to larger-than-Corrsin eddies in HSF.

\begin{figure}
\centerline{%
\includegraphics[width=.95\textwidth,clip]{\arpath 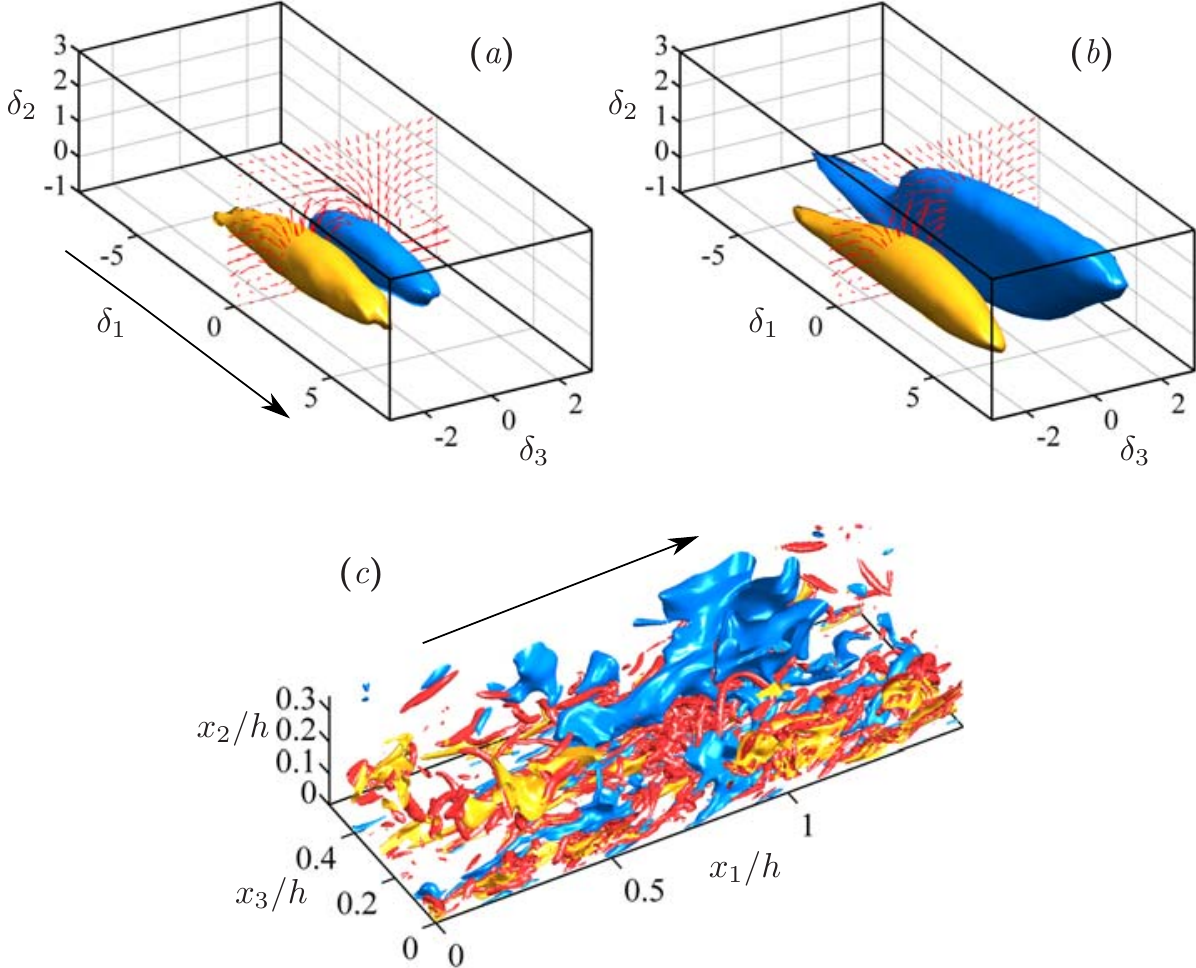}%
}%
\caption{%
Flow fields conditioned to attached Q2--Q4 pairs reaching into the logarithmic layer. 
Channel CH2000. Flow is from top-left to bottom-right,
and axes are scaled with the distance from the centre of gravity of the pair to the wall. 
(\aaa) P.d.f.s of the points belonging to the Q2 (yellow at left) and Q4 (blue at right). 
The isosurfaces plotted are $0.75$ times the
maximum value of each p.d.f.
(\bbb) Conditional streamwise perturbation velocity. The yellow object at left is the
low-speed isosurface, $u^+_1=-0.5$. The blue one is $u^+_1=0.5$. 
The arrows in (\aaa,\bbb) are transverse velocities in the cross-flow plane, with the longest one approximately $0.5 \utau$.
(\ccc) Instantaneous Q2--Q4 pair, with its associated vortex tangle. Yellow is
the ejection, blue is the sweep, and red are the vortices.  Flow is
from left to right.
}
\la{fig:condQ}
\end{figure}

Conditional flow fields around pairs of attached Q$^-$s in a channel are shown in figures
\ref{fig:condQ}(\aaa,\bbb). Since we saw in figure \ref{fig:uvpdf}(\ccc,\ddd) that pairs
come in all sizes with self-similar aspect ratios, the conditional average is compiled after
rescaling each pair to a common size and centring it on the centre of gravity of the pair.
Therefore, the coordinates in figures \ref{fig:condQ}(\aaa,\bbb), $\delta_i= x_i/\bar{x}_2$,
are multiples of the distance, $\bar{x}_2$, from the wall to the centre of gravity of each
individual pair. Also, because the equations are symmetric with respect to reflections along
$x_3$, the ejection is always defined to lie to the left of the picture. Figure
\ref{fig:condQ}(\aaa) displays the mean conditional geometry of the Q-pair, and is actually
an isosurface of the probability of finding a point belonging to a structure in the
similarity coordinates. The object to the left is the ejection, and the one to the right is
the sweep. The arrows in the central cross-section are conditional velocities in that plane.
They reveal an approximately streamwise roller, rising out of the ejection and sinking into
the sweep. There are secondary counter-rollers to both sides of the pair, but they are much
weaker than the primary central one, and there is little evidence of symmetric `hairpin
legs'. The Qs in figure \ref{fig:condQ}(\aaa) are in the logarithmic layer
$(x_{2,max}^+>100)$. Their rollers can only be identified as conditional objects, and it is
difficult to relate different rollers to one another, but the arrangement of streamwise
vortices in the buffer layer has been studied extensively. They are arranged
antisymmetrically, with vortices of opposite sign alternatively staggered along each side of
a streak \citep{stretch90,Schoppa02}. The lack of strong counter-rollers in figure
\ref{fig:condQ}(\aaa) suggests that the average arrangement in the logarithmic layer is
similar. The aspect ratio of the Q-pair in figure \ref{fig:condQ}(\aaa) is approximately
4:1:1.5 along the three coordinate directions \citep{loz:flo:jim:12}.

Because $u_1<0$ in ejections, and $u_1>0$ in sweeps, the ejection sits in a low-velocity
streak, and the sweep in a high-velocity one. The pair sits at the border between the two,
and the intervening roller has the direction of rotation required to reinforce both streaks.
Figure \ref{fig:condQ}(\bbb) shows the conditional streaks associated with the pair in
figure \ref{fig:condQ}(\aaa). The high-velocity streak is larger than the low-velocity one.
The early experimental perception was the opposite, probably because ejections collect
tracers from the wall, and are easier to visualise, but high-speed regions are fed from
above, while low-speed ones are blocked by the wall. This intrinsic asymmetry was shown by
\cite{orlandi94} to be at the root of why the skin friction is higher in turbulent flows
than in laminar ones. The size relation between the two streaks in figure
\ref{fig:condQ}(\bbb) is statistically robust, and is evident both in instantaneous pictures
and in conditional ones. Both conditional streaks are longer than the Qs, especially the
high-velocity one, but not as much as the very long structures implied by the correlations
of $u_1$ in figure \ref{fig:corrheight}(\eee). In fact, the dimensions of the low-speed streak in figure
\ref{fig:condQ}(\bbb) are very similar to those of the conditional ejection in figure
\ref{fig:condQ}(\aaa). The high-speed streak is longer, but also taller, and its
length-to-height ratio, approximately 5:1, is also not very different from either the Qs or
from the low-velocity streak, although shorter than the 20:1 ratios for
attached eddies implied by figure \ref{fig:corrheight}(\eee). This point will be further
discussed in \S\ref{sec:streaks}, after we see examples of instantaneous streaks.

It should be made clear that the rollers in figures \ref{fig:condQ}(\aaa,\bbb) are not
vortices, and that individual structures are not as smooth as the conditional ones. A
typical instantaneous sweep-ejection pair is shown in figure \ref{fig:condQ}(\ccc). Its
width is $0.5h$, or 1000 wall units. The much smaller vortices in the vorticity cluster
associated with the pair have been added to the figure, for comparison. The dimensions of
the conditional rollers in figures \ref{fig:condQ}(\aaa,\bbb) are similar to those of the
correlation rollers discussed in connection with figure \ref{fig:corrcuts}(\aaa,\bbb).

\begin{figure}
\centerline{%
\includegraphics[width=0.95\textwidth,clip]{\arpath 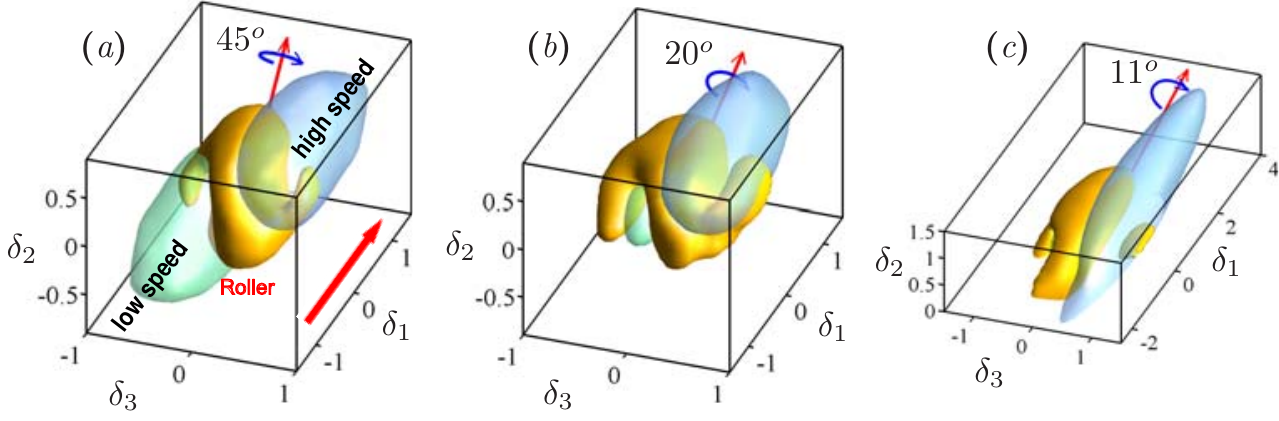}%
}%
\caption{%
Effect of the inhomogeneity of the mean velocity profile on the flow field conditioned to Q2--Q4
pairs. The central opaque S-shaped object is the isosurface of the magnitude of the
conditional perturbation vorticity, at 25\% of its maximum. The two translucent objects are
isosurfaces of the conditional perturbation streamwise velocity, $u_1^+=\pm 0.6$.
(\aaa) Homogeneous shear turbulence, HSF100.
(\bbb) Detached Qs in channel CH950.
(\ccc) Attached Qs in CH950.
Adapted with permission from \cite{dong17}.
}
\label{fig:condatt}
\end{figure}

The three-dimensional geometry of the conditional rollers associated with the Q$^-$ pairs is
displayed in figures \ref{fig:condatt}. The three figures correspond to different flow
configurations. They are reproduced from \cite{dong17}, which should be consulted for
details. Figure \ref{fig:condatt}(\aaa) corresponds to homogeneous shear turbulence, and is
the clearest. Because of the statistical symmetries of this flow, the conditional sweep and
ejection (not shown in the figure, for clarity) are equivalent, and are arranged
antisymmetrically with respect to the centre of gravity of the pair. The roller is located
between them, approximately aligned to the $(x_1, x_2)$ plane, and inclined at 45\degree\
with respect to the mean flow velocity. This is the direction of maximum extension by the
shear \citep{rog:moi:87}. At both sides of the roller are the high- and low-velocity
streaks, each one extending downstream from the roller: the high-speed streak extends
towards the right of the figure, and the low-speed streak extends to the left. The
implication is that the roller moves at an intermediate velocity from that of the two
streaks, so that the faster high-speed streak moves ahead of it while the slower low-speed
one is left behind. It is known that the streamwise advection velocities of the sweeps,
which are located in high-speed streaks, are approximately $3\utau$ faster than those of the
ejections, which are located in low-speed ones \citep{loz:jim:14}. The roller in figure
\ref{fig:condatt}(\aaa) is capped at its top and bottom by two `hooks' resembling
incomplete hairpins. These are also conditional structures.

Figure \ref{fig:condatt}(\bbb) is plotted with the same parameters as figure
\ref{fig:condatt}(\aaa), but for detached pairs in the logarithmic layer of a channel. The
arrangement is similar, but the roller is less symmetric, and less inclined to the free
stream (20\degree ). The high-speed streak, which is now farther from the wall than the
low-speed one, is also substantially larger. The conditional geometry of wall-attached pairs
is displayed in figure \ref{fig:condatt}(\ccc). It can best be understood as a further
evolution of the transition from figure \ref{fig:condatt}(\aaa) to figure
\ref{fig:condatt}(\bbb). The roller is almost parallel to the wall, and the low-speed streak
has almost disappeared underneath the surviving upper hook of the roller, which now looks as
an essentially full asymmetric hairpin. Note that the lower hook is still visible in the figure,
although almost completely truncated by the wall and hidden underneath the high-speed
streak.

As we have already seen several times in this article, the conclusion has to be that the
structures depicted in figures \ref{fig:condatt}(\aaa--\ccc) are aspects of the same
process, although both the low-speed streak and the lower hook of the roller get
increasingly modified by the nonuniformity of the mean shear. They are also damped by the
impermeability of the wall as they approach it.

\subsection{The velocity structures}\la{sec:streaks}

\begin{figure}
\vspace*{3mm}%
\centerline{%
\includegraphics[width=.90\textwidth,clip]{\arpath 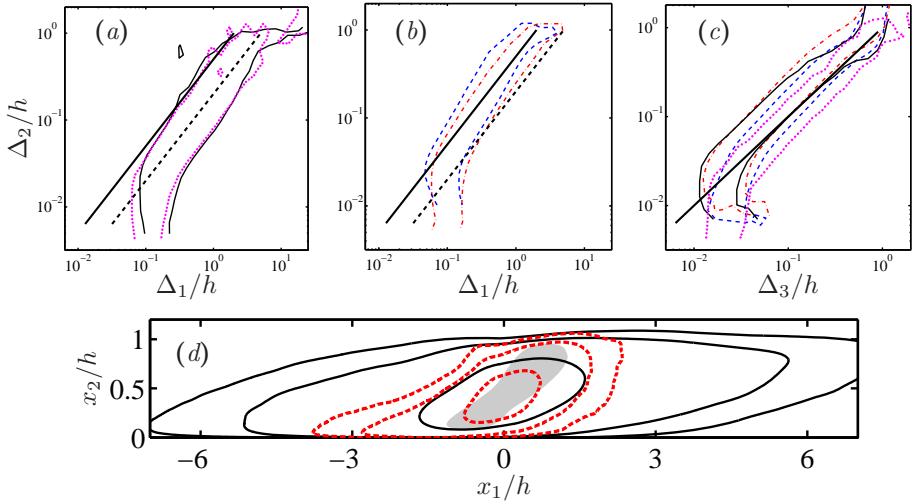}%
}%
\caption{%
(\aaa) One-dimensional p.d.f.s of the logarithm of the streamwise length, $\Delta_1$,
of intense attached structures of $u_1$, as functions of $\Delta_2$. 
(\bbb) Same for the transverse velocities.
(\ccc) Same for the spanwise length, $\Delta_3$. In all cases:
\solid,  $|u_1|^+>2$ in BL6600 at $\retau\approx 1600$ \citep{juan:phd:14};
\dotted, same for CH2000 (Lozano--Dur\'an, private communication);
\dashed, $|u_2|^+>1.3$ in BL6600;
\chndot, $|u_3|^+>1.6$ in BL6600. 
The diagonals in (\aaa,\bbb) are: \solid, $\Delta_1=2 \Delta_2$;  \dashed, $\Delta_1=5 \Delta_2$. 
The diagonal in (\ccc) is, $\Delta_3=\Delta_2$.
(\ddd) Two-dimensional streamwise section through the reference point $(\tx_2/h\!=\!0.4)$ of
the two-point autocorrelation function of $u_1$. Contours are $C_{11}=[0.05, 0.1, 0.3]$.
\solid, Channel CH2000; \dashed, boundary layer BL6600 at $\retau\!=\!1530$; the grey patch
is the $C_{33}=0.05$ contour of the $u_3$ correlation in the channel.
}
\la{fig:pdf_uvwsters}
\end{figure}

The intense structures of individual velocity components were studied by \cite{juan:phd:14}.
They also separate into attached and detached families, and are self-similar in the
logarithmic layer. Figure \ref{fig:pdf_uvwsters} displays the p.d.f.s of their aspect
ratios. The elongation of the structures of $u_1$ in figure \ref{fig:pdf_uvwsters}(\aaa),
$\Delta_1/\Delta_2 \approx 5$, is higher than for the transverse velocities in figure
\ref{fig:pdf_uvwsters}(\bbb), $\Delta_1/\Delta_2 \approx 2$, although, somewhat
surprisingly, not much more so. The latter value is also the aspect ratio of
the Qs discussed in figure \ref{fig:uvpdf}(\bbb), showing that the elongation of the Qs is that of the
transverse velocity components. The spanwise aspect ratio is shown in figure
\ref{fig:pdf_uvwsters}(\ccc), and approximately agrees for all variables, including $u_1$ and the Qs.

More interesting is the comparison in figure \ref{fig:pdf_uvwsters}(\aaa) between the $u_1$
structures of channels and of boundary layers. Figure \ref{fig:pdf_uvwsters}(\ddd\/) shows
that, as already discussed in \S\ref{sec:correl}, the correlations of $u_1$ in boundary
layers are considerably shorter than those in the channels, but it turns out that the
intense $u_1$ structures in the logarithmic layer are very similar in the two flows. If we
interpret the reference height of the two-point correlation as the centre of gravity of the
structures, the p.d.f.s in figure \ref{fig:pdf_uvwsters}(\aaa) would imply that the structures of
$u_1$ responsible for the correlation in figure \ref{fig:pdf_uvwsters}(\ddd) would have
$\Delta_1/h \approx 2$--6. This is in reasonable agreement with the correlations of the
boundary layer, and with the approximate size of the individual velocity substructures in
figures \ref{fig:specky}(\bbb) and \ref{fig:streaksall}(\bbb), but is too short for the
correlations of the channel. Figure \ref{fig:pdf_uvwsters}(\aaa) suggests that the
difference between the correlations of boundary layers and channels is mostly due to the
very long and tall `global' modes that dominate the outer part of the flow, and which appear
in the p.d.f.s as `overhangs' at the upper end of the self-similar region. These structures
contribute strongly to the correlations in the channel because they are large, but figure
\ref{fig:pdf_uvwsters}(\aaa) shows that they are not particularly intense, or at least that
they do not strongly affect the smaller structures in the logarithmic layer.
\cite{loz:flo:jim:12} mention that there is at least one very large Q4 that crosses the
whole computational box in most flow fields, but that those objects are concatenations of
smaller ones. The suggestion is that the difference between the correlation of $u_1$ in
boundary layers and channels lies in the details of the concatenation mechanism, rather in
the substructures themselves. Given our discussion in \S\ref{sec:correl}, it would be
interesting to repeat the present analysis for Couette flow.

\begin{figure}
\vspace*{3mm}%
\centerline{%
\includegraphics[width=.90\textwidth,clip]{\arpath 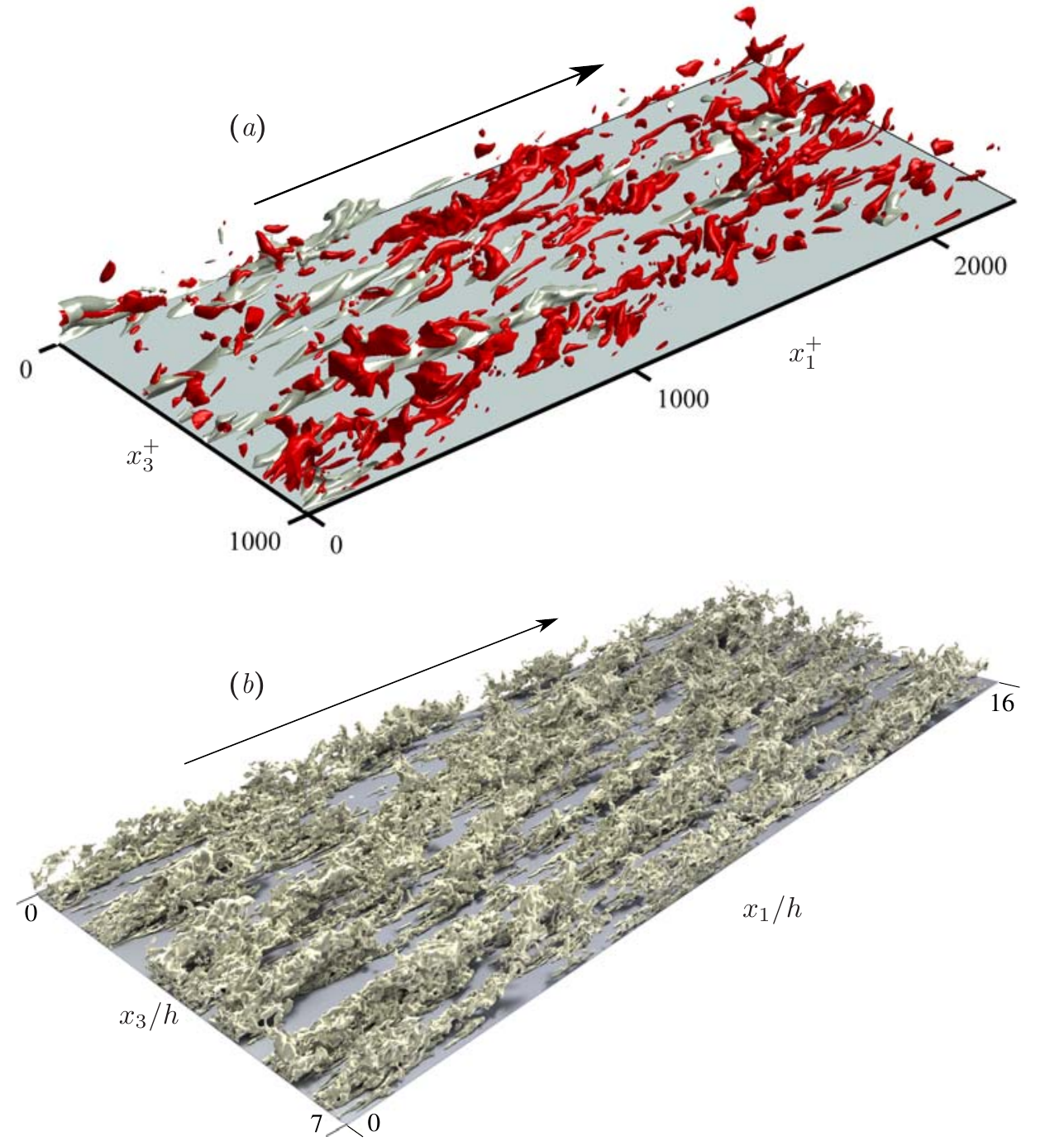}%
}%
\caption{%
Perturbation velocity isosurfaces, with the flow from lower left to upper right.
(\aaa) Buffer-layer low-velocity streaks, $u_1^+=-2$ (light grey). Channel CH2000. 
Structures higher than $x_2^+=150$ and smaller than $10^6$ cubic
wall units have been removed for clarity. The plot spans $(L_1^+\times L_3^+ =2300 \times 1000)$.
The darker (red) objects are $|u_2|^+=2$.
(\bbb) Outer-flow high-velocity streaks, $u_1^+=2$. Boundary layer BL6600 at $\retau\approx
1200$--1400. Structures shorter than $\Delta_1=h/2$ have been removed.
$(L_1^+\times L_3^+ =21000 \times 9000)$ (picture credits, J.A. Sillero).
}
\la{fig:streaksall}
\end{figure}

The long and narrow structures of $u_1$ are the well-known streamwise streaks, which are
found at all sizes and at all distances from the wall. Illustrating their range of scales,
figure \ref{fig:streaksall}(\aaa) displays isosurfaces of the perturbation velocity in the
buffer layer, and figure \ref{fig:streaksall}(\bbb) displays the large outer streaks
spanning most of the thickness of the flow. The geometry of $u_1$ is similar in the two
figures, but the sizes are very different. While the streaks in figure
\ref{fig:streaksall}(\aaa) have widths of the order of $\Delta_3^+\approx 100$, those in
figure \ref{fig:streaksall}(\bbb), have $\Delta_3^+ \approx h^+ \approx 1500$. The main
difference between them is that the larger structures have a higher intrinsic Reynolds
number, which results in a wider range of scales and in a rougher appearance. Another
difference is that the near-wall structures are less uniformly distributed than the outer
ones. This reflects the modulation of the near-wall layer by the outer flow. The isosurfaces
in figures \ref{fig:streaksall}(\aaa,\bbb) are defined in wall units that use a uniform
friction velocity, while different parts of the wall `live' in the local environment created
by the larger outer eddies. It was shown by \cite{jim12_arfm} that some of the nonuniformity
of the wall can be absorbed by scaling the fluctuations with a `local' friction velocity.
Similar models have been used as boundary conditions in LES for a long time
\citep{dear:70,PioBal02}, and a related idea has been expressed as a modulation of the inner
flow by the outer one in \cite{Mar_Sci10}. This works well for the intensity of the
fluctuations, but it does not fully describe the effect of the outer flow on the wall
structures. Continuity requires that large-scale ejections $(u_2>0)$ should be regions of
local lateral convergence, which tend to concentrate the streaks underneath. The
opposite is true underneath large-scale sweeps for which $u_2<0$ \citep{TohItan05}. Although
figures \ref{fig:streaksall}(\aaa,\bbb) correspond to different flows, the spanwise width of
the emptier regions with few streaks in figure \ref{fig:streaksall}(\aaa) is approximately
the same, $O(h)$, as the width of the large streaks in figure \ref{fig:streaksall}(\bbb). We
saw in figure \ref{fig:pdf_uvwsters}(\ccc) that the width of all the velocity structures is
$\Delta_3 \approx h$ when $\Delta_2\approx h$.

It is interesting that the outer flow in figure \ref{fig:streaksall}(\bbb) is relatively
well organised, suggesting that, in the absence of exogenous modulation by still larger
structures, and when stripped of small-scale instabilities, the largest scales of the
channel tend to form a relatively organised pattern which could perhaps be described by
some simple theoretical model of coherence. This agrees with the narrowing of the energy
spectra in figure \ref{fig:sp2} away from the wall, and with the evidence from the advection
velocity discussed in \S\ref{sec:advect}. It is also interesting that, independently of the
differences between small and large streaks, the absolute magnitude of the $u_1$ isocontours
in figures \ref{fig:streaksall}(\aaa,\bbb) is the same in wall units, in agreement with the role of
$\utau$ as a universal velocity scale.

Figure \ref{fig:streaksall}(\aaa) includes isosurfaces of the wall-normal velocity. They are
located near the streaks, but they are much shorter than them, as already implied by the
correlations in \S\ref{sec:correl}. The spanwise velocity and the strong vortices near the
wall (not shown) have sizes comparable to $u_2$, but the structures of $u_3$ tend to be flat
`flakes' parallel to the wall, in agreement with the transverse correlation in figure
\ref{fig:corrcuts}(\aaa). The vortices are quasi-streamwise `worms' with sizes comparable to
the width of the streaks. This suggest that the vortices are responsible for the wall-normal
velocities in the buffer layer, at least for short stretches comparable to the length of the
$u_2$ structures in \ref{fig:streaksall}(\aaa), but the same is not true farther from the
wall. The large central structure in figure \ref{fig:specky}(\bbb) is a sub-structure of the
velocity streaks in the flow in figure \ref{fig:streaksall}(\bbb). Figure
\ref{fig:specky}(\bbb) also includes vortices, but they are much smaller than the velocity
structures, and cannot create wall-normal velocities on a scale appropriate to modify $u_1$
(see also figure \ref{fig:condQ}\ccc). It was shown by \cite{del:jim:zan:mos:06} that the
ejections of the logarithmic layer are associated with large vortex clusters, rather than
with individual vortices, and by \cite{jim:13b} that the vortices away from the wall are
essentially isotropically oriented, and generate little net large-scale velocity. The
counterpart of the near-wall vortices far from the wall are the conditional rollers
discussed in \S\ref{sec:Qs}. They are the collective effect of the residual anisotropy of
the vorticity organisation in the presence of the shear of the mean velocity profile.

On the other hand, examples of well-organised hairpin forests have been shown to exist in
turbulent flows \citep{adr07}, although they tend to be more common at relatively low
Reynolds numbers, and to become disorganised when the Reynolds number increases.
Somewhat confusingly, this disorganised vorticity is sometimes referred to as hairpin `fragments', which
makes it difficult to distinguish it from regular vortices. An exception may be the
structures reported in \cite{wu_17}. Hairpin forests have always been known to form in the
turbulent spots that mediate bypass laminar-turbulent transition. In this sense they are
low-Reynolds number structures. The observation in \cite{wu_17} is that very similar spots
and hairpins form intermittently underneath fully turbulent boundary layers ($x_2^+\lesssim
100$ and $\retau\lesssim 1000$), although they also tend to disorganise as they move farther
from the wall. In this sense they may be important contributors to the regeneration of
turbulence in the buffer layer.

It is striking that neither far nor near the wall are there indications of the long
streamwise rollers that are often important ingredients of low-order models of the turbulence
energy cycle. The observations support shorter structures that would not, by themselves,
create a long streak. On the other hand, we saw in \S\ref{sec:Qs} that the conditional
rollers are arranged with respect to the streaks in such a way as to always help to sustain
them: clockwise rollers are located on the left-hand edge of a high-velocity streaks, and
counter-clockwise ones are located on the right hand. Thus, the streamwise average of all the
shorter rollers can be modelled as a long average roller that creates a long average streak, even if
the flow is better described locally as a concatenation of the smaller aligned units in
figure \ref{fig:streaksall}(\bbb). The mechanism for this alignment is unclear, and we will
come back to it in \S\ref{sec:open} as one of the open problems in modelling wall-bounded
turbulence.

Note that the streaks are different from the `hairpin packets' often proposed as building
blocks for the logarithmic layer \citep{adr07}. Even if hairpins, rather than just isolated
inclined vortices or conditional rollers, could be shown to exist at high Reynolds numbers,
the individual hairpins would be equivalent to the $u_2$--$u_3$ structures whose aspect
ratio is displayed in figure \ref{fig:pdf_uvwsters}(\bbb). The packets would be the somewhat
longer $u_1$ structures in figure \ref{fig:pdf_uvwsters}(\aaa), whose dimensions suggest
that each of them contain at most three or four shorter structures. The streaks themselves are the
overhang at the top of the p.d.f.s in figure \ref{fig:pdf_uvwsters}(\aaa), and correspond to
the approximately linear arrangement of substructures spanning the full thickness of the
boundary layer in figure \ref{fig:streaksall}(\bbb).

\subsection{Other variables}\la{sec:oops}

The main arbitrariness in the definition of the structures described in the previous
sections is the choice of the variable being thresholded. For example, while the streamwise
velocity can be considered a surrogate for kinetic energy, it is unclear why the velocity
fluctuations, $u_1$, rather than the full velocity, $\tu_1$, should be considered as the
variable of interest. In terms of the total energy equation, the fluctuating $u_1$ is
negligible, and the transverse velocity fluctuations even more so. On the other hand, the
total kinetic energy is not a Galilean invariant quantity. Unfortunately, the geometry of the two
fields is quite different. While the streaks of $u_1$ can be attached or detached from
the wall \citep{juan:phd:14}, the instantaneous profile of the full velocity is almost
always monotonic. Thresholding $\tu_1$ always results in an irregular layer attached to the
wall, and the streaks become uniform-momentum regions in which the mean velocity profile has
been mixed by the tangential stresses. These are the very-large uniform-momentum zones
mentioned by \cite{adr:mei:tom:00}. They are not part of the self-similar organisation of
the logarithmic layer, and are at least ten times longer than individual $u_1$ structures.

The Reynolds stresses discussed in \S\ref{sec:Qs} appear to be a more objective choice,
because they are in the equation for momentum transfer, and ultimately determine drag.
However, it was noted by \cite{jim16} that the quantity in the momentum equation is the
divergence of the stresses, and that the stresses themselves can be modified by adding any
symmetric tensor of zero divergence without changing anything. As an example, he introduced
an `optimal' stress tensor, $\phi_{ij}$, which minimises the integrated squared magnitude of the
stresses over the channel. This tensor field is an arbitrary modification of the classical
stresses, but not more arbitrary than the classical one, and it turns out that its
statistics are rather different from those of $u_iu_j$. In particular, the p.d.f. of $\phi_{12}$ in a channel is
much closer to Gaussian than that of $u_1 u_2$, and the amount of `counter-gradient'
momentum flux in figure \ref{fig:severalpdfs}(\bbb) is much reduced. Conversely, the product
$\tu_1 u_2$, which uses the full streamwise velocity rather than the perturbations, and
which correspond to a different rearrangement of the momentum equation, is much more
intermittent than the classical Reynolds stress, and involves much more
counter-gradient backscatter \citep{jim16}.

None of this should be fundamentally troubling, because these alternative fluxes are just
different choices of the gauge used to represent the Reynolds-stress field, and they leave the equations
invariant. But it raises the question of what is the real significance of the quadrant
analysis in \S\ref{sec:Qs}, and of whether the properties that we have discussed for the Qs
might be artefacts linked to a particular form of the equations.

There are two options. The first one is to renounce to the Reynolds stresses, and to consider
the geometry of its divergence. This `Lamb vector' has been studied, for example, by
\cite{wu:etal:99} in the context of modelling, and it changes the character of the stress
field completely. The stresses are large-scale quantities, while their divergence is
associated with much smaller scales. The question is akin to whether the quantity of interest
in a flow is the pressure or the pressure gradient. The gradient is the only term in the
equations of motion, and determines the acceleration of the fluid; but the pressure, which
is the integral of the gradient, is more directly related to the velocities. Think of Bernouilli's
equation.

Similarly, while the Lamb vector is the quantity directly related to the acceleration
of the fluid particles, its integral determines the overall velocity profile. However, this
leaves open the question of which gauge to use, and of whether we should study the
properties of the Reynolds-stress Qs, as in \S\ref{sec:Qs}, the thresholded regions of
strong $\phi_{12}$, or something else. The most direct way of answering this question is
to repeat the analysis in \S\ref{sec:Qs} for $\phi_{12}$. This was done recently by
\cite{osawa16}, and the tentative result is that the streamwise elongation of the optimal
flux structures (`op-sters') is similar to that of the Qs, while their spanwise aspect ratio
is twice as wide. Other correlations suggest that, since opsters do not differentiate between
sweeps and ejections, they approximately correspond to the combined Q2--Q4 pairs.

Note that, because the difference between $\phi_{ij}$ and $u_i u_j$ is divergence-free, the
integral of the two quantities over any sufficiently large volume only differs by a small
boundary term. In particular, the ensemble-averaged stresses of the two representations are
identical. The preliminary analysis just discussed suggests that their structures are also
essentially similar, giving some hope that they are not artefacts, and that they are
physically significant in both cases.

\subsection{Small and minimal flow units}\la{sec:minimal}

Minimal simulations began to appear in the 1980's as devices to isolate invariant
(initially meaning `simple') solutions in two-dimensional channels
\citep{herbert76,jim2dchan:87}. Turbulent channels are typically simulated in numerical
domains that are periodic in the two wall-parallel directions, so that a fundamental flow
unit repeats itself in a doubly infinite sequence of identical copies. Usually, simulation
boxes are chosen large enough for the periodicity not to interfere with the solution,
making sure that copies are `far enough not to see each other'.

A different approach is to do the opposite, choosing a periodicity so short that, if
individual structures exist, the fundamental cell contains a single structure with just
enough dynamics to sustain itself. The original hope was that, if the flow was constrained
enough, individual structures could be made to be steady, or otherwise simpler to describe
than the chaotic ones in real flows. This, for example, was found to be the case in the
two-dimensional turbulent channels mentioned above. Even if that flow is different enough
from three-dimensional turbulence to be irrelevant to the present article, those results proved that
single nonlinear structures with a well-defined characteristic size could survive in a
sheared environment, and that incrementally releasing the computational constraints could
lead to bifurcations into temporally periodic orbits, chaos, and spatially localised states.

\begin{figure}
\centerline{%
\includegraphics[width=.95\textwidth,clip]{\arpath 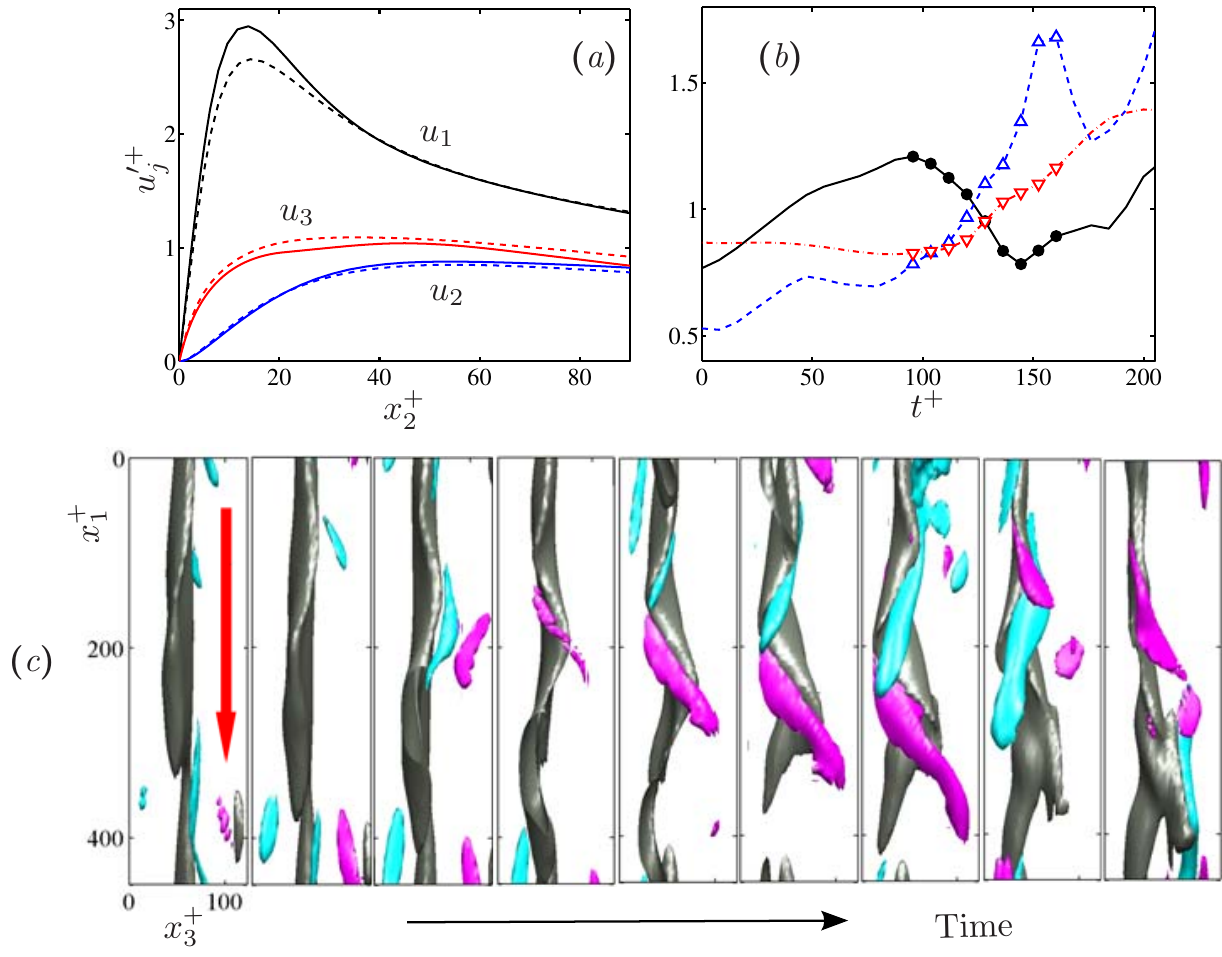}%
}%
\caption{%
(\aaa) Velocity fluctuation profiles of: \solid, a minimal channel at $\retau=181$, $L_1^+
=460$, $L_3^+ =127$; \dashed, a full-sized channel at $\retau=180$.
(\bbb) Evolution during a burst in the minimal channel in (\aaa) of the instantaneous: 
\chndot, box-averaged velocity gradient at the wall;
\solid, root-mean-squared intensity of the streamwise-averaged value of $u_1$ in $x_2^+\in
(25-50)$, representing the strength of the streak;
\dashed, same for $\omega_1$, representing the strength of the streamwise roller.
All quantities are normalized with their long-time average. The symbols correspond to the
snapshots in 
(\ccc), which show the evolution of the buffer layer before and during the burst. The
grey central object is the low-velocity streak, $u_1^+=-4$, and the shorter coloured ones
are the vortices, $\omega_1^+=\pm 0.25$. Objects taller than $x_2^+=80$ have been removed
for clarity. The view looks towards the wall, with the flow from top to bottom; time moves
from left to right, with an interval between frames $\Delta t^+\approx 8$. Axes move
downstream with a velocity $c^+=7.6$, to keep the wave in the streak approximately fixed.
}%
\la{fig:minmovie}
\end{figure}

Minimal simulations of three-dimensional channels appeared soon after \citep{jim:moi:91}.
They were at first restricted to the viscous layer near the wall, where they provided the
first evidence that the empirical spanwise streak separation, $\Delta_3^+\approx 100$, is
essentially a critical Reynolds number below which turbulence cannot be sustained. These
solutions were not steady, but they contained a wavy streamwise-velocity streak with two flanking
staggered quasi-streamwise vortices (see the individual frames in figure
\ref{fig:minmovie}\ccc), strongly reminiscent of the structures that had been educed from visual
inspection of boundary layer simulations by \cite{rob:91}, or by machine processing of
large-box channels by \cite{stretch90}.

An intuitive interpretation of minimal flow simulations is that they substitute the disordered
arrangement of the structures in real turbulence by a periodic `crystal' in which the
structures can be studied more easily. The surprising observation is that, even after what
is clearly a major change in the dynamics, the low-order statistics of the minimal flow are
essentially correct (figure \ref{fig:minmovie}\aaa). This strongly suggests that the
structures isolated by a minimal cell are fundamentally autonomous, with dynamics that
depend only weakly on the interaction with their neighbours, or at least on the detailed
geometry of those interactions. They therefore satisfy the basic criterion for a coherent structure, 
as defined in the introduction to this article.

A striking characteristic of minimal turbulent solutions is that they burst intermittently
and irregularly. The streak is approximately straight and steady most of the time, and
occasionally meanders and breaks down into a burst of vorticity and wall-normal velocity
that recalls the observations of tracers in early experiments with boundary layers
\citep{kli:rey:sch:run:67}. The cycle then restarts \citep{jim:moi:91}. The evolution of
several box-integrated quantities during a burst is presented in figure
\ref{fig:minmovie}(\bbb). Different quantities peak at different moments during the burst.
The streamwise-averaged streamwise velocity grows first, followed by the formation of a
roller indicated by the growth of the streamwise-averaged $\omega_1$, and finally by the
velocity gradient at the wall. These quantities are compiled in figure
\ref{fig:minmovie}(\bbb) over a flow slab near $x_2^+ = 40$, which is where the flow in
figure \ref{fig:minmovie}(\ccc) suggests that most of the bursting activity is concentrated.
Averages at different distances from the wall result in slightly different timing relations,
and it is risky to draw conclusions from a single flow realisation, but the general sequence
of events is consistent with the picture derived by \cite{jim:13a} using the temporal
cross-correlation of different variables in bursts farther from the wall. It is interesting
that the wall friction is a trailing indicator of bursting, and that the velocity gradient
at the wall is still increasing when most of the burst appears to have subsided. This
suggests that the burst originates far from the wall and migrates downwards. The evolution
of the corresponding flow fields is given in figure \ref{fig:minmovie}(\ccc), which only
includes structures fully contained below $x_2^+=80$. Soon after the last frame in figure
\ref{fig:minmovie}(\ccc), most of the vorticity fluctuations migrate above the buffer layer,
and disappear from our visualisation box. The streak disorganises for a while, to reform
later.

Refinements of the idea that an  instability of the streak originates streamwise vortices,
which in turn reinforce the streak, led to the codification of a self-sustaining process
(SSP), first in minimal low-Reynolds number flows by \cite{Hamilton95}, and
eventually in the near-wall layer of larger-scale channels by \cite{Schoppa02}.
   
\begin{figure}
\vspace*{3mm}%
\centerline{%
\includegraphics[width=.85\textwidth,clip]{\arpath 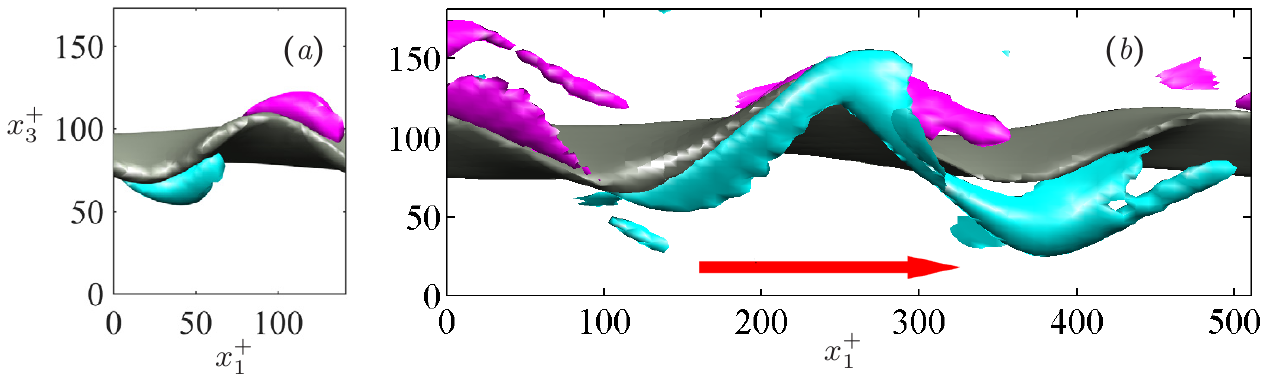}%
}%
\caption{%
Two examples of simple solutions in a minimal box that has been stabilised by filtering the
simulation above a distance $\delta$ from the wall \citep{jim:sim:01}. View and notation,
as in figure \ref{fig:minmovie}(\ccc), but the flow is from left to right, and the
isolevels are, $u_1^+=-3.5$ and $\omega_1^+=\pm 0.18$.
(\aaa) A steady wave with $\retau=118$, $L_1^+ =146$, $L_3^+ =177$, $\delta^+=49$. 
(\bbb) An oscillatory wave. The two wavelengths exchange amplitudes with a period of
$T^+\approx 370$ and an amplitude of the friction coefficient, $c'_f/\bra c_f\ket \approx
0.05$. $\retau=124$, $L_1^+ =520$, $L_3^+ =185$, $\delta^+=51$.
}%
\la{fig:jimsim}
\end{figure}

At about the same time, numerically exact three-dimensional permanent-wave solutions of the
Navier--Stokes equations were found for minimal Couette flow by \cite{Nagata90}. They also
contain a streak flanked by staggered vortices, and have dimensions that, when expressed in
wall units, agree with those of minimal turbulent flows. These solutions are unstable to
perturbations, and can only be computed by numerically imposing stationarity or temporal
periodicity, although enough numerical constraints stabilises them in the same way as in the
two-dimensional experiments mentioned at the beginning of this section. As in that case,
progressively releasing the constraints leads to bifurcations into limit cycles and chaos
\citep{jim:sim:01}. Two examples are given in figure \ref{fig:jimsim}. The one in figure
\ref{fig:jimsim}(\aaa) is a permanent wave, and the one in figure \ref{fig:jimsim}(\bbb) is
an oscillatory solution in which the two wavelengths in the box exchange amplitudes
periodically. The geometry of the two solutions is similar, and also similar to the chaotic
minimal channel in figure \ref{fig:minmovie}(\ccc), but the unsteadiness in figure
\ref{fig:jimsim}(\bbb) is relatively weak. The temporal oscillation of its friction
coefficient is about 5\%, compared to about 20\% in the minimal channel in figure
\ref{fig:minmovie}(\ccc). Most known exact periodic solutions share this weakness, but some
of the homoclinic orbits that have also been found do not. A quantitative comparison of the
properties of simple flow solutions, chaotic minimal flows, and large-scale wall turbulence
can be found in \cite{jim:kaw:sim:nag:shi:05}.

These separate lines of research eventually merged into a model in which the steady
solutions are understood as `frozen' versions of the SSP \citep{Waleffe97}, while related
periodic and homoclinic or heteroclinic orbits represent the bursts. A review can be found
in \cite{KawEtal12}, but this is a rapidly growing field, and interested readers are
encouraged to check recent individual research papers.

\begin{figure}
\centerline{%
\includegraphics[width=.95\textwidth,clip]{\arpath 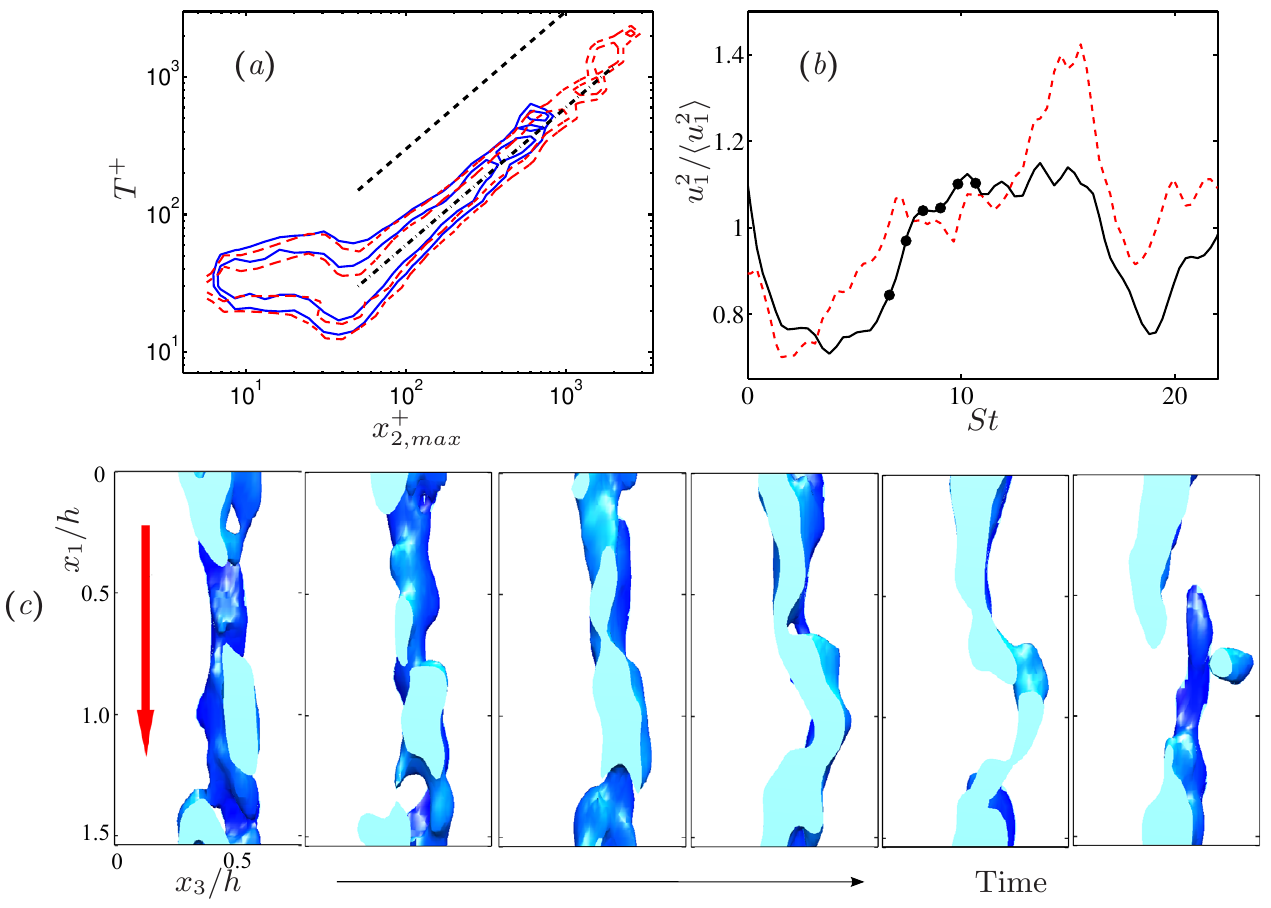}%
}%
\caption{%
(\aaa) Lifetime of sweeps and ejections, as function of the average distance from the top of the
structure to the wall \citep{loz:jim:14}. \solid, CH950; \dashed, CH2000. The contours
contain 50\% and 75\% of the data. The chain-dotted diagonal is $T^+=0.6 x_{2,max}^+$; the
dashed one is $T^+=3 x_{2,max}^+$.
(\bbb) Evolution of the kinetic energy of the streamwise velocity during a burst in a small
channel $(\retau=1850,\, L_1/h=\pi/2,\, L_3/h=\pi/4)$ which is minimal at $x_2/h\approx
0.25$ \citep{flo:jim:10}. \dashed, Averaged over $x_2/h\in (0.05, 0.1)$; \solid, $x_2/h\in
(0.1, 0.25)$. Each curve is normalised with its long-time average, and time is scaled with
the shear of the mean velocity profile at $x_2/h=0.2$. The symbols correspond to the
snapshots in
(\ccc), which show the evolution of the low-velocity streak, $u_1^+=-2$, leading to the burst.
Only the layer $x_2/h\in (0.1, 0.25)$ is represented. The view looks towards the wall, with
the flow from top to bottom. Time moves from left to right, with an interval between frames
$S \Delta t\approx 0.7$ $(\Delta t^+\approx 106)$. Axes move streamwise with
$c^+=20$, approximately equal to $U_1$ at $x_2/h=0.2$.
}%
\la{fig:logmovie}
\end{figure}

Although the first minimal simulations were truly minimal, in the sense of having the
smallest dimensions that can sustain turbulence, their main advantage from the point of view
of turbulence research is that they contain a single structure that can be easily studied
and followed in time without the need for tracking it in space. When the structure crosses
the downstream end of the computational box, an identical copy enters upstream. Minimal
boxes also allow low-order quantities, such as the intensity of the bursting structures, to be
studied by integration over the whole computational box, as in figure
\ref{fig:minmovie}(\bbb). It was eventually realised that what could be called `small-box'
simulations can be used in similar ways to study the larger structures of the logarithmic
layer. If the box size is large enough to contain several buffer-layer structures, but only one
of the larger structures that reach up to, say, $x_2=0.2 h$, that simulation can be used as
a tool to study the logarithmic-layer structure in isolation. \cite{flo:jim:10} found that
the critical box dimension is the spanwise period, $L_3$, and that boxes are minimal with
respect to the flow at $x_2\approx L_3/3$. If we identify this height with the centre of
gravity of the largest structure fitting in the computational box, the resulting aspect
ratio, $\Delta_3=L_3\approx 3 x_2\approx 1.5 \Delta_2$, is consistent with the p.d.f.s in
figure \ref{fig:pdf_uvwsters}(\ccc) and with the cross-plane correlations in figure
\ref{fig:corrcuts}. These maximal structures are not as smooth as the buffer-layer ones in
figure \ref{fig:minmovie}(\ccc), and the geometry of the burst is correspondingly more
complex, but they can be identified as coherent by neglecting the superimposed small scales,
in the same sense as for the streaks in figure \ref{fig:streaksall}(\bbb). In particular,
\cite{flo:jim:10} showed that logarithmic-layer structures also burst intermittently, and
that the process is a larger-scale version of the one in the buffer layer. The largest
structure that fits in the box is a streamwise streak flanked by a sweep and an ejection.
The streak occasionally meanders and destroys itself while creating a pulse of wall-normal
velocity, Reynolds stresses, and dissipation. It eventually reforms.

These logarithmic-layer bursts were studied in large-box simulations by \cite{loz:jim:14} as
part of their detailed study of wall-attached sweeps and ejections. The dependence of their
lifetimes on the maximum height of the structure is given in figure
\ref{fig:logmovie}(\aaa). There is a self-similar regime in which the lifetime is
proportional to the height, $T^+\approx 0.6 x_{2,max}^+$, and a lower range in which the
structures stay within the buffer layer, and their lifetime is $T^+\approx 30$. The two
regimes appear as continuations of each other, giving credence to the idea that they are
different aspects of the same phenomenon. If we empirically identify $x_{2,max}\approx 2
x_{2,c}$, where $x_{2,c}$ is the average height of the centre of gravity of the structures
during their lifetime, the self-similar relation in figure \ref{fig:logmovie}(\aaa) is
equivalent to $T^+\approx 1.5 x_{2,c}^+$, or $S_c T\approx 4$, where $S_c^+=1/\kappa
x_{2,c}^+$ is the mean shear. This is the order of magnitude of the lifetimes deduced by
\cite{jim:13a} for any shear flow, on the basis of the time it takes for a structure to be
sheared. The history of the box-averaged intensities of one such logarithmic-layer burst is
given in figure \ref{fig:logmovie}(\bbb), and the width of the burst approximately agrees
with the above estimate. The duration of the buffer-layer structures in figure
\ref{fig:logmovie}(\aaa) also agrees approximately with the width of the peaks in figure
\ref{fig:minmovie}(\bbb).

A different definition of the bursting period was used by \cite{flo:jim:10}, as the
spectral peak of the frequency spectrum of box-integrated quantities in small-box
simulations. Their result, $T^+=6 x_2^+\approx 3 x_{2,max}^+$ is approximately four times
longer than the tracking result in figure \ref{fig:logmovie}(\aaa). The two trend lines have
been included in that figure. They refer to different quantities: the Fourier result measures
the time interval between bursts, while the tracking one measures the duration of its intense
phase. The difference between the two periods measures the fraction of time (20\%) that the
flow is bursting. This minimal-box result probably overestimates the prevalence of bursting in
large-scale flows; \cite{loz:flo:jim:12} find that attached Q$^-$s only cover about 8\% of
the wall.

The meandering streak corresponding to figure \ref{fig:logmovie}(\bbb), filtered to the same
relative resolution as the buffer-layer figures \ref{fig:minmovie} and \ref{fig:jimsim}, is
shown in figure \ref{fig:logmovie}(\ccc).

\subsection{Causality}\la{sec:causal}

\begin{figure}
\centerline{%
\includegraphics[width=.90\textwidth,clip]{\arpath 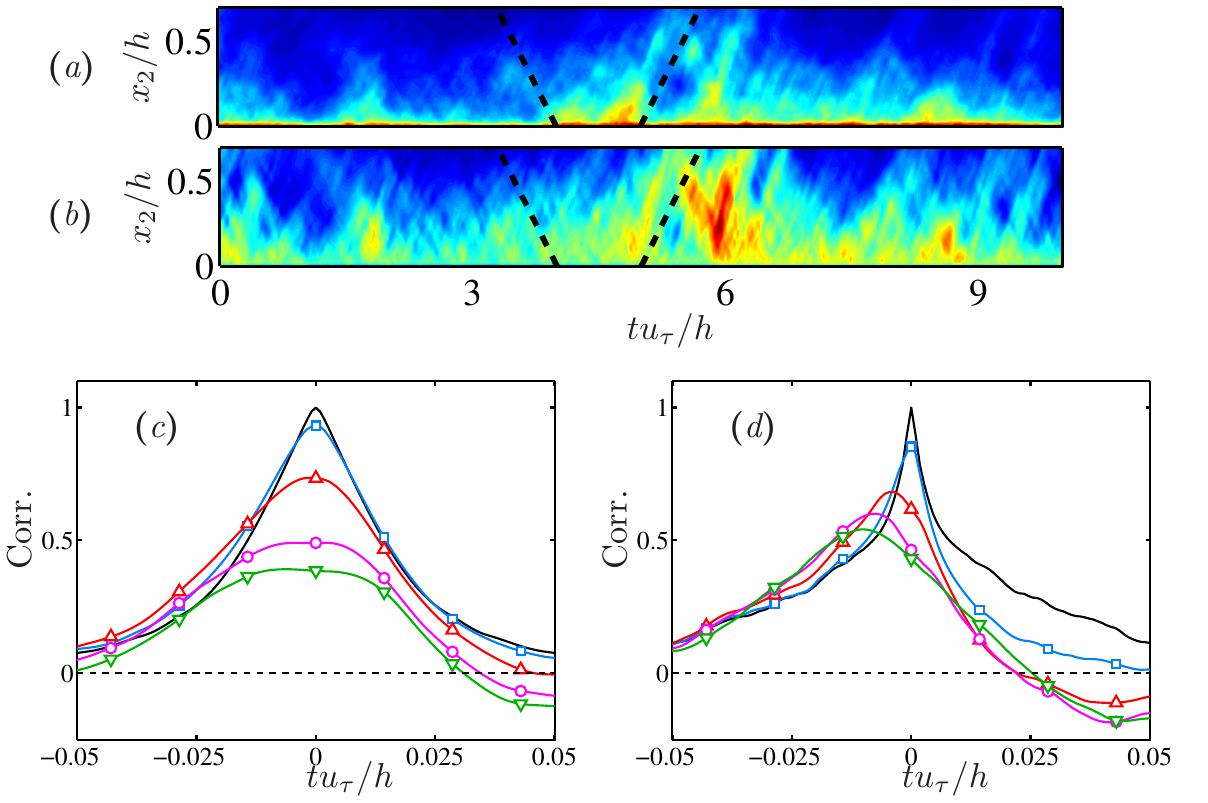}%
}%
\caption{%
Temporal evolution of the Reynolds stresses in the small channel in figures \ref{fig:logmovie}(\bbb,\ccc).
(\aaa) Evolution in the $(x_2,\, t)$ plane of the streamwise velocity fluctuations, averaged
over wall-parallel planes. The colour range (dark to light) is $\bra u_1^2\ket_{13}^+ \in
(0, 11)$.
The two dashed diagonals have slope $\dr x_2/\dr t = \pm \utau$.
(\bbb) As in (\aaa), for $-\bra u_1u_2\ket_{13}^+\in (0, 1.8)$. 
(\ccc) Temporal cross-correlation coefficient between $\bra u_1^2\ket_{13}$ at  $\tx_2^+=25$ and:
\solid, $x_2^+=25$; \squar, 70; \trian, 140; \circle, 275; \dtrian, 460 or $x_2/h=0.25$.
(\ddd)  As in (\ccc), for $\bra u_1 u_2\ket_{13}$. 
}
\la{fig:causes}
\end{figure}

We have mentioned in several occasions that turbulent structures can form far from the wall,
and that they appear to only require the presence of shear. Of course, this does not rule
out the possibility that in wall-bounded turbulence, where a wall is present, structures may
predominantly form near it, and it is interesting to ascertain whether, under those
circumstances, the structures of wall-bounded flows originate at the wall, or away from it.
At first sight, it looks that the wall should be the origin. We mentioned in
\S\ref{sec:balances} that the highest turbulence intensities are in the buffer layer, and
that so is the strongest shear, which is the source from which turbulence draws its energy.
Equation \r{eq:chan_prodplus} shows that the balance between dissipation and the production,
both of which are proportional to the shear, requires that $\omega'^{2+} \sim 1/x_2^+$ in
the logarithmic layer, so that vorticity is also concentrated near the wall. We discussed in
\S\ref{sec:balances} experiments that show that the dynamics of the near-wall region wall is
essentially independent of the outer flow, and it is tempting to hypothesise, from this
evidence, that turbulent fluctuations are created in the buffer layer and diffuse outwards
in some unspecified manner \citep[see, for example,][]{adr07}.

However, none of these results prove that the converse cannot also be true, and that the
outer flow can survive independently of the wall. It is clear that the wall is required to
create a shear, but it is possible that this is its main role, and that structures are
created everywhere, with each wall distance basically independent from others. In this
alternative model, turbulence draws its energy from the local shear, and either decays
locally or diffuses away from its origin, including towards the wall. The arguments for this view have
also been repeated often. The earliest one is the lack of influence of wall roughness on
the properties of the logarithmic and outer flows \citep{tow:76}. Roughness destroys the
details of the near wall region, including the near-wall peak of the turbulence intensity,
but it only has minor effects above a layer of the order of a few roughness heights
\citep{jim:rough:04}. Similarly, large-eddy simulations, which tend to represent poorly the
near-wall layer, reproduce features of the outer flow relatively well
\citep{moi:kim:82,kravMM96,PioBal02}. \cite{miz:jim:13} present computations of a turbulent
channel in which the wall is substituted by an off-wall boundary condition that mimics the
logarithmic, rather than the buffer layer, with relatively few deleterious effects, and we
saw in figure \ref{fig:condatt} that there is a smooth transition between the attached
eddies of wall-bounded flows and the structures of homogeneous shear turbulence, in which
there is shear but no walls.

All these experiments strongly suggests that the dominant root cause of the structures of
wall turbulence is the shear rather than the viscous wall, and it was indeed shown by
\cite{tue:jim:13} that even minor artificial changes of the mean shear produce major effects
in the fluctuations. However, the most direct evidence for the direction of causality comes
from simulations in the small channels discussed in the previous section, in which
individual structures manifest themselves in the global averages. The channels in
\cite{flo:jim:10} are minimal with respect to structures of height $x_2/h\approx 0.2$--0.3,
within the logarithmic-layer, and the corresponding structures can be detected by averaging
the intensities over wall-parallel planes. These planar averages are only functions of the
wall distance and of time, and are represented in the $x_2$--$t$ maps in figures
\ref{fig:causes}(\aaa,\bbb). The first of these two figures displays the evolution of the
plane-averaged intensity of the streamwise velocity, and figure \ref{fig:causes}(\bbb) displays
the tangential Reynolds stress. In both cases, it is evident that eddies move both towards
and away from the wall, with wall-normal velocities of the order of $\pm\utau$
\citep{flo:jim:10}. The dominant direction of causality is tested in figures
\ref{fig:causes}(\ccc,\ddd), which plot the cross-correlation of the planar averages of the
fluctuations in the buffer layer with those at different distances from the wall. Within the
distances plotted in the figure, the peak of the cross-correlation of the planar average
$\bra u_1^2\ket_{13}$ stays centred near $t=0$, but it gets wider and develops a flat top as
$x_2$ moves away from the wall, probably reflecting that the taller eddies are related to
those at the wall either at earlier or at later times. On the other hand, the correlations
of the mean Reynolds stress, $\bra u_1 u_2\ket_{13}$, drift towards {\em earlier} times as
$x_2$ increases, implying that the dominant evolution of these eddies in the logarithmic
layer progresses from the outside towards the wall, rather than the other way around.
Different variables behave differently, although most tend to behave like $u_1$ in figure
\ref{fig:causes}(\ccc). We could not find any case in which correlations moved outwards on
average at a rate similar to the inwards velocity of the Reynolds stress in figure
\ref{fig:causes}(\ddd).
 
It should be cautioned that these results could be affected by the small computational box. It
can indeed be shown that they are corrupted by the box above $x_2/h\approx 0.5$, where causality
reverses, but the wall-normal propagation velocity of individual strong $(u_1u_2)$ structures
was measured by \cite{loz:jim:14} in large channels, free of minimal-box effects. They found
that ejections $(u_2>0)$ move away from the wall with a distribution of velocities centred
around $\utau$, while sweeps $(u_2<0)$ move towards the wall at approximately the same rate.
Since the two types of structures tend to forms pairs of one sweep and one ejection, the net
wall-normal velocity of the pair approximately vanishes. Note that, in this description, the
wall only plays an essential role in creating the shear, but not in the generation of either
sweeps or ejections.

\section{Theoretical models}\la{sec:theory}

Turbulence has defied a theoretical solution for almost 200 years. Traditionally, the main
problem has been the description of the inertial scales, with their large number of degrees
of freedom and the intuitively paradoxical property that the viscous energy dissipation is
empirically believed to remain finite even when viscosity tends to zero. In this article, we
have been interested in the apparently simpler problem of the large scales that extract
energy from a mean shear and transfer it to the cascade, but even they are not fully
understood. Because they draw energy from the shear, large scales have to be, up to a point,
`linear' in the sense that their dynamics is tightly coupled to the mean flow. This led to
the hope that the large eddies of turbulent shear flows would be related to those of
transition, with whom they share the energy-extracting role. This hope was partly realised
by the discovery of large coherent structures in free-shear flows, such as shear layers,
jets and wakes, which were soon understood to originate from the Kelvin--Helmholtz
instability of the inflectional mean velocity profile \citep{brownr}. In free-shear layers,
the correspondence between linear theory and empirical observations remains quantitative
even when the amplitude of the turbulent fluctuations is not infinitesimal, suggesting that
the energy-containing structures can be described linearly as long as the flow is unstable.
Only when the flow grows to be thick enough to make the linear instability neutral (itself a
nonlinear effect), does nonlinearity become relevant to the structures \citep{gkw85}. This
is the model encapsulated by the observation in \S\ref{sec:production} that the Corrsin
parameter $S^*$ is large in shear flows.

Linearised models of wall-bounded flows predate those in free-shear layers. The early hope
was that their mean velocity profile would be determined by the requirement that it should
be marginally stable with respect to linearised instabilities similar to those in
boundary-layer transition \citep{malkus56}. Similar models later became popular in other
areas of physics under the name of self-organised criticality \citep{BakTanWie:88}, and
posit that any deviation from the equilibrium profile triggers an instability that restores
the marginal state. Unfortunately, \cite{reytied67} showed that the mean profile of a
turbulent channel is far from being linearly unstable. As we will see below, this does not
necessarily mean that linear processes are irrelevant, but it makes wall-bounded turbulence
harder to model than the free-shear case. In the latter, the role of nonlinearity is mostly
restricted to limiting the amplitude of linear instabilities. In wall-bounded turbulence,
the \cite{corrsin58} argument still holds: the shear is the fastest dynamical process for
the large scales, and linear mechanisms dominate, but the lack of linearly unstable modes
implies that some form of nonlinearity has to be an integral part of the energy extraction
process.

Large scales do not typically have a dimensionality problem, because their size is of the
order of the thickness of the flow, and they carry most of its energy. However, wall-bounded
turbulence is special in this respect because the number of structures involved in
energy generation increases without bound with the Reynolds number. The reason is that the
integral scale is proportional to $x_2$ in the logarithmic layer. The number of
integral-scale structures per unit volume is then proportional to $x_2^{-3}$, whose integral
over $x_2$ is dominated by the lower limit at the edge of the buffer layer. The result is
that the number of integral-scale structures per unit projected area is
proportional to ${\retau}^2$ \citep{miz:jim:13} and, while the large scales of free-shear
turbulence can approximately be treated as a relatively low-dimensional dynamical system in
which most degrees of freedom are relegated to the dissipative cascade, the logarithmic
layer is intrinsically high-dimensional and multi-scale.

\subsection{Linear approximations}\la{sec:linear}
 
The key dichotomy in the linear behaviour of shear flows is between the modal instabilities
characteristic of self-adjoint operators, and the transient growth in non-normal ones. In
the former, fluctuations can be expanded in a set of orthogonal eigenfunctions, and
stability analysis reduces to determining the properties of the corresponding eigenvalues.
If a temporal eigenvalue has a positive real part, the associated eigenfunction grows
exponentially and eventually dominates. Stable eigenfunctions die exponentially, and
orthogonality ensures that individual eigenfunctions can be treated as essentially
independent of one another as long as the system remains linear and autonomous.
 
On the other hand, the eigenfunctions of non-normal operators are not necessarily
orthogonal. In general, there will be cancellations among some of the eigenfunctions
contributing to a given initial condition and, even if the evolution of the system is such
that all the eigenvalues are stable and decay exponentially, the balance of those
cancellations may change during the decay. The result is that some initial conditions grow
for a while, typically algebraically, even if they eventually decay exponentially when all
the involved eigenfunctions do so. If a group of eigenfunctions are almost parallel to each
other, the effect can be large. A recent survey of applications to hydrodynamic transition
is \cite{schm07}, and a textbook account is \cite{schmid01}.
 
For parallel flows, the linearised version of the Navier--Stokes equations is
\beq
(\p_t + U_1 \p_1)\, u_i = -u_2 U'_1 \delta_i^1 -\p_i p +\nu \nabla^2 u_i,
\la{eq:LNS}
\eeq
where primed capitals denote derivatives with respect to $x_2$, and $\delta_i^j$ is
Kronecker's delta. It can be reworked into the Orr--Sommerfeld equation
for $u_2$ \citep{dra:rei:81},
\beq
(\p_t + U_1 \p_1)\underbrace{\nabla^2}_{\mbox{\scriptsize Orr}} u_2 =
    \underbrace{U''_1 \p_1 u_2}_{\mbox{\scriptsize K-H}} +\nu \nabla^2 \nabla^2 u_2,
\la{eq:OS}
\eeq
and the Squires equation for the wall-normal component of the vorticity,
\beq
(\p_t + U_1 \p_1)\, \omega_2  = - \underbrace{U'_1 \p_3 u_2}_{\mbox{\scriptsize lift-up}} + 
       \nu \nabla^2 \omega_2.
\la{eq:Sq}
\eeq
Several terms in these equations can amplify fluctuations. The leading operator in the
left-hand side of \r{eq:LNS}--\r{eq:Sq} is advection by the mean profile. It does not
amplify fluctuations by itself, but, because $U_1$ is a function of $x_2$, it deforms them,
generally tilting them forward. Because \r{eq:OS} is autonomous in $u_2$, it
determines whether the flow is stable or not. The term marked `K--H' on its right-hand side
is the deformation of the mean shear by the cross-shear velocity fluctuations. It rearranges
the distribution of vorticity without amplifying it, but it is responsible for the
Kelvin--Helmholtz modal instability when the mean velocity profile has a shear maximum
(i.e., an inflection point) because it moves vorticity fluctuations to where originally
there were none. Because wall-bounded flows are not inflectional in the absence of strong
pressure gradients, this term is generally not important for them.

\begin{figure}
\vspace*{.03\textwidth}%
\centerline{%
\includegraphics[width=.95\textwidth,clip]{\arpath 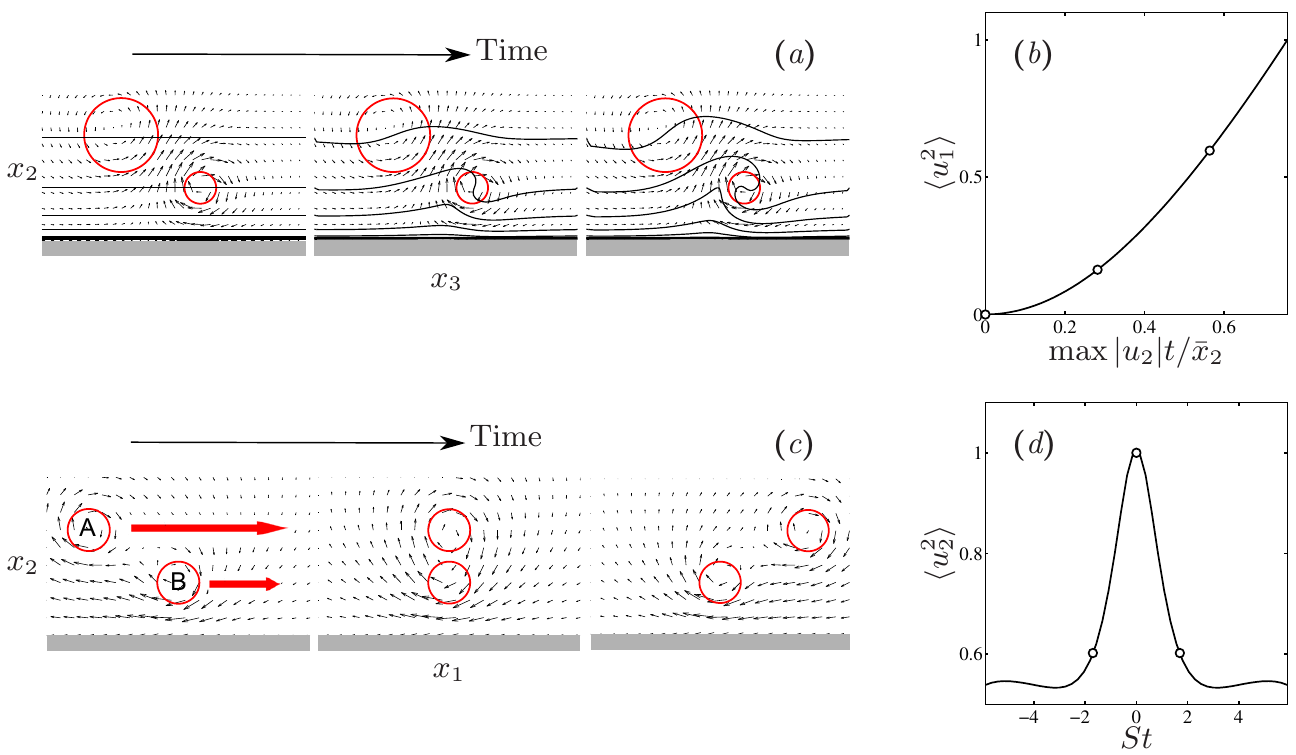}%
}%
\caption{%
(\aaa) Sketch of the evolution of the lift-up mechanism in the cross plane $(x_3, x_2)$. A
pair of stationary vortices with zero net circulation, represented as red circles, deforms the
background shear, represented by the solid isolines. Time increases from left to right.
(\bbb) Evolution of the perturbation energy $\bra u_1^2\ket$ during (\aaa).
Time is normalised with the maximum of $u_2$ and with the distance from the wall to
the centre of gravity of the vortex pair.
(\ccc) Sketch of the evolution of the Orr mechanism in the streamwise plane $(x_1, x_2)$.
Time increases from left to right. As the upper vortex is advected by the ambient shear
above the lower one, their wall-normal velocities reinforce each other. When the vortices
are not vertically aligned, their velocities tend to cancel.
(\ddd) Evolution of the integrated energy of the wall-normal velocity during the evolution
in (\ccc). Time is normalised with the shear advecting the vortices.  The symbols in (\bbb,\ddd)
are the times of the snapshots on the left. The units for the amplitudes are arbitrary.
}
\la{fig:linsketch}
\end{figure}

The advection of $\omega_2$ in Squires equation \r{eq:Sq} is forced by the wall-normal
fluctuations generated by \r{eq:OS}. This does not create instability by itself, but it can
amplify fluctuations considerably. The relevant term is marked as `lift-up', and represents
the deformation of the mean velocity profile by the spanwise variations of $u_2$. This is
one of the terms responsible for non-orthogonal eigenvectors of the evolution operator
\r{eq:OS}--\r{eq:Sq}. It leads to the formation of the streaks of $u_1$, because it acts
most strongly on long narrow features for which $\omega_2\approx \p_3 u_1$. A sketch is
given in figure \ref{fig:linsketch}(\aaa), drawn in the $(x_3, x_2)$ cross-flow
plane for simplicity. The solid contours are $U_1$, which in this case increases from
bottom to top. The arrows are the field of transverse velocities due to a pair of streamwise
vortices, shown here as stationary. Lift-up works by moving low-velocity fluid from the wall
upwards and vice versa. In the particular case of figure \ref{fig:linsketch}(\aaa), it is
creating a low-velocity streak in the centre of the sketch, where the $U_1$ contours move
away from the wall. Figure \ref{fig:linsketch}(\bbb) shows the growth of the perturbation
energy, which is typically algebraic. Unless the mean velocity profile stops growing far
from the wall, there is no obvious way to limit the growth of the lifted streak, because
high-velocity fluid keeps being moved towards the wall, and vice versa. However, the growth
rate typically slows after some time, because fast fluid has to be drawn from the upper
layers where the transverse velocities created by vortices located near the wall are weaker.
Nonlinearity also limits the rate of growth by mixing laterally the fluid in the streak,
as is beginning to happen in the right-most snapshot in figure \ref{fig:linsketch}(\aaa).
The net effect is to rearrange the mean velocity profile. In the case of the figure, it
flattens the profile near the wall, and creates a sharper shear layer above the vortex pair.
  
The other non-normal part of the evolution is the Laplacian marked as `Orr' in the left-hand
side of \r{eq:OS}. If the advection in \r{eq:OS} acted on $u_2$ instead of on $\nabla^2
u_2$, its effect would just be to reorganise the velocity fluctuations without changing
their amplitude. However this deforms their shape, violating continuity. The Laplacian
represents the reaction of the pressure to restore continuity. The effect of this term,
first discussed by \cite{orr07a}, is to amplify backwards-tilting packets of $u_2$ as they
are carried towards the vertical by the shear. The effect is transient, with a time scale
$O(S^{-1})$, because $u_2$ is eventually damped again as the packets are tilted past the
vertical. However, the net effect of a transient burst of $u_2$ is to create a streak of
$u_1$ through the lift-up mechanism mentioned above. The damping effect of the tilting on
$u_2$ is due to the change of $\p_2 u_2$ as the structure becomes vertically stacked. The
streaks created by the lift-up are also tilted and vertically thinned by the shear, but
$\p_2 u_1$ is not in the continuity equation, and pressure does not damp them. As a
consequence, the effect of the Orr bursts of $u_2$ is long-lasting.

The intuitive mechanism behind the Orr amplification is sketched in figure
\ref{fig:linsketch}(\ccc), which portraits, for simplicity, a two-dimensional case in the
longitudinal $(x_1, x_2)$ plane. The figure shows two corrotating
vortices being advected by the mean flow in such a way that vortex `A' overtakes `B'. When
the two vortices are far apart, their velocity fields are independent. As they get closer,
the down-wash of `A' initially cancels the up-wash of `B', and the total energy of the pair
decreases slightly. Later, when they pass above each other, their velocities mutually
reinforce, the maximum wall-normal velocity doubles and, even if the integral of $u_2^2$ is
concentrated over a smaller area, the integrated energy doubles. As the vortices separate
again, the reinforcement decreases and the energy amplification is lost. The evolution of
the integrated energy during the events in figure \ref{fig:linsketch}(\ccc) is plotted in
figure \ref{fig:linsketch}(\ddd). It should be clear from this explanation that Orr's is a
robust mechanism which does not depend on the details of the velocity profile responsible
for the overtaking of the two structures, and that similar results would be obtained if the
two vortices were substituted by a tilting elliptical vorticity patch, or by a
three-dimensional vorticity distribution. Note that pressure is not explicitly invoked by
this second explanation, but that the expression for the velocity in terms of the vorticity
distribution implies the Biot--Savart law, which incorporates continuity.

It is important to stress that neither lift-up nor Orr are intrinsically linear processes,
although they subsist in the linear approximation, and are most easily analysed in that
limit. The effects depicted in figure \ref{fig:linsketch} work equally well for strong
perturbations as for weak ones. In particular, the Orr superposition of flow fields in
figure \ref{fig:linsketch}(\ccc) only depends on having eddies moving on average at
different speeds at different heights, and it does not require that the mean velocity
profile exists as such at any moment. Of course, if the overtaking eddies interact with
other perturbations of comparable intensity, the effect of Orr superposition might be
difficult to isolate. Linearisation does not add anything to the equations, and the effects
surviving the removal of the nonlinear terms constitute the core of the evolution operator under the
conditions in which linearisation makes sense. The linearity to which we allude in this subsection is a
simplification that recognises that, for the large scales in the presence of shear, the
mechanisms in the linearised equations are fast enough to remain relevant even in the
presence of nonlinearity, and strong enough to be responsible for many of the locally
intense structures.

The viscous terms on the right-hand side of \r{eq:OS} and \r{eq:Sq} act to damp the
fluctuations, specially when the eddies are thinned by tilting at the beginning or at the end of a
burst. Viscosity can be responsible for instabilities that are important
in transition, but the times involved are much longer than the shear time, and they
are not usually relevant in turbulence.

Because we are considering linear dynamics in this section, the evolution of any initial
condition subject to \r{eq:OS}--\r{eq:Sq} is most easily analysed for individual spatial
Fourier modes. An appendix in \cite{schmid01} provides a practical computer code to optimise
the amplification history of some chosen norm,
\beq
A_\chi = \| \chi(t) \|^2 / \|\chi(0)\|^2, 
\la{eq:amp}
\eeq
where $\|\chi\|^2 =\int |\chi|^2\dd V$ is integrated over the whole flow as a function of
time. The analysis provides the most amplified initial condition and its temporal evolution,
including the time for maximum amplification. The only input required is the mean velocity
profile and the viscosity. The most amplified initial conditions play the same role as the
most unstable eigenfunctions of modal instabilities. For any random initial condition, the
system tends to select the projection over the most amplified direction, whose evolution
eventually dominates the flow. The difference is that, while modal eigenfunctions grow while
retaining their shape during their evolution, non-normal initial conditions change as they
evolve. The shape of the most dangerous initial condition is in general very different from
the final amplified perturbation that dominates the statistics. Also, because the growth in
the amplitude of each initial condition is not a simple exponential, the dominant
perturbation at different times may correspond to different initial conditions, as each
perturbation grows and decays, to be substituted by another one with a slower evolution
time.

This analysis was first applied to the velocity profile of a turbulent channel by
\cite{but:far:93}, who found that the most amplified initial condition is a
backwards-leaning set of oblique rollers with weak $u_1$, which evolve into strong streaky
structures of the streamwise velocity. The aspect ratio of the different perturbations is
fixed by the wavelengths chosen for the analysis, but the most amplified initial conditions
are those which evolve into infinitely long structures with $\lambda_3/h\approx 3$, filling
most of the height of the channel. The buffer-layer streaks, with $\lambda_3^+\approx 100$,
only become prominent if the time during which the system is allowed to evolve is
artificially restricted.

\begin{figure}
\centerline{%
\includegraphics[width=.99\textwidth,clip]{\arpath 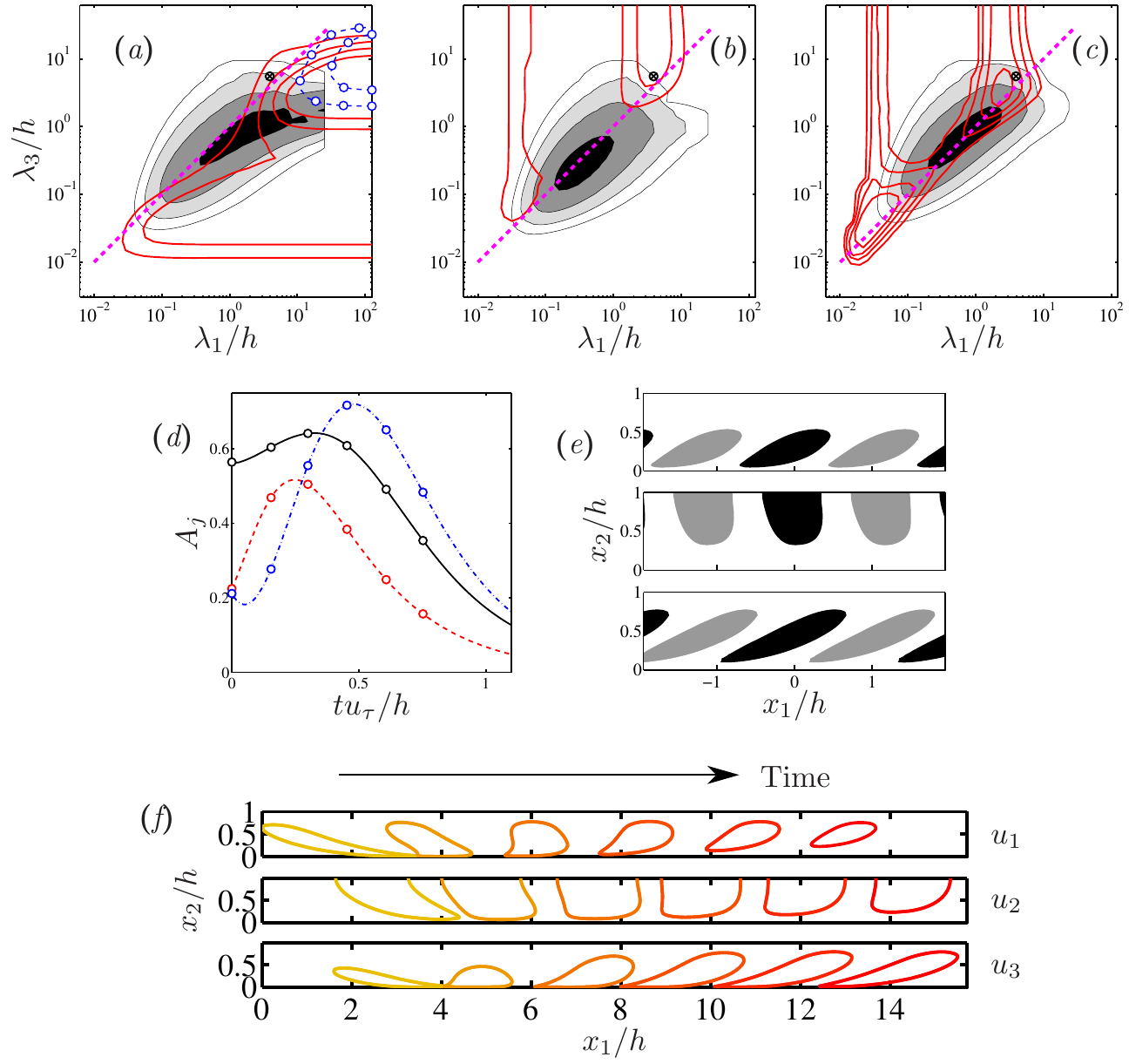}%
}%
\caption{%
Maximum amplification under linearised dynamics of the energy integrated over the whole
channel height, as a function of the wall-parallel wavelengths. Full eddy viscosity and
$\retau=2000$.
(\aaa) \solid, Kinetic energy amplification, $A_{q,max}= \sqrt{2}, 2, 2\sqrt{2}, \ldots$;
\dashcir, time for maximum amplification, $\utau t_{max}/h=1$ and 2; the shaded contours are
the vertically integrated energy spectrum, $\Phi_{qq}$, for CH2000. The dotted diagonal is
$\lambda_1=\lambda_3$.
(\bbb) Amplification of $u_2$ for the optimal histories in (\aaa), and $\Phi_{22}.$ 
(\ccc)  For $u_3$ and $\Phi_{33}$.
%
%
%
(\ddd) Amplification history of the three velocity components in the linearised case marked
by a symbol in (\aaa--\ccc): $\lambda_1/h=3.9$, $\lambda_3/h=5.55$. \solid,
$u_1$; \dashed, $u_2$; \chndot, $u_3$. Symbols are the snapshots in (\fff--\hhh).
(\eee) Leading PODs of the three velocity components for CH2000 and $\lambda_1/h=3.2$. From
top to bottom, $u_1$, $u_2$ and $u_3$. Black patches are $u_j>0.5$, and grey ones are
$u_j<-0.5$.
(\fff) Evolution of the most amplified linearised solution for the case in (\ddd). Time and
flow run from left to right, with $\Delta t\utau/h\approx 0.15$. The advection velocity is
adjusted arbitrarily to fit all the snapshots within the plot. From top to bottom, the contours
are $u_1=4$, $u_2=1$ and $u_3=3$, in arbitrary units.
}
\la{fig:tgrowthis}
\end{figure}

\Citet{ala:jim:06} repeated the analysis adding to the right-hand side of
\r{eq:OS}--\r{eq:Sq} the eddy viscosity required to maintain the mean velocity profile
\citep[for the extra terms arising in the equations, see][]{puj:etal:09}. Because the eddy
viscosity depends on $x_2$, the results are roughly comparable to introducing a damping time
that varies with the distance from the wall, and both the outer and the buffer-layer streaks
appear naturally. An example of the results is given in figures
\ref{fig:tgrowthis}(\aaa--\ccc), where the maximum amplification of each velocity component
is compared to the vertically integrated spectral density of a turbulent channel at the same
Reynolds number. The three figures refer to the histories and initial conditions that
maximise the amplification, $A_q$, of the kinetic energy. The maximum energy amplification is
given in figure \ref{fig:tgrowthis}(\aaa) as a function of the wall-parallel wavelengths,
and reflects mostly the growth of $u_1$. The outer streaks appear at long wavelengths
$(\lambda_1\gg h)$ and $\lambda_3\approx 3h$, and it is interesting that they approximately
agree with the results of \cite{but:far:93} using a very different viscosity model. This
supports our previous observation that viscosity is secondary to the mechanism of energy
production. In fact, the analysis of the evolution of $u_2$ can be done using an inviscid
code, because the impenetrability boundary condition is inviscid, but $u_1$ and $u_3$
require some viscosity to enforce no-slip at the wall \citep{jim:13a}.

As in \cite{but:far:93} the intensity profiles of the large amplified structures in
\cite{ala:jim:06} fill most of the channel height. They have dimensions that scale in outer
units for different Reynolds numbers, and appear to be fairly universal features of
equilibrium shear flows. The spectrum in figure \ref{fig:tgrowthis}(\aaa) shows that the
spanwise wavelength of this outer amplification peak agrees well with that of the longest
turbulent energy in channels.

There is a secondary, weaker, amplification ridge near $\lambda_3^+=100$, which corresponds
to the buffer-layer streaks. It has intensity profiles concentrated near the wall, and
scales in wall units with the Reynolds number. A curious feature of both ridges is that the
maximum amplification occurs for infinitely long structures, while we have seen that the
streaks in real flows are long, but probably not infinite. This discrepancy should not be
taken too seriously. The use of a wavelength-independent eddy viscosity for all the
fluctuations should only be understood as a rough
approximation to the energy dissipation by smaller-scale turbulence. Actual nonlinear processes
can be expected to act over times of the order of the eddy turnover, which is $O(h/\utau)$ for the
scales represented by the outer ridge. It is intuitively clear that the higher amplification
of the longest structures can only be achieved at the expense of longer amplification times.
This is confirmed by the analysis. Isolines of the time required to reach maximum
amplification are included in figure \ref{fig:tgrowthis}(\aaa), and suggest that the length
of the streaks is probably limited by evolution times longer than a few turnovers. We gave a
similar argument in \S\ref{sec:correl} to justify the shorter streamwise correlations of
boundary layers, and we mentioned above that \cite{but:far:93} used a time limit to retrieve
the inner amplification ridge. Note that the turnover time, as well as the eddy viscosity,
are nonlinear effects that depend on the finite amplitude of the fluctuations, and have 
been artificially superimposed on the linear analysis.

Transient-growth analysis selects the maximum singular value from the spectrum of the
evolution operator, but there often are other singular values which are close to the first
one. Because of the symmetry of the velocity profile in channels, the spectrum is formed by
pairs of very similar magnitude, corresponding to perturbations that are symmetric or
antisymmetric with respect to the centreline. For short wavelengths, the amplification of
the two members of the pair is almost the same, because they represent essentially
independent solutions evolving near one of the walls. This symmetry breaks for longer or
wider wavelengths, and solutions with a symmetric $u_2$ spanning the whole channel tend to
be slightly more amplified. The optimal solutions represented in figure
\ref{fig:tgrowthis}(\aaa) mostly have this symmetry, but both symmetric and antisymmetric
solutions can be expected to occur in real cases.

This mix of symmetry classes is found both in the correlations and in the PODs. For example,
the correlation of $u_1$ in figure \ref{fig:corrcuts}(\ccc) has a negative lobe on the
opposite half of the channel, even if it is compiled relatively near the lower wall,
suggesting an antisymmetric $u_1$. The issue was avoided for the filtered PODs in
\S\ref{sec:eddies} by restricting the analysis to the lower half of the channel, because
using the whole channel interferes with the compactification of asymmetric variables whose
integral vanishes. This is why the PODs of all variables in figure \ref{fig:pod} appear to
be restricted to a half channel, but we will see below that full-channel structures reappear
when the correlation of long wavelengths is analysed over the whole flow.
 
Figures \ref{fig:tgrowthis}(\bbb,\ccc) display the maximum amplification of the two
transverse velocity components for the same optimal histories used in figure
\ref{fig:tgrowthis}(\aaa). The strongest amplification of the different components does
not take place for the same wavelengths, nor at the same times. For example, the maximum
amplification of $u_2$ occurs at the two ends of the $\phi_{22}$ spectrum, but it is also large at very
wide wavelengths for which the kinetic energy is not amplified, and where the spectrum of
turbulence has no energy. The linear evolution of $u_2$ is controlled by the Orr mechanism,
which is most efficient for wide $(\lambda_3\gg \lambda_1)$ waves. But the lift-up depends
on the spanwise derivative in the right-hand side of \r{eq:Sq}. Very wide $u_2$
perturbations, even if they are amplified, cannot create a streak and do not result in
long-lasting energy amplification. This also means that these perturbations are not found in
turbulence, because there are no initial conditions in that range of wavelengths to serve as
seeds for two-dimensional waves of $u_2$. A similar analysis applies to $u_3$ in figure
\ref{fig:tgrowthis}(\ccc).
    
Figure \ref{fig:tgrowthis}(\ddd) shows the amplification history of the wavelength
combination marked by a symbol in figures \ref{fig:tgrowthis}(\aaa--\ccc). The different
evolution of the three velocity components is clear, as is the transient nature of their
amplification. Six snapshots of the evolution of the perturbation field of each of the three
velocity components are shown in figure \ref{fig:tgrowthis}(\fff). Note the difference
between the initial and final shape of the perturbations, as mentioned when discussing
transient growth at the beginning of this subsection. All the structures are progressively
tilted forward during their evolution, and it is interesting to note the similarity between
the shape of the structures at their most amplified stage (between snapshots three and four)
and the PODs of a real channel at a similar $\lambda_1$, displayed in figure
\ref{fig:tgrowthis}(\eee).

A quantitative comparison of the correspondence between bursting in fully nonlinear
small-box channel simulations and linear analysis is given in \cite{jimenez:2015}. While that
reference should be consulted for details, a short summary of its conclusions is that the
evolution of intense `minimal' Fourier modes of the wall-normal velocity (i.e. those with
wavelengths similar to the box dimensions) can be described well as a linearised transient
Orr burst. For the simulations in that paper, which are minimal around $x_2/h=0.25$, linear
prediction works for time intervals of the order of $t\utau/h \approx 0.15$, which is
approximately half the bursting lifetime. The time fraction of the flow history that can
thus be described is approximately 65--70\%, and accounts for an even higher fraction of the
total Reynolds stress. Bursts are essentially inviscid, and the addition of an eddy
viscosity does not improve predictions.

Linearisation fails during periods of weak perturbations because the flow cannot be
described by a few wall-normal eigenfunctions, but nonlinearity is overwhelmed by the linear
process while bursting is active. This recalls the observation at the
beginning of this section that nonlinearity in free-shear layers only matters when the
thickening of the layer pushes the prevailing wavelength outside the instability range of
the linear Kelvin--Helmholtz mechanism. In both cases, nonlinearity only becomes important for
the large scales after the flow decouples from the shear, and it is striking and contrary to
common intuition, that the strongest events are those best described by linearised models.

On the other hand, it should be emphasized that these conclusions refer to the dynamics of a
few Fourier modes chosen to represents coherent structures. For example, each snapshot in
figure \ref{fig:tgrowthis}(\fff) represents a single period of a uniform wave train, and
the PODs in figure \ref{fig:tgrowthis}(\eee) are spatially periodic. Any discussion of the
evolution of realistic initial conditions should be able to deal with wave packets that include
a range of wavelengths, in the same spirit as the compact eddies discussed in
\S\ref{sec:eddies}.

A summary of the ongoing work on the generalisation to full-sized simulations, in which
several large scales coexist, is \cite{encinar16}. Early indications are that
the above results continue to hold for individual wavelengths, with relatively little
interaction among sufficiently different scales.
      
\subsection{Nonlinearity}\la{sec:nonlinear}

\subsubsection{The relevance of streaks}\la{sec:streaks2}

Even if transient linear amplification explains most of the dynamics of shear flows, it
cannot constitute a complete theory for them, because perturbations do not survive after a long time.
Any initial condition eventual dies, and the flow laminarises. The piece missing from the
puzzle is what closes the self-sustaining process qualitatively described in
\S\ref{sec:minimal}. The linear mechanism in the previous section describes the formation of
the streaks by the vortices (or by the rollers in the logarithmic layer), and the
strengthening of the rollers once they have been initiated, but it lacks a way of initiating
them. This initiation mechanism cannot be linear, because we know from \cite{reytied67} that
the linearised equations for a channel are stable. Nonlinearity is required as an
integral part of the regeneration cycle.

Most nonlinear proposals centre on the influence of the streaks. This is probably unavoidable, 
because we have seen that the rollers, or their associated sweeps and ejections, are damped at
the end of a burst, and that the mean profile is stable. Only the streak is left. Note that,
even if many of the analyses mentioned below are linear with respect to the streaky flow,
they all require that the streak have non-infinitesimal amplitude, and are therefore
nonlinear with respect to the mean flow.

Early models centred on inflectional instabilities associated with the flanks of the streaks
\citep{swe:black:87}, and many later studies examined the structure of these
instabilities. The details depend on the velocity distribution within the streak, and
\cite{Schoppa02} performed the analysis using empirical flow fields obtained from direct
simulations of channels. Their results were twofold. In the first place, they found many
streaks whose intensity was too low to be modally unstable, and many fewer that were strong
enough to be unstable. In the second place, they noted that even the stable streaks were
subject to non-modal growth. They concluded that modal instability was irrelevant to streak
breakdown, and that transient growth driven by the streak profile was the dominant process.
While suggestive, there are two problems with this conclusion. Firstly, the absence of unstable
streaks can be equally interpreted as an indication that instability is important. Unstable
flow patterns would not be found precisely because instability destroys them. One may
think of the low probability of finding upright pencils on a shaking table.

The second objection is more subtle. Transient growth is presented in \cite{Schoppa02} as a
property of the streak profile, implying that the energy of the fluctuations is drawn from
the energy of the streak. This is not a problem near the wall, where the energy of the
cross-flow velocities $(u'^2_2+u'^2_3)$ is a small fraction of $u'^2_1$, but it is more
problematic farther from the wall, where we saw in \S\ref{sec:anisotropy} that both energies
are comparable. If the transverse velocities had to obtain their energy from the streak, one
would expect a negative correlation between the two energies. The opposite is true: although
the streamwise energy leads the transverse one in the temporal cross-correlation of the two
quantities, the correlation is always positive, and both energies grow together over most of
the burst \citep[see figure 6 in][]{jim:13a}.

We saw in the previous section that the alternative model in which the transverse velocities
draw their energy directly from the mean shear explains the dynamics of the burst well,
independently of the presence of a streak, but the two models need not be incompatible. The
streak profiles used by \cite{Schoppa02} include the mean shear, and it is possible that
their transient growth is driven by the shear instead of by the presence of a streak. This
is made more plausible by their observation that the transient amplification is relatively
independent of the streak intensity, even in cases in which viscosity damps the initial
streak fast enough to essentially erase it during the growth of the perturbation (see their
figure 11). A possibility is that the role of the streak is to be a catalyst for the rollers, rather 
than their engine.

In fact, the nonlinear effect of the streaks is important for a more fundamental reason. We
have seen that the linearised Orr mechanism followed by lift-up explains a lot of the
dynamics of the velocity. It creates a transient burst followed by a more permanent streak.
Because the streak has a longer lifetime, it can also be sheared by the mean profile for a
longer time, resulting in somewhat longer structures than those of the transverse
velocities. But this cannot explain the large observed differences in the length of the
three velocity components, and we concluded in \S\ref{sec:correl} that each streak should
contain several bursts. This observation has been popularised elsewhere as that hairpins
collect into packets \citep{adr07}.
 
Linear perturbations of the mean velocity profile cannot explain this organisation, because
a linear process has no amplitude selection mechanism, and in particular has no definite
sign. The streak created by a given burst can equally be positive or negative. Most often,
it forms pairs of a high- and a low-speed streak, as seen in the examples in figure
\ref{fig:condatt}, and the newly created pair can either reinforce or weaken pre-existing
ones. The problem is the spanwise homogeneity of the mean profile, which lets bursts be
created at any spanwise location with equal probability, while what is needed is a mechanism
to ensure that new rollers are created approximately aligned to pre-existing ones, in such a
way as to reinforce their streaks. The most important nonlinear effect of streaks is
probably to break spanwise homogeneity and to localise the transient modes. On the other
hand, streaks are also important in triggering the generation of new bursts.
\cite{jim:pin:99} showed that, when the streaks in the buffer layer are filtered to lengths
shorter than $L_x^+\approx 600$, turbulence decays. The quasi-streamwise vortices were not
explicitly filtered in those experiments, but they stopped being created, and decayed
viscously.

\subsubsection{Random forcing}\la{sec:random}

A possibility that has attracted a lot of attention is to represent nonlinearity as a random
force acting on the linear part of the Navier--Stokes operator. The underlying assumption is
that the linear and nonlinear components of the equations are statistically independent. We
will pay relatively little attention to this possibility on the grounds that it contradicts
the rules about chance that we imposed on ourselves in the abstract and in the introduction to this article,
but the approach has to be discussed. Randomness bypasses the need to restart transient
growth by constantly seeding it. It is clear that any random forcing would
occasionally contain components along some initial condition leading to growth, and
that, if the most dangerous initial condition is dominant enough, the result of forcing
would be reasonably close to the optimally growing perturbation. The most elaborate
applications of this idea are probably those by the group of \cite{McKSha2010}, which are
broadly based on the theory of control. With some physically motivated assumptions about the
forcing noise, they report forced solutions which are very similar to the optimally growing ones
discussed above, and, therefore, to the statistics of nonlinear flows. This approach has
been the subject of a recent \emph{Perspective} by \cite{McKPers17}.
 
A related question is how much of those results are due to the choice of forcing, and
how much to the structure of the system. Equivalently, the question is whether we should
worry about nonlinearity, or whether it is enough to rely on the linear kernel of the
Navier--Stokes equations to describe, and eventually to control, turbulence. A partial answer
is due to \cite{zar_jova17}, who look at the theoretical question of which are the
requirements on the forcing noise to reproduce the second-order statistics of the
flow. They show that some constraints on the noise are necessary.

From our point of view, as discussed in the introduction, randomness is a choice
rather than a property, and these models do not answer the question of whether some
approximately autonomous set of structures can be identified in the flow, or of whether it
can be used as a basis for control. There is a well-developed theory of optimal control in
the presence of noise, which motivates many of these models, but our interest here is to
avoid noise as much as possible, rather than to minimise its deleterious effects.

Moreover, we know that simple deterministic solutions exist (although it is unclear whether
they are the ones that predominate in real turbulence), because the unforced nonlinear invariant
solutions discussed in \S\ref{sec:minimal} contain both the growth and the
trigger.

There is a more serious objection to substituting noise for the nonlinearity of turbulence.
It is well known that turbulence, in common with most high-dimensional nonlinear systems, is
characteristically sensitive to initial conditions, which results in its tendency towards a chaotic
attractor. This is a property that no autonomous statistically steady linear system can
reproduce. Noise therefore substitutes a fundamental property of turbulence by an external
input. This may not be very relevant in the short term for the large scales, whose Lyapunov
exponents are typically much slower than those of the small scales, but it raises the
question of what exactly is being reproduced. This is specially relevant regarding the
multiscale nature of the flow, which is most likely intrinsically unsteady.

\subsubsection{Lower-order nonlinear models}\la{sec:ROM}

More interesting from our point of view are models which discard as much of the nonlinearity
of the equations as possible, while retaining enough to preserve some property of interest.
The best-known examples are the large-eddy simulations (LES) to which we have referred
occasionally in the course of this article, in which the nonlinear inertial range is
substituted by a simpler sub-grid model. LES is an engineering tool whose aim is mainly
to reproduce the low-order statistics of the flow (e.g., the friction
coefficient). As such, it is mostly an unintended bonus that some sub-grid models produce
essentially correct multipoint statistics for the resolved scales
\citep{moi:kim:82,kravMM96}, although this agrees with the canonical cascade theory in which
energy and causality flow from large to small scales \citep{rich20}. More recently, LES has
been used as the base equation from which to compute invariant flow solutions that strongly
resemble those of direct simulations \citep{hwa:wil:cos:16}. Ideally, these solutions would
include part of the inertial range of scales, and provide some indication of how multiscale
turbulence works. In practice, they resemble more the single-scale solutions of low-Reynolds
number flows.

Another popular line of enquiry is the class of reduced-order models in which the
energy-generation cycle is represented by the interaction of a few modes, usually loosely
motivated by the observation of low-Reynolds number minimal flows \citep{Waleffe97}.
Contrary to LES, these models are not directly derived from the Navier--Stokes equations,
and their aim is not to reproduce the flow statistics, which they typically do not,
but to clarify characteristics and mechanisms of the flow that are deemed     
important by their originators. The best of them include components corresponding to
streaks and rollers, and result in bursting and, most importantly, in a self-sustaining
cycle.

Because of their relative simplicity, reduced-order models are often used as proxies for
true wall-bounded turbulence when developing diagnostic or control schemes. They are
attractive for such purposes, particularly as indications of which aspects of the problem
are fundamental for some particular purpose, and which ones are accessory. However, their
lack of a clear connection with the original equations raises the question of whether the
result of such exercises can be used for predictive purposes. A recent review of the
systematic use of model reduction for flow analysis is \cite{row:arfm17}.

\subsubsection{Quasi-linear models}\la{sec:QL}

An attractive variation of reduced-order models is the work of the group of \cite{far_ioa12}
on quasi-linear approximations to the wall-bounded Navier--Stokes equations. The simplifying
assumption is the division of the flow into an infinitely long streak, defined as all the
streamwise Fourier components with $k_1=0$, and everything else, which is treated as small
scales \citep{GayEtal:10}. The streak itself is nonlinear, although two-dimensional in the
cross-flow plane, and driven by the nonlinear Reynolds stresses created by the small scales.
The latter evolve within the nonuniform flow field of the streak, but all the nonlinear
interactions among themselves are neglected. In its original version, only the statistical
second-order moments of the small scales are computed, including the stresses that feed back
into the streak. An equilibrium is attained in which the highest Lyapunov exponent of the streak
vanishes, reflecting the absence of secular growth or decay.
 
The method itself is an evolution of classical multiple-scale averaging schemes, whose use
in celestial mechanics dates back to Gauss and Poincar\'e. In these procedures, the system
is also separated into slow and fast time scales. The slow scale is linear, or otherwise
simple to solve, but is forced by a nonlinear combination of the fast variables. The latter
satisfy linearised equations, but feel the nonlinearity through slowly varying coefficients
that depend on the slow scale. There is typically a parameter in the solution of the slow
equations, equivalent to the amplitude of the fluctuations in the channel, that has to be
adjusted to prevent secular terms. Two accounts of the averaging method, with very different
flavour, are \S 3 in \cite{cole68} and \S 4 in \cite{arnold83}.

The original version of the quasi-linear approximation of \cite{far_ioa12} was
stochastically forced. At small forcing amplitudes, the only stable equilibrium of the
model system is laminar, but, as the forcing increases, there is first a bifurcation to a
steady streak and roller, and later another one to an unsteady `bursting' state. These
unsteady solutions connect to a turbulent branch in which forcing can be
removed.

In a latter version of the same idea, a direct simulation of a channel flow is separated
into `long' $(k_1=0)$ and `short' modes $(|k_1|>0)$. As in the stochastic version, the long
modes are fully nonlinear and see the Reynolds stresses of the short ones, but the latter
are nonlinearly coupled only to the long modes. There are no nonlinear interactions among
short scales, and there is no stochastic forcing. Somewhat surprisingly, considering that
most of the nonlinear machinery of turbulence has been suppressed, the system self-sustains and
settles into a chaotic bursting behaviour that is strongly reminiscent of true turbulence
\citep{Farr:etal:16}. Even more interestingly, all but a few short modes spontaneously decay
after a while, leaving a self-sustaining system with only the nonlinear infinitely long
streak and a few relatively long linearised streamwise modes $(\lambda_1/h\gtrsim 0.5)$,
although with full resolution in the spanwise and wall-normal directions.
 
Even if, as with most reduced-order models, the statistics of this truncated turbulence only
qualitatively approximate those of real flows, the mechanism by which such a simple system
self-sustains is intriguing, and could give clues about which is the key self-sustaining event in
real turbulence. It is clear from the original papers that the amplification processes
involved are non-modal, and it is indicative that the stochastic system only becomes
independent of the forcing after it has bifurcated to unsteadiness. In fact,
\cite{far_ioa96} had argued that non-stationarity is a fundamental ingredient in turning
transient growth into permanent one. In essence, the growth of the short scales forces the
temporal oscillation of the long ones, and the non-stationarity of the latter continuously
restarts the transient growth.

It is clear that the model just described is minimal in the sense that all the nonlinear
modes have a single length scale (infinitely long), and that it assumes a separation of
scales that does not exist in reality. There are no gaps in the energy spectra in
\S\ref{sec:classical}. The definition of large scales as a streamwise mean value $(k_1=0)$
is also troubling because it depends on the size of the computational domain. On the other
hand, the model is rigorously derived from the Navier--Stokes equations, and we saw in
figure \ref{fig:corrcuts} that there is a relatively large difference between the average
length of the correlations of $u_1$ and those of the two transverse components. The
quasi-linear approximation can probably best be understood as a model for the very
large-scale structures in the central part of the channel, which it treats as simplified and lumped
into a single infinitely long proxy.

This may be a good point to reflect on how much of what we believe that we understand about
the dynamics of coherent structures is derived from minimal or small-box simulations. These
reduced systems have undoubtedly been a boon for our understanding of the dynamics of
wall-bounded turbulence, but an excessive reliance on them is a potential problem. Minimal
units are, by definition, single-scale flows, and we should ask ourselves whether they
really represent multiscale turbulence. For example, we have seen that the time fraction
during which structures burst in minimal boxes is approximately twice higher than the
corresponding area ratio in full-sized ones, and we had to warn in \S\ref{sec:causal} that
the causality relations among different wall distances reverse when they interfere with the
box size. Moreover, even if it turns out that the structures in minimal flows are truly
representative of those in fully turbulent large-scale flows, minimal simulations give us
little information on how to address multiscale dynamics.

Although a fully multiscale model of turbulence may be challenging at our present stage of
knowledge, what we could call `weak multiscaling' need not be impossible. For example, some
of the randomly forced solutions in \cite{McKPers17} contain two discrete scales, and the
same is true of some of the equilibrium solutions in \cite{sek:jim:17}. In another example,
it is hard not to speculate whether the temporal variability of the infinitely long
streaks discussed in this section could more realistically be substituted by a long-wave
multiscale spatial modulation.
  
\section{Discussion and open problems}\la{sec:conc}

We have tried to summarise in this article what is known about coherent structures in
wall-bounded turbulent flows, particularly regarding the large scales responsible for the
conversion of the velocity difference across the mean shear into the kinetic energy of the
turbulence fluctuations. Structures are important for turbulence, and we have given
in \S\ref{sec:lorenz} a simple example of how they appear naturally when otherwise chaotic
systems are examined over limited times. The same is presumably true of spatially extended
systems, such as turbulence, when examined locally in space. An obvious example where this
might be required is when considering control strategies. 

Over longer times or larger regions, statistics, in the sense of probability distributions
rather than specific events, are probably a more natural representation of the system, and
they have to be taken into account. One-point statistics measure the intensity of the
fluctuations, and two-point statistics, such as spectra and correlations, give an idea of
their spatial scales. They define the underlying physics that has to be described correctly
by any model of turbulence. For many applications, they are also the important quantities to
be predicted. But they are not enough to describe a functioning turbulent flow. For example,
it is well known that a negative skewness of the velocity increments is required for
turbulence dissipation \citep{betc56}, but skewness is unrelated to either intensity or
spectra, and is only found in the presence of structure. A Gaussian noise, even with a
coloured spectrum, has zero skewness. In spite of this, we have dedicated \S
\ref{sec:classical} to reviewing the basic facts about correlations and spectra in
wall-bounded turbulence because, if structures exist, correlations measure their size. The
conclusion of this section is that there are at least two kinds of correlations: short ones
for the two transverse velocity components, and long ones for the streamwise velocities. The
transverse dimensions of these correlations are very similar, of the order of the distance
from the wall, but their length is not. The correlation of the streamwise velocity in
channels is at least ten times longer than those of the transverse velocities.

The meaning of these correlations is explored in the next two sections. Section
\S\ref{sec:eddies} looks at advection velocities as indicators of coherence. Again, we find
that eddies can be divided in two groups. Long coherent eddies are found either in the viscous layer
near the wall or in the central part of the channel. They move as units
with a relatively uniform velocity. Although this property applies to all the velocity
components, most of the kinetic energy of these long structures is in the streamwise
velocity. The other class of eddies is closely associated with the logarithmic layer. They
exist at all length scales with self-similar aspect ratios, $\Delta_1$:$\Delta_2$:$\Delta_3
\approx 4$:1:1.5, and are not coherent enough to maintain a uniform advection velocity. They
are advected by the local flow, and are deformed by it. As such, they cannot be expected to have
lifetimes much longer than the shear time, $S^{-1}$. An interesting conclusion from this
section is that deep eddies, $\Delta_2=O(h)$, are not necessarily attached to the
wall. Long and wide eddies are generically deep, and they are found at all distances from
the wall. Only when they become so large that their depth is larger than their distance from
the wall, they attach to it, but there appear to be no statistical differences between
attached and detached large eddies. Finally, eddies smaller than the local Corrsin scale,
$L_c\approx x_2$, are isotropic and decoupled from the shear; they form the local Kolmogorov
inertial range.

Section \S\ref{sec:struct} examines the evidence from structures that are strong enough to
be isolated from the rest of the flow by thresholding. Several flow variables are examined
this way. In most cases, structures are found to correspond to the eddies discussed in
\S\ref{sec:eddies}. In particular, the transverse-velocity eddies correspond to structures
defined by strong $u_2$, $u_3$, or by the Reynolds stress, $-u_1 u_2$. Because intense
structures can be identified and measured individually, a lot is known about them, including
their temporal evolution. In this way, for example, we show that the lifetime of the
logarithmic-layer eddies is indeed a low multiple of the shear time, as suggested by the
wall-normal variation of their advection velocity. Somewhat surprisingly, the long
correlations of the streamwise velocity do not correspond to particularly intense
structures. The thresholded structures of $u_1$ are longer than those of $u_2$ or $u_3$, but
not by much. The very long correlations of $u_1$ appear to be due to the concatenation of
smaller units with individual aspect ratio $\Delta_1/\Delta_2\approx 5$, instead of the
$\Delta_1/\Delta_2\approx 2$ of the transverse velocities. Boundary layers have shorter
$u_1$ correlations than channels, but the difference can be traced to the details of how
units are concatenated. The basic units appear to be the same in both cases.

Structures similar to those in channels are found in turbulent homogeneous shear flows
(HSF), allowing us to distinguish the effect of the shear from that of the wall. In
all cases, figure \ref{fig:condatt} shows that the conditional flow field around structures
of intense $u_1u_2$ is an inclined roller located between a high- and a low-velocity
streak of $u_1$. The roller corresponds to the transverse-velocity eddies described above,
and the streaks to the streamwise-velocity ones. The streaks appear to be causally
associated with the roller, because they extend downstream from it: the high-speed streak
forward, and the low-speed one backwards. The streak-roller structure is symmetric in the
HSF, with the roller aligned to the most extensive direction of the shear, at 45\degree\
from the flow direction. This symmetry is progressively lost as we move from HSF to channels
far from the wall, and to channel structures attached to the wall. In the process, the
high-speed streak becomes stronger than the low-speed one, the inclination of the roller
decreases, and the lower end of the roller is truncated by the wall. Far from the wall, both
ends of the conditional roller are capped by hooks reminiscent of incomplete forward and
backwards hairpins. In the case of the attached structures, only the upper (forward) hook
survives.

Section \S\ref{sec:minimal} reviews the evidence obtained from minimal or otherwise
small simulation boxes, defined as those which contain a single structure of some particular
size. The most important information derived from these simulations is that rollers burst
intermittently with a time scale of the order of the shear time, in agreement with the
lifetime found by tracking intense Reynolds-stress structures. We argue that minimal
structures, attached structures in channels, and eddies larger than the Corrsin scale in HSF,
are different manifestations of the same phenomenon. When the simulation box is chosen small
enough to contain a single logarithmic-layer roller, but much larger than the viscous structures
near the wall, it can be used to study the dynamics of the lower part of the logarithmic
layer. We use it in \S\ref{sec:causal} to show that the temporal correlation between the
Reynolds stress at different distances from the wall moves downwards across the logarithmic
layer, from the outside to the wall, rather than the other way around. Other quantities move
both ways, in agreement with tracking results for individual structures in larger simulation
boxes. We find no example of quantities moving predominantly from the wall outwards.

Finally, \S\ref{sec:theory} briefly surveys present theoretical approaches to the
description of the structure of wall turbulence. We conclude that the Orr and lift-up
mechanisms, best known from the linearised stability equations, explain most of the
formation of bursting rollers and streaks, and we remark that neither process is
intrinsically linear. Together with the Kelvin--Helmholtz instability mechanism, they form
the `core' of the interaction of velocity fluctuations with the ambient shear, and are
therefore robust enough to survive linearisation. They should be expected to remain relevant
at all intensities in most shear flows. However, we note that the combination of Orr and
lift-up is not enough to explain the complete self-sustaining cycle of wall-bounded
turbulence, because both processes are transient. Something else is required to restart the cycle, and
that extra process is not contained in linear approximations. We argue that this is the main
difference between free-shear flows and wall-bounded ones. The former are modally unstable,
and linear stability controls most of their dynamics. The latter are modally stable, and
their persistence requires an intrinsically nonlinear component. In other important
respects, they are more similar. In particular, as long as some linearisable process is
active, it dominates the energy-production cycle, because the shear is the fastest time
scale. Somewhat counter-intuitively, these linearisable periods are the most active ones in
the flow. Only when they become inactive, for reasons that differ among flows, does
nonlinearity have a chance to act.

We review in \S\ref{sec:nonlinear} the different proposals for the nonlinear closure of the
generation cycle in wall-bounded flows. Most of them centre on the role of the streaks,
although we argue that their most important role is not probably to destabilise the flow,
but to catalyse the formation of pre-existing non-modal instabilities due to the ambient
shear, and to guide their location. Most likely, this is the reason for the streamwise
concatenation of smaller structures into longer ones.

To conclude, we remark that most of the available structural theories for wall-bounded flows
refer to minimal or single-harmonic situations. The common reference to a `self sustaining
cycle' (in singular) indicate that most theories address single structures. We argued in the
introduction that equilibrium thermodynamics is not a good model for turbulence, but this
should not be taken to mean that statistical mechanics has no role to play, particularly
regarding the interaction among many coherent structures of different sizes. Stretching an
analogy, minimal flow units are the molecules of wall-bounded turbulence. A lot can be
learned about materials by studying their molecules, and minimal units have allowed us to do
the equivalent of chemistry. The next step of turbulence theory should be to move from
chemistry to condensed-matter physics or to materials science.

\subsection{Some open problems}\la{sec:open}

It could be concluded from the previous summary that almost everything is known about the
mechanics of individual structures in turbulent wall-bounded flows. This is probably true,
although there are bound to be differences in interpretation. But this does not mean that we
understand everything about wall-bounded turbulence, and it is the nature of articles like
the present one to consider some of the open problems left in the field. The following are a
few representative examples:

\begin{enumerate}

\item\la{Q:SSP} The simplest question left open by the discussions above is how to precisely
characterise what is the nonlinear event that restarts the self-sustaining cycle. We listed
in \S\ref{sec:nonlinear} some of the possibilities that have been discussed in the
literature, but none of them is definitive enough to provide a predictive criterion for the
location and time where a new cycle is about to begin. Since one of the most important
applications of such a criterion is to inform active control strategies, it would be most useful if
it could be reduced to a few variables, preferably at the wall. Given how much information
we have on individual structures, this is the sort of problem that can probably be solved
from existing data and in the next few years.

\item\la{Q:longu} How does a short roller become a long streak? We have already discussed
this question in \S\ref{sec:streaks2}, where we distinguish between the factor of two or
three between the length of the intense structures of the streamwise and of the transverse
velocities in figure \ref{fig:pdf_uvwsters}, and the much larger difference among the
correlation functions in figure \ref{fig:corrcuts}. We have seen that the geometry of the
structures of $u_1$ is consistent with the concatenation of shorter units. The p.d.f. of the
length of the sublayer streaks was examined by \cite{jim:kaw:13}, who showed that it has an
exponential tail, suggesting the concatenation of individual units of length $L_c^+\approx
500$--1000. This is consistent with the length of minimal boxes \citep{jim:moi:91}, but the
study was limited to $x_2^+\le 60$, and we are not aware of similar studies in the
logarithmic layer. Minimal simulations are not very informative in this respect, because
their streaks are always found to cross the box, and are therefore infinitely long.
Simulations in longer boxes of minimal span do not result in longer rollers, but in several
rollers along a longer streak (see figure \ref{fig:jimsim}\bbb). 

\quad The question is how the concatenation takes place. Equivalently, why sweeps and
ejections tend to form on the correct side of the streak to reinforce it, as in figure
\ref{fig:condatt}. We argued above that one of the effects of pre-existing streaks is to
break the spanwise translational symmetry for the generation of new bursts, but the precise
mechanism is unclear, as are its quantitative aspects. For example, how often are new bursts
created in the `correct' location? It could be the case that the long streaks that we observe are an
example of observational bias: streaks grow by random superposition, and we are distracted
by the streaks that happen to be long. The exponential p.d.f.s mentioned above would be
consistent with this accretion model. The infinitely long streaks of minimal boxes would not
be, but they could be an artefact of minimality.

\item\la{Q:multi} Perhaps the problem farthest from solution, and arguably the most
important one, is how the multiscale nature of the flow is organised. We have a reasonable
understanding of individual structures, but we know very little about their multiscale
interactions. For example, each of the conditional structures in figures \ref{fig:condQ} and
\ref{fig:condatt} is associated with a given lengthscale and with a given distance from the
wall, but we saw in figures \ref{fig:uvpdf}(\ccc,\ddd) that these structures exist at all
scales in the logarithmic layer, and that they form a self-similar hierarchy. Geometrically,
structures of different sizes coexist everywhere overlapping each other, and we
have mentioned that they carry about 60\% of the tangential Reynolds stress. 

\quad Presumably, the requirement that they should provide a given mean stress sets their
velocity scale. Momentum conservation requires that the average of the Reynolds stress of
all the structures that intersect a given plane has to be $\utau^2$, but, how is this
information communicated to individual structures? What happens to structures which are too
weak or too strong? Do they interact with each other to `reach a consensus' on the right
intensity, or do they interact directly with the mean flow?

\quad It is easy to construct feed-back models in which the shear of the mean velocity
profile controls the intensity of the bursts, specially if we admit, as we have argued, that
bursts form at all heights instead of growing from the wall. Qualitatively, if the mean
intensity of the structures in some layer is too weak, the shear increases, and the
structures intensify. However, such `one-stage' models are unlikely to be the whole story.
Consider the structures reaching up to a given $x_2$ (e.g. one meter) within the logarithmic
region of the atmospheric surface layer, whose thickness $h$ is $O(100)$ meters. How do
these structures receive the information of the mean profile, which can only be defined by
averaging over distances of several boundary-layer thicknesses? In this example, the problem
of setting the velocity scale is not restricted to structures whose size is one meter, but
also to everything bigger, because momentum conservation requires that the mean Reynolds
stress of all the structures intersecting that plane has the right value. But the ratio of
the turnover times of the largest and the smallest structures intersecting our plane is
$h/x_2=O(100)$. If the small scales adjust to the stress missing from the largest ones, how
do they measure it? If the large scales adjust to the small ones, how do they do the
average?

\quad The problem is not only that we do not know the answer to this question, but that we
do not have the right tool to investigate it, and that we  need to develop a
system equivalent to minimal simulations for multiscale problems.

\item\la{Q:malkus} A final unsolved problem is the one with which we opened
\S\ref{sec:theory}: what determines the mean velocity profile in wall-bounded turbulence?
Motivated by it, a lot of progress has been made on the mechanisms of elemental structures,
but very few things in \S\ref{sec:theory} refer to how the mean profile is put together. We
mentioned in (\ref{Q:multi}) that part of the problem is that we lack a multiscale theory, but it is not
immediately clear how would we use such a theory if we had it. It may be time to
revisit \cite{malkus56} idea of criticality. The quasilinear model in \S\ref{sec:QL} can
be interpreted as a requirement of marginal stability for a base flow that now includes the
infinitely long streak (Ioannou, private communication). The null Lyapunov exponent in that
model takes the place of the neutral eigenvalue in \cite{malkus56}.

\quad In a simpler modal setting, we have mentioned that linear theory can be used to
quantitatively describe unstable free shear layers, and that nonlinearity only takes effect
when the linear modes become neutral. But that is not the whole story. The Reynolds stress
is a nonlinear effect that depends on the amplitude of the fluctuations, and it controls the
growth of the shear-layer thickness that eventually drives the Kelvin--Helmholtz eigenvalue
to become neutral. A Malkusian interpretation would be that nonlinearity adjusts itself to
keep the mean velocity profile at a state of (self-organised) criticality.

\end{enumerate}

There are many more open questions that do not fit in the present article, and probably
still more that I am not able to see at the present time, but, if I may finish the article on a
personal note, this is what still makes wall-bounded turbulence fun.

\begin{acknowledgements}
This work was supported by the European Research Council under the Coturb grant
ERC-2014.AdG-669505. I am indebted to discussions with too many colleagues and students to
cite individually, particularly during an extended stay at the Transturb17 programme of the
Kavli Inst. of Theoretical physics at the U. California at Santa Barbara, supported in part
by the National Science Foundation under grant NSF PHY11-25915t. I am especially grateful to
M. K. Lee and R. D. Moser for early access to the data of their high-Reynolds number channel
simulation.
\end{acknowledgements}

\appendix
\section{Similarity solutions and invariances}\la{sec:appA} 

We mentioned in \S\ref{sec:scales} that not just the existence, but the functional form of
some similarity solutions, can be derived from symmetry considerations. Two examples that
are important in turbulence are the power law of the inertial energy spectrum, and the
logarithmic velocity profile of wall-bounded flows. In this appendix we outline how this is
done, but remark that obtaining the specific exponents and coefficients usually depends on
physics beyond simple invariance. A more rigorous presentation of this material is
\cite{oberlack.01}.

\subsection{The logarithmic law}

Consider first the logarithmic law \r{eq:loglaw} for a mean velocity profile that depends
only on the transverse coordinate $x_2$, and assume that the Reynolds number is high enough
for viscosity to be negligible. The equations for inviscid flow are invariant under
independent scalings of the velocities and of the space coordinates, as well as under
coordinate translations and the Galilean addition of a uniform velocity. Unless these
symmetries are broken, either spontaneously or by boundary conditions, the velocity $U$
should be expressible in the form
\beq
(U - V)/B = f(\xi),\quad\mbox{with}\quad \xi=x_2/\delta + H,
\la{appc:eq1}
\eeq
where $B$ and $\delta$ are associated respectively with the scaling of the velocities and of
lengths, and the origins $V$ and $H$ are associated with Galilean invariance and with space
translation. Invariance means that the value of $U$ resulting from applying \r{appc:eq1}
should not depend on our particular choice of the values of the parameters, but it is
important to understand that, although the parameters are arbitrary, they are not
necessarily independent from one another. For example, a coordinate shift $x_2 \ra x_2+H$ in a
linear velocity profile $U= Sx_2$ implies a velocity increment $U \ra U+SH$. In general, the
problem of finding laws that are invariant to all the {\em independent} symmetries above is
too restrictive, and we have to look for laws that satisfy the invariances of \r{appc:eq1}
together with the functional dependences among the parameters.

In most situations, some scales and origins are imposed by the boundary conditions, and not
all the parameters in \r{appc:eq1} can be chosen arbitrarily, but the free parameters
impose restrictions on the form of the function $f$. In essence, if there is no
physical reason to fix the value of some parameters, the form of $f$ can be
determined by assuming arbitrary values for them, expressing the law in the most general
form compatible with its invariances, and requiring the final expression to be independent
of the arbitrarily chosen parameters.

In the case of the logarithmic velocity profile \r{eq:loglaw}, dynamics provides a velocity
scale, $\utau$, and a preferred origin at the wall, $x_2=0$. These assumptions determine $B$
and $H$ in \r{appc:eq1} but not $V$ and $\delta$, and the general form of the profile can be
written as
\beq
U = \utau f(x_2/\delta) + V.
\la{appc:eq2}
\eeq
The requirement that $U$ cannot depend on the choice of parameters can be expressed as
\beq
{\dd U \over \dd \delta} =
   - {\utau \over \delta} \xi f_\xi (\xi) + V_\delta = 0,
\la{appc:eq3}
\eeq
where subindices indicate differentiation. This can be rearranged into
\beq
\xi f_\xi =  {\delta V_\delta \over \utau},
\la{appc:eq4}
\eeq
where the right-hand side has to be a constant, independent of $\delta$ and $\xi$. Integration of
\r{appc:eq4} leads to the logarithmic law \r{eq:loglaw}.  

Note that the law itself is a direct consequence of the assumed invariances, but that the
information of which parameters should be treated as fixed is a physical argument
that depends on the existence of a constant-stress layer near the wall, and on the
importance of wall distance as the relevant coordinate. The value of the K\'arm\'an constant,
$\kappa= \utau/(\delta V_\delta)$ also has to be determined from arguments beyond
invariance. On the other hand, the result that the arbitrariness in the length scale
should be linked to the Galilean invariance $(V_\delta\ne 0)$ reminds us that the
logarithmic law cannot be extended to the wall, and that it does not contain the no-slip
boundary condition.

\subsection{Power laws}

It is easy to see the need for a linkage between $V$ and $\delta$ in the
above derivation, since making them independent would lead to a zero
right-hand side, and to a constant velocity as the unique solution.  This is
generally true, and at least a pair of linked symmetries are needed to
avoid trivial solutions.

Power laws occur when there is neither a velocity or a length scale, but the 
origins are fixed. Consider the \cite{kol41} argument for the dependence of the velocity
increment $\Delta u$ on the length of the interval $\ell$ across which it is measured. Since both 
$\Delta u=0$ and $\ell=0$ have physical meaning, $V$ and $H$ can be set to zero
in \r{appc:eq1}, which becomes
\beq
\Delta u = B f(\ell/\delta),
\la{appc:eq5}
\eeq
where $B$ and $\delta$ are arbitrary. Differentiation with respect to
$B$ leads to
\beq
f + \frac{\delta}{B \delta_B}\, \xi f_\xi = 0 ,
\la{appc:eq6}
\eeq
and to $f\sim \xi^\alpha$, with $\alpha=\delta/B\delta_B$. Note that, as with the
logarithmic profile, the exponent $\alpha$ cannot be found from the invariance properties, and
depends on the physical argument that the energy transfer rate, $\Delta u^3/\ell$, is
conserved \citep{kol41}. 

\subsection{Fourier expansions}\la{sec:Fourier}

A case of particular interest in parallel shear flows is that of functions or vectors
defined by operators which are invariant to translations along some coordinate direction. An
example are the PODs used in \S\ref{sec:eddies}, which are defined in appendix
\ref{sec:appB} as eigenvectors, $\phivec$, of the two-point covariance,
\beq
\int R(\xvec,\tilde{\xvec}) \phivec(\tilde{\xvec}) \dd \tilde{\xvec} = \mu  \phivec(\xvec).
\la{eq:pod0}
\eeq
Assume that the flow is homogeneous along $x_1$, so that the covariance is $R(\xvec,\tilde{\xvec})=
R(x_1-\tilde{x}_1,\yvec,\tilde{\yvec})$, where $\yvec$ stands for those directions in $\xvec$ which are
not $x_1$. The covariance is then invariant to translations, $x_1\to x_1+c$,
and the eigenvector of a given eigenvalue can at most change by a normalisation
factor, $\phivec(x_1+c,\yvec) = C(c) \phivec(x_1,\yvec)$. Differentiating with respect to
$c$ at $c=0$ yields
\beq
\p\phivec(x_1,\yvec)/\p{x_1}  = (\dr C/\dr c)_{c=0}\, \phivec(x_1,\yvec),
\la{eq:pod1}
\eeq
which integrates to an exponential in $x_1$. If we also require that $\phivec$ remains bounded
at $|x_1|\to\infty$, the only options are the Fourier basis functions,
\beq
\phivec(x_1,\yvec) = \widehat{\phivec}(\yvec) \exp (\ii k_1 x_1).
\la{eq:CED0}
\eeq
This is inconvenient when considering individual structures in statistically homogeneous
systems, because it requires a method to combine several PODs into a compact wave packet (see
\S\ref{sec:eddies} and appendix \ref{sec:MCE}). There are many variants of this result,
and perhaps the most interesting are those in which the homogeneous direction is time. For
example, the evolution of linear autonomous dynamical systems can be expanded in terms of
exponentials, which have to be Fourier functions for statistically steady systems. Even in
nonlinear dynamical systems, it turns out that the evolution operator acting on {\em all}
possible observables is linear. If the underlying system is invariant to temporal
translations, so is this `Koopman' operator, and its eigenfunctions (Koopman modes) are also
exponential in time. As with their spatial counterparts, several Koopman modes have to be
combined, with mutually correct phases, to represent temporally localised events such as bursts. Recent
reviews of the use in fluid mechanics of the spectral properties of the Koopman operator are
\cite{mezic:arfm13} and \cite{row:arfm17}.

\section{The statistical representation of eddies}\la{sec:appB} 

This appendix reviews the theory of optimal representation of flows in terms of `eddies'. It
briefly surveys such subjects as proper orthogonal decomposition, compact eddies, and linear
stochastic estimation. This is a well-trodden field, much of which was
initially developed in the context of the theory of communication. As such, it has
relatively little to do with fluid mechanics or with the Navier--Stokes equations. In
particular, it is indifferent to the dynamics of the physical system which is being represented, including to
whether it is linear of not. In the simplest cases, it reduces to Fourier
analysis. Those interested in the early history of the subject may consult \cite{shannon} or
\cite{wiener}. Those seeking to apply any of these techniques should study the original references
mentioned below. 

Communication theory deals with how to represent and send information as economically as
possible. The requirements of physical modelling may not necessarily be the same, and optimal
representations in the sense of this appendix may not always be ideal for the purpose of
physics. To begin with, we will see that most reduced-order representations are linear
transformations. When they are used to create reduced models, they result in a restriction
to a linear subspace. However, the attractor of turbulent flows is usually not a hyperplane
and, even if the subspace is made `fat' (i.e. high-dimensional) enough to include the
curvature of the attractor, it may still miss important physics. For example, a projection
chosen to optimally represent the energy of the flow would probably miss most of the
dissipation, because the dissipative scales contain very little energy. However, both energy
and dissipation are important to model turbulence \citep[on the other hand, see][for ways to balance two
metrics.]{row:arfm17}

Even so, there may be some advantages to representing a system compactly. Even if,
as we have just discussed, retaining the nonlinearity may require that the approximation has
to be made `fatter' than it otherwise should be, any reasonably optimal representation will
probably eliminate many irrelevant details, and can be considered as an optimised filter.
Communication theory was originally developed in parallel with applications to control,
which share with turbulence the importance of nonlinearity. The mismatch between the
nonlinearity of the turbulence attractor and the geometry of linear subspaces was understood
from the very beginning, and has been discussed often \citep[e.g.,][]{Ber:Hol:Lum:93}.

The approximation theory described in this appendix is essentially statistical. It applies
to `ensembles' of individual functions, which are sets of functions with a probability
distribution \citep{shannon}. For a given cost (e.g. number of coefficients), the goal of
these approximations is not to find the best representation of a particular flow field, or
even the most probable model for a given ensemble, but to find the model that maximises the
probability that a member of the ensemble agrees with its predictions. Therefore, models are
not adjusted to sets of functions (flow fields) but to their probability distribution. In
practice, they are typically adjusted to match their statistical moments.

It is easy to show that the best model for the first-order moments of an ensemble is the
mean. This is what is being done when flow velocities are represented by their mean
profiles. The next step, having more to do with eddies and structures, is to take into
account the second-order quantities.

To fix ideas, consider a discrete set of observations, each of which is represented by a
finite dimensional vector $\uvec_{(j)}$, where $(j)$ labels observations. This discretisation
is not as restrictive as it may appear. One of the central sampling theorems of Fourier
analysis is that any function of time, $u(t)$, whose spectrum is band-limited to frequencies
below $\Omega$ (e.g., by viscosity), can be exactly represented by discrete samples at
uniformly spaced times, $t_m=m/(2\Omega)$ \citep{shannon,gasquet98},
\beq
u(t) = \sum_{m=-\infty}^\infty u(t_m) \frac{\sin 2\pi\Omega(t - t_m)}{2\pi\Omega(t -t_m)}.
\la{eq:shannon1}
\eeq
Thus, for a flow field $u(\xvec, t)$, the number of independent flow snapshots per unit time
is $2 \Omega$. If the flow is also band-limited in space to wavenumbers below $\Xi$, each
three-dimensional snapshot can be similarly represented by $(2\Xi)^3$ discrete samples per
unit volume.

Although each snapshot corresponds to an instant in time, we will initially treat them as
independent samples, and only model their spatial statistics. The second-order structure of
a scalar field is expressed by its covariance
\beq
R_{uu}(\xvec,\tilde{\xvec}) =\bra u(\xvec) u^*(\tilde{\xvec})\ket. 
\la{eq:corr1a}
\eeq
If continuous functions are discretised as explained above, all operations reduce to
algebraic manipulations with vectors and matrices. This is the operating mode in simulations
and experiments, and we will use it in the rest of the appendix.

Form the $m\times n$ matrix $\Umat=[u_{ij}]$, whose columns are observations, and whose
$i$-th row contains the $i$-th component of each sample. Assume that all the rows have zero
mean. The two-point covariance of the ensemble $\uvec_{(j)}$ is the Hermitian $m\times m$
matrix
\beq
\Rmat = \sum_j \uvec_{(j)}\uvec_{(j)}^*  = \Umat\Umat^*,
\la{eq:POD1}
\eeq
where averaging has been substituted by summation over all samples, and the asterisk denotes
the Hermitian transpose. Note that the right-hand side of \r{eq:POD1} should have been
divided by $n$, to make it closer to an average. In fact, there usually are other weighting
factors in the inner product implied by \r{eq:POD1}, which are needed to reduce it to
whatever is the desired definition of the norm (e.g., some discrete integration formula). Such
weights can always be incorporated into $\Umat$, and will not be explicitly displayed here.
The same will be true of the normalisation by $n$.

\subsection{Proper orthogonal decomposition}\la{sec:POD}

The goal of optimal stochastic decomposition is to find a set of orthonormal (column) basis vectors,
$\phimat=[\phivec_{(k)}]=[\phi_{ik}]$, such that, for example, projecting over the first one
explains as much as possible of the variance of the observations. The projection of
$\uvec_{(j)}$ on a basis vector $\phivec_{(1)}$ is the inner product
$\phivec_{(1)}^*\uvec_{(j)}= u_{ij}\phi_{i1}$. Projecting all the snapshots over
$\phivec_{(1)}$ results in the row vector $\phivec_{(1)}^*\Umat$, whose norm,
$\phivec_{(1)}^*\Umat\Umat^*\phivec_{(1)}$, is what we want to maximise. This is the
classical characterisation of an eigenvector of the Hermitian matrix $\Rmat=\Umat\Umat^*$.
Briefly, given a flow field whose covariance function is $\Rmat$, the optimum expansion
basis is given by the set of $m$ eigenvectors defined by
\beq
\Rmat\phivec_{(k)} = \mu^2_{(k)}\phivec_{(k)}, 
\la{eq:POD2}
\eeq
where repeated indices do not imply summation. The covariance matrix can then be expressed as
\beq
\Rmat = \phimat\Mmat^2\phimat^*,
\la{eq:POD3}
\eeq
where $\Mmat$ is the diagonal matrix of the $\mu_{(k)}$, and $\phimat=[\phi_{ik}]$ is
a unitary matrix whose columns are the eigenvectors.

The optimal expansion of an arbitrary flow field is 
\beq
\uvec = \sum_k \hu_k \phivec_{(k)},  
\la{eq:POD3.1}
\eeq
and it follows from \r{eq:POD3} that the variance of the expansion coefficients for fields
in the same statistical ensemble as the observations, is $\bra|\hu_k|^2\ket =
\mu^2_{(k)}$. The sum of all the eigenvalues is the variance (or energy) of the original
ensemble, and how many eigenvalues are required to represent a given fraction of the total
variance measures the compression efficiency of the expansion. Although a typical flow field
from DNS can have several billion components, and the matrix $\Rmat$ cannot typically be
constructed or diagonalised, it is found empirically that a few eigenvalues often contain
most of the energy. The covariance in \r{eq:POD3} can then be approximated by retaining a
few eigenvectors, which can be computed by some variant of Arnoldi's method \citep{press}.
This expansion has been rediscovered several times, initially as Principal Component
Analysis by \cite{pearson}. In fluid mechanics, it takes the name of Proper Orthogonal
Decomposition \citep[POD,][]{Ber:Hol:Lum:93}.

There is an interpretation of \r{eq:POD3} that suggests an alternative technique for the
computation of the leading PODs. The covariance in \r{eq:POD1} can be understood as the sum
of the covariances, $\uvec_{(j)}\uvec_{(j)}^*$, of the flow fields of individual snapshots.
Similarly, the expansion in \r{eq:POD3} is the sum of the covariances,
$\phivec_{(k)}\phivec_{(k)}^*$, of the individual PODs, weighted by their eigenvalues. It is
then natural to interpret each eigenvector as representing a group of flow fields, and its
eigenvalue as a measure of `how often' that eigenvector has been used in computing the
statistics. This suggests that the leading eigenvectors are those found most often in the
flow, and that, if it were possible to isolate them within individual samples, it should be
possible to compute them using less snapshots. What would be neglected by this procedure
would be the less important eigenvectors, of which there are many, but which appear only
seldom. Moreover, \r{eq:POD2} generates $m$ eigenvectors, while \r{eq:POD1} shows that the
rank of $\Rmat$ is at most $n$. In most situations, the number of samples is much smaller
than the number of degrees of freedom, $n\ll m$, and the majority of the eigenvalues in
\r{eq:POD3} are zero.

This is the basis for the `method of snapshots' \citep{sirov87}, which starts by factoring
both \r{eq:POD1} and \r{eq:POD3}. Consider a set, $\Umath$, of $\bar{n}\ll m$ snapshots,
where $\bar{n}$ is typically of the order of the number of PODs to be retained. Perform the
singular-value decomposition of $\Umath$ \citep[SVD,][]{press},
 \beq
 \Umath = \phimath \Mmath \Amat^*,   
 \la{eq:POD4}
 \eeq
where $\Amat$ and $\phimath$ are unitary $(\Amat\Amat^*= \phimath\phimath^*=\Imat)$, and
$\Mmath$ is a reduced, $\bar{n}\times \bar{n}$, diagonal matrix of singular values,
each of which is real and non-negative. Note that, because $\Umath$ only has a few columns, the
SVD is relatively cheap, and $\Mmath$ only contains a few singular values. Substituting
the decomposition \r{eq:POD4} in \r{eq:POD1}, we obtain   
 \beq
 \Rmath =  \phimath \Mmath \Amat^* \Amat \Mmath\phimath^* =
  \phimath \Mmath^2 \phimath^*.  
 \la{eq:POD5}
 \eeq
Comparing \r{eq:POD5} with \r{eq:POD3} shows that $\Mmath^2$ is a statistical
estimate of a few eigenvalues of \r{eq:POD3}, hopefully the most significant ones, while the columns of
$\phimath$ contain the principal PODs. The method of snapshots does not provide a full POD
representation, and assumes that the chosen snapshots are representative
of the flow statistics, but it is typically much cheaper than the diagonalisation of the
full covariance, and works well when a few PODs are clearly dominant over the rest.

\subsection{Most compact eddies}\la{sec:MCE}

Even if the PODs form an optimal expansion basis, they are not very good models for
localised eddies, because we saw in appendix \ref{sec:appA} that they are Fourier
modes along the homogeneous directions of the flow. 
If the flow is statistically homogeneous along $x_1$, the covariance depends only on the
distance increment along that direction, $R_{uu}(x_1, \tx_1, \yvec,\tilde{\yvec}) =
R_{uu}(x_1-\tx_1, \yvec,\tilde{\yvec})$, where $\yvec$ represents the remaining coordinates
in $\xvec$. The covariance and the spectrum then form a Fourier-transform pair,
\beq
R_{uu}(x_1-\tx_1) =
       \int \widehat{R}(k_1,\yvec,\tilde{\yvec})  \exp\,[ \ii k_1(x_1-\tx_1)] \dd k_1, 
\la{eq:corr2a}
\eeq
where $\widehat{R}(k_1,\yvec,\tilde{\yvec})=\bra \hu(k_1, \yvec)
\hu^*(k_1,\tilde{\yvec})\ket$, and $\hu(k_1, \yvec)$ are the coefficients of the Fourier
expansion of $u$ along $x_1$. The POD modes can also be written as
\beq
\phivec_{(k)}(x_1,\yvec) = \widehat{\phivec}_{(k)}(\yvec) \exp (\ii k_1 x_1)
\la{eq:CED1}
\eeq
for any wavenumber $k_1$, where $\widehat{\phivec}_{(k)}$ is an eigenfunction of the Fourier
coefficient of the covariance, $\widehat{R}_{uu}(k_1,\yvec, \tilde{\yvec})$. This simplifies
the calculation of the PODs, because $\widehat{\phivec}_{(k)}$ can be computed from the
covariances of the Fourier coefficients of $u$, but it also makes $\phivec_{(k)}(x_1,\yvec)$
a bad eddy model. As already discussed in the body of the paper, Fourier functions are
unlocalised, while dynamically significant structures should have some degree of
localisation because the Navier--Stokes equations are differential equations in physical
space, not in Fourier space. A summary of early attempts of how to construct localised
eddies out of the PODs is \cite{Ber:Hol:Lum:93}. Here we just discuss the particularly
intuitive method of \cite{moi:mos:89}, which was used to generate figure \ref{fig:pod} in
the body of the article.

\begin{figure}
\centerline{%
\includegraphics[width=.80\textwidth,clip]{\arpath 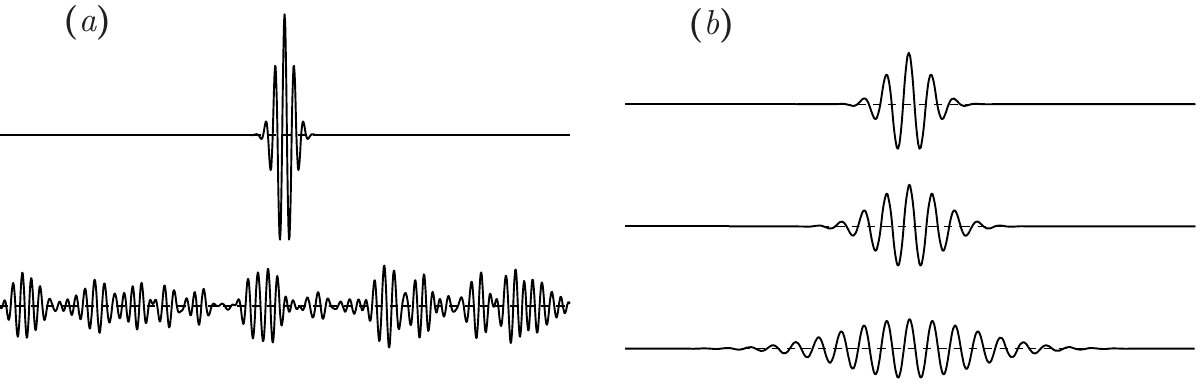}%
}%
\caption{%
(\aaa) The two signals have the same intensity and an identical Gaussian spectrum with a
standard deviation equal to $\sigma/k_0=0.17$ with respect to the central wavenumber (see
text). The Fourier components of the bottom line have random phases. All the phases of the
upper line are zero at the centre of the plot.
(\bbb) As in the upper signal in (\aaa), with $\sigma/k_0=0.17,\,0.11,\,0.055$, top to bottom.
}
\la{fig:eddies}
\end{figure}

A localised eddy can be constructed from Fourier PODs by adding a band of wavenumbers. The
amplitude of each Fourier component is known, because we saw after \r{eq:POD3.1} that
$\bra|\hu|^2_k\ket = \mu^2_{(k)}$, but the derivation of the PODs says nothing about the
phase of the coefficients. This is crucial. Adding Fourier wavetrains with random phases
typically results in functions without recognisable structure (see the bottom part of figure
\ref{fig:eddies}\aaa). This has nothing to do with the uncertainty relation that links the
width of the spectrum with the length of the signal in physical space. The two signals in
figure \ref{fig:eddies}(\aaa) have the same spectral content and the same total energy. The
only difference is the relative phase of their harmonics. \cite{moi:mos:89} reasoned that,
since the integral of the square of the velocity is given by its spectrum, a 
localised eddy would also be tall, and they introduced the condition that the phase
of all the harmonics should vanish at some chosen `central' location. In that way, all the
wavetrains contribute as much as possible to the function at that point. For example, if we
wish our eddy to be centred around $x_1=0$, the `most compact' eddy (MCE) would be
\beq
u_{mc}(x_1) =\sum_k |\hu_k| \exp(\ii k x_1).
\la{eq:MLE1}
\eeq
This is how the upper pulse in figure \ref{fig:eddies}(\aaa) is generated, and is always
probably close to the narrowest possible signal for a chosen spectrum $|\hu_k|$. In a case
with several dimensions, such as in figure \ref{fig:pod} in the body of the article, it is
usually not possible to zero the phases at all wall distances. The solution of
\cite{moi:mos:89} is to zero the phase of each vertically averaged wavetrain, $\gamma_k=\int
\hu_k (\yvec) \dd \yvec$,
\beq
u_{mc}(x_1,\yvec) =\sum_k \hu_k(\yvec) (\gamma^*_k/|\gamma_k|) \exp(\ii k x_1).
\la{eq:MLE2}
\eeq
This construction results in reasonably looking flow fields, but  its wealth of adjustable
parameters makes it less than general. The most obvious arbitrariness is the position of the
pulse, which can be located anywhere. Next is the bandwidth with which to construct it, and
the shape of the spectral filter used to isolate the desired wavenumbers. Figure
\ref{fig:eddies}(\aaa) is generated with a Gaussian spectrum, $|\hu_k| =
\exp[-(k-k_0)^2/2\sigma^2]$. The central wavenumber, $k_0$, defines the wavelengths being
considered, and the width of the pulse, measured in terms of the basic wavelength, is
determined by the ratio $\sigma/k_0$ (see figure \ref{fig:eddies}\bbb). A consequence of
this freedom is that there are too many possible localised eddies to serve as basis
functions for an orthogonal or complete expansion in the sense of the PODs. The MCEs do not
necessarily describe the mean structure of the flow, because they do not form a basis
in which to expand the covariance, and they are usually not orthogonal. However, if there are
strong localised structures in the flow which are common enough to influence the
statistics, the MCEs suggest their form.

A more rigorous procedure for isolating eddies with some desirable characteristics is the
use of wavelets, which can be seen as Fourier packets with a predefined shape. The examples
in figure \ref{fig:eddies} are wavelets of a particular kind (Morlet), and so are
(approximately) the base functions used in the Shannon expansion \r{eq:shannon1}. Once a
`mother' wavelet (i.e., its shape) is defined, signals can be expanded into weighted sums or
integrals of resized and translated versions of that shape, and, if certain characteristics
are satisfied, the expansion can be inverted. Because of the multiple freedoms in choosing a
wavelet shape, position and scale, wavelets share with compact eddies the property of not
forming a unique expansion basis, and there is usually a wide latitude on how to compute the
wavelet coefficients, and on which family of wavelets to use for reconstruction. Not all of
these freedoms are necessarily useful in fluid mechanics. This often makes the use of
wavelets too restrictive for the advantages gained, but many common operations can be
expressed in terms of wavelets. For example, convolution with a family of translated
single-scale wavelets is a band-pass filter, and there are decimated families of wavelets
that form an orthogonal basis which can be used as an alternative to Fourier analysis in the
identification of localised flow features. The mathematical theory of wavelets is very well
developed. An elementary textbook introduction can be found in \cite{gasquet98}, and an
excellent account of the use of wavelets in fluid mechanics is \cite{farge92_arfm}.

\subsection{Conditional averages and Linear stochastic estimation}\la{sec:LSE}

A further method of local statistical approximation deserves a short comment because of its
historical importance, and because we have used its results at several points of our
discussion. It is intuitively obvious that the only way to defeat statistical homogeneity is
to choose a particular location in the flow as more important than others.

For example, we have done this when computing the conditional flow fields in figures
\ref{fig:condQ}(\aaa, \bbb) and in figure \ref{fig:condatt} in \S\ref{sec:Qs}. If the
conditioning event is physically motivated and if enough statistics are available, the best
conditional approximation is the average of all the events satisfying the condition. Both
requirements are met in \S\ref{sec:Qs}, where Qs have a specific physical meaning, and
several hundreds of thousands of samples are available for each type of structure.

An alternative in more poorly specified cases is linear stochastic estimation (LSE) which
seeks to find the best {\em linear} approximation of the second-order moments of the
statistics of the sample. The details are beyond the scope of this appendix, but a good
introduction of its applications to turbulence is \cite{adr:moi:88}. Stochastic estimation
has a long history in statistics, and is a generalisation of linear least-square
approximation. As such, it is intimately connected with the two-point correlation function,
from which it inherits the geometrical structure and the symmetries. For example, the
optimum LSE flow field in isotropic turbulence, given the velocity vector at one point, is a
vortex ring. In homogeneous shear turbulence, the ring deforms into a hairpin, either
trailing or leading with respect to the flow. In wall bounded turbulence, one of the two
hairpin orientations dominates. A similar evolution can be seen in figure \ref{fig:condatt}, but it
should be remembered that, both in LSE and in the conditional flow, the symmetries or the
result are inherited from the symmetries of the statistics.


\end{document}